\documentclass[
twocolumn,
amsmath,amssymb,
aps,prl
]{revtex4-2}

\usepackage{graphicx}
\usepackage{dcolumn}
\usepackage{bm}

\usepackage[none]{hyphenat}
\usepackage{lipsum}
\usepackage{float}
\usepackage{xcolor}
\usepackage{cancel}
\usepackage[colorlinks=true, 
linkcolor=blue,          
citecolor=blue,        
filecolor=blue,      
urlcolor=blue]{hyperref}

\usepackage[all]{hypcap}

\begin{document}
	
\preprint{APS/123-QED}
	
\title{Quantum vacuum amplification in time-varying media with arbitrary temporal profiles}

\author{A. Ganfornina-Andrades}
\author{J. E. Vázquez-Lozano}
\author{I. Liberal}
\affiliation{Department of Electrical, Electronic and Communications Engineering,\\
Institute of Smart Cities (ISC), Public University of Navarre (UPNA), 31006 Pamplona, Spain.
}
	
\begin{abstract}
In this work we address quantum vacuum amplification effects in time-varying media with an arbitrary time-modulation profile. To this end, we propose a theoretical formalism based on the concept of conjugated harmonic oscillators, evaluating the impact on the transition time in temporal boundaries, shedding light into the practical requirements to observe quantum effects at them. In addition, we find nontrivial effects in pulsed-modulations, where the swiftest and strongest modulation does not lead to the highest photon production. Thus, our results provide key insights for the design of temporal modulation sequences to enhance quantum phenomena.
\end{abstract}

	\maketitle
	\sloppy
	

Time-varying media (often referred to as temporal metamaterials) are emerging as a groundbreaking paradigm for the manipulation of electromagnetic waves \cite{galiffi2022photonics,engheta2023four}. In contrast to traditional approaches based on the spatial engineering of photonic nanostructures, time-varying media rely on the possibility of exploiting time as an additional degree of freedom \cite{engheta2020metamaterials}, thus allowing for an active and dynamical control of the optical properties of matter (typically, dielectric permittivity $\varepsilon$ and/or magnetic permeability $\mu$). Besides enriching the physics by breaking temporal symmetries \cite{engheta2020metamaterials,engheta2023four,caloz2019spacetime,caloz2019spacetime2}, this new approach provides with some important practical benefits, such as access to nonreciprocity, lifting restrictions on energy conservation, and a higher design flexibility and reconfigurability.

Within the context of quantum electrodynamics, time-varying media resonate with the dynamical Casimir effect and other vacuum amplifications effects \cite{nation2012colloquium}, which have been observed at microwave frequencies using superconducting circuits \cite{wilson2011observation,lahteenmaki2013dynamical}. In addition, this technological platform aligns with experiments on transmission line metamaterials (TL-MTMs) \cite{caloz2005electromagnetic,eleftheriades2005negative}, and the recently reported first experimental observation of a temporal boundary at radio frequencies \cite{moussa2023observation}, and momentum band-gaps in time-varying transmission lines \cite{reyes2015observation} and metasurfaces \cite{wang2023metasurface}. The implementation of quantum temporal metamaterials at optical frequencies is expected to be more challenging, as it requires ultrafast temporal scales. Nevertheless, the feasibility for realizing temporal metamaterial concepts is spurring a recent upsurge of experiments on ultra-fast modulation based on doped semiconductors \cite{zhou2020broadband,bohn2021spatiotemporal,tirole2023double}.

On the theoretical ground, time-varying media have inspired the discussion of new quantum optical phenomena \cite{mendoncca2000quantum}, including the directional amplification of vacuum fluctuations with anisotropic temporal boundaries \cite{vazquez2022shaping}, ultra-fast quantum state frequency shifting with antireflection temporal coatings \cite{liberal2023quantum,pacheco2020antireflection}, new amplification schemes \cite{pendry2022photon,Pendry2021gain}, modified quantum light emission from two-level systems \cite{lyubarov2022amplified} and free electrons \cite{dikopoltsev2022light} in photonic time crystals, and spontaneous photon generation in transluminal gratings \cite{horsley2023quantum}.

\begin{figure}[t!]
	\includegraphics[scale=0.36]{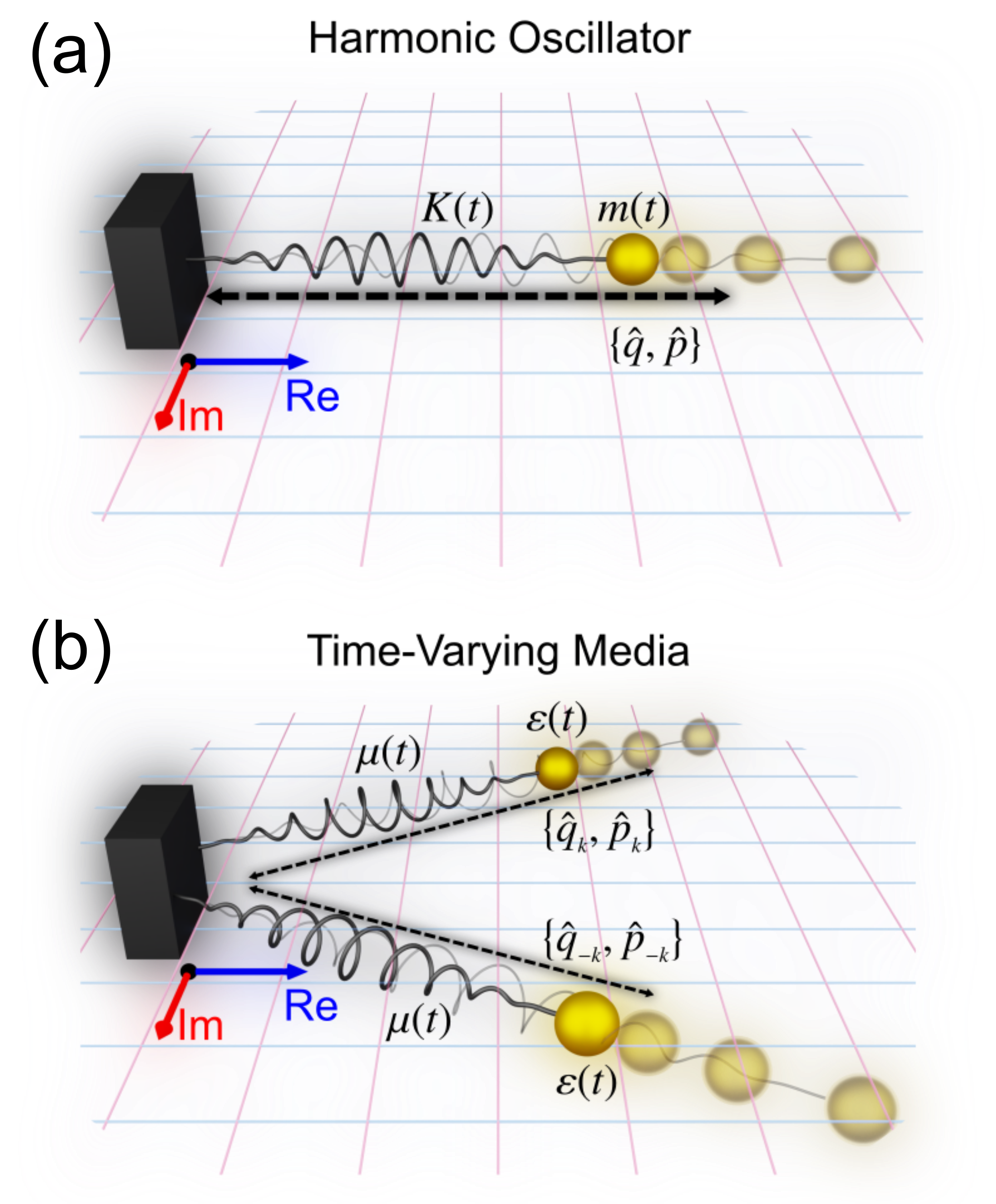}
	\caption{\label{HO_Conjugated_oscillator_picture} (a) Harmonic oscillator vs. (b) time-varying media. In upper panel, operators are forced to move on the real axis. In contrast, on the time-varying media scenario, operators linked to the forward  ($+\bold{k}$) and backward ($-\bold{k}$) waves describe conjugate trajectories in the complex plane. Moreover, $\varepsilon(t)$ and $\mu(t)$ portray the role of the time-varying mass $m(t)$ and the time-dependent spring force $K(t)$, respectively.}
\end{figure}

\indent All these works are based either on instantaneous temporal boundaries \cite{vazquez2022shaping,liberal2023quantum,pacheco2020antireflection} or sinusoidal periodic modulations \cite{pendry2022photon,Pendry2021gain,lyubarov2022amplified,dikopoltsev2022light,horsley2023quantum}. In the present work, we introduce a theoretical formalism based on the concept of time-varying ``conjugated" harmonic oscillators that enables the evaluation of vacuum amplification effects on time-varying media with arbitrary time-modulated dielectric profiles [see \hyperref[HO_Conjugated_oscillator_picture]{Fig. \ref*{HO_Conjugated_oscillator_picture}}]. Additionally, we show how such theoretical framework allow us to tackle important time-varying media quantum mechanical problems such as temporal tapering, exponential relaxation processes, and nontrivial interactions with short pulses. 

\paragraph*{Canonical quantization in time-varying media.-}

We start by considering a linear, isotropic, and homogeneous medium whose optical properties, dielectric permittivity $\varepsilon(t)$ and magnetic permeability $\mu(t)$, vary in time. Without loss of generality, we consider the modes propagating along the z-axis. Therefore, the Lagrangian $L=\frac{1}{2}\int dV\,\left[\varepsilon_{0}\varepsilon\left(t\right)\left(\partial_{t}A_{x}\left(z,t\right)\right)^{2}-{1}/({\mu_{0}\mu\left(t\right)})\left(\partial_{z}A_{x}\left(z,t\right)\right)^{2}\right]$ is described by the  dynamical variable $A_x(z,t)$ and its conjugated momentum $\Pi_x(z,t)=\partial\mathcal{L}/\partial\dot{A_{x}}(z,t)$. Thus, fields are linked with these canonical variables as $\partial_z{{{A}}}_x(z,t)={{{B}}}_y(z,t)$ and ${\Pi}_x(z,t)=-{{{D}}}_x(z,t)$. Quantization is encompassed by promoting the canonical variables and their conjugate momenta into operators obeying equal-time commutation relations $\left[\widehat{A}_{x}\left(z,t\right),\widehat{\Pi}_{x}^{\dagger}\left(z',t\right)\right]=i\hslash\,\delta\left(z-z'\right)$. Hereunder, it is convenient to carry out a plane-wave decomposition: 
$\widehat{{{A}}}_x(z,t)= (1/\sqrt{\varepsilon_0})\int dk \,\,\widehat{q}_k(t)e^{ikz}+\text{h.c}$ 
and 
$\widehat{{{\Pi}}}_x(z,t)=\sqrt{\varepsilon_0}\int dk \,\,\widehat{p}_k(t)e^{ikz}+\text{h.c}$. In addition, since both quantities are observables, the reality condition \cite{cohen1986quantum} $\widehat{p}_k(t)=\widehat{p}^{\dagger}_{-k}(t)$ and  $\widehat{q}_k(t)=\widehat{q}^{\dagger}_{-k}(t)$  shall be satisfied, leading to the dynamics illustrated in \hyperref[HO_Conjugated_oscillator_picture]{Fig. \ref*{HO_Conjugated_oscillator_picture}}. Finally, after some mathematical rearrangements (see Supplementary Material), the Hamiltonian of the electromagnetic field in time-varying media can be compactly written as:
\begin{align}
\nonumber\widehat{\mathcal{H}}(t)&=\frac{1}{2}\int_{0}^{\infty}dk\frac{1}{\varepsilon(t)}\big(\widehat{p}_k(t)\widehat{p}^{\dagger}_k(t)+\widehat{p}^{\dagger}_k(t)\widehat{p}_k(t)\big)\\
&+\frac{k^2c^2}{\mu(t)}\big(\widehat{q}_k(t)\widehat{q}^{\dagger}_k(t)+\widehat{q}^{\dagger}_k(t)\widehat{q}_k(t)\big)=\int_{0}^{\infty}dk\widehat{\mathcal{H}}_k(t). 
\label{Quantum_time_varying_media_Hamiltonian}
\end{align}

\paragraph*{Conjugated harmonic oscillators.-}

The Hamiltonian (\ref{Quantum_time_varying_media_Hamiltonian}) could be alternatively written in terms of the usual creation and destruction operators (see Supplementary Materials). However, its present form is convenient as $\widehat{\mathcal{H}}_k(t)$ presents clear similarities with the Hamiltonian of a harmonic oscillator (HO) with time-varying mass $m(t)$, spring constant $K(t)$, and frequency $\omega(t)=\sqrt{K(t)/m(t)}$
\begin{equation}
\widehat{\mathcal{H}}(t)=\frac{\widehat{p}^2(t)}{2m(t)}+\frac{1}{2}m(t)\omega^2(t)\widehat{q}^2(t).
\label{Quantum_harmonic_oscillator_Hamiltonian}
\end{equation}

By inspecting Eqs.\,(\ref{Quantum_time_varying_media_Hamiltonian}) and (\ref{Quantum_harmonic_oscillator_Hamiltonian}) one can identify the analogies, providing additional insight into the dynamics of the system. This analogy is not surprising, since tutorial introductions to the quantization of electromagnetic fields are usually described in terms of cavity modes acting as harmonic oscillators \cite{loudon2000quantum,Miller2008quantum}. At the same time, both systems actually present important differences. First, while the position and momentum operators $\widehat{p}$ and $\widehat{q}$ of the HO are Hermitian operators, the $\widehat{p}_k$  and $\widehat{q}_k$ represent complex field amplitudes and they are not necessarily Hermitian. In fact, a Hermitian Hamiltonian is only recovered through the reality condition  $\widehat{p}_k(t)=\widehat{p}^{\dagger}_{-k}(t)$ and  $\widehat{q}_k(t)=\widehat{q}^{\dagger}_{-k}(t)$. As shown in \hyperref[HO_Conjugated_oscillator_picture]{Fig. \ref*{HO_Conjugated_oscillator_picture}}, the time-varying media Hamiltonian is depicted by two conjugated oscillators, i.e., the operators linked to both modes, the forward wave ($+\bold{k}$) and backward wave ($-\bold{k}$), describe complex conjugate trajectories. This model highlights that while the forward ($+\bold{k}$) and backward ($-\bold{k}$) modes are seemly independent, their dynamics are actually deeply intertwined, to the point that the Hamiltonian can be effectively written in terms of the forward modes only. This representation contrasts with the time-varying HO in  (\ref{Quantum_harmonic_oscillator_Hamiltonian}), differentiated just by the absence of an explicit distinction of forward-backward mechanism and the constrain of both $\widehat{q}$ and $\widehat{p}$ to move on exclusively along the real axis. 

\paragraph*{SU(2) Lie algebra.-}
The analogy between time-varying media and time-varying HOs can be more rigorously derived by comparing the algebra that determines their time evolution. As known for the time-varying HO \cite{cheng1988evolution,baskoutas1992study}, it can be demonstrated (see Supplementary Material) that $\widehat{\mathcal{H}}_k(t)$ also obeys a SU(2) Lie algebra. Thus, both Hamiltonians (\ref{Quantum_time_varying_media_Hamiltonian}) and (\ref{Quantum_harmonic_oscillator_Hamiltonian}) can be written as $\widehat{\mathcal{H}}_k(t)=a_1(t)\widehat{J}_+ + a_2(t)\widehat{J}_0 + a_3\widehat{J}_-$, where $\widehat{J}_+$, $\widehat{J}_0$, and $\widehat{J}_-$ are total angular momentum SU(2) group operators, obeying the commutation relations $\left[\widehat{J}_{+},\widehat{J}_{-}\right]=2\widehat{J}_{0}$ and $\left[\widehat{J}_{0},\widehat{J}_{\pm}\right]=\pm\widehat{J}_{\pm}$. The time-dependent terms are given by  $a_1(t)=\hbar\varepsilon(t)\omega_k^2(t)$, $a_2(t)=0$ and $a_3(t)=\hbar/\varepsilon(t)$ for time-varying media in (\ref{Quantum_time_varying_media_Hamiltonian}), whereas they are given by $a_1(t)=\hbar m(t)\omega^2(t)$, $a_2(t)=0$ and $a_3(t)=\hbar/m(t)$ for the generalized parametric oscillator (\ref{Quantum_harmonic_oscillator_Hamiltonian}) with both time-dependent mass and frequency. Thus, it can be established an analogy, between the time-varying mass and permittivity, 
$m(t)\Rightarrow \varepsilon(t)$, and time-varying spring force and permeability, 
$K(t)=m(t)\omega^2(t)\Rightarrow k^2c^2/\mu(t)$, that holds even in reproducing the dynamics of the system.

There has been an extensive body of work devoted to obtaining precise analytical solutions of the time-varying HO based on its SU(2) Lie algebra \cite{cheng1988evolution,baskoutas1992study,husimi1953miscellanea,popov1969parametric,dodonov2018coherent,
gerry1990evolution,pedrosa1997exact,de2008quantum,colegrave1982harmonic}. 
However, these descriptions often result in limitations, such as narrow scope applications \cite{ben2008harmonic} (e.g., relying on adiabatic approximations), or integral solutions with oscillatory functions in the denominator, i.e., prone to rapid divergence issues \cite{cheng1988evolution,baskoutas1992study,pedrosa1997exact} in cases where both mass and frequency (consequently, $\varepsilon(t)$ and $\mu(t)$ in our case), are time-modulated functions that exhibit a significant non-linear behavior. For all these reasons, we address the problem under minimal assumptions, and directly from Heisenberg equations of motion.

\paragraph*{Heisenberg equations.-}
Heisenberg equations of motion yield to evolution of both position ${q}_k(t)$ and momentum ${p}_k(t)$ operators:
\begin{align}
\frac{d\widehat{q}_k(t)}{dt}=\frac{1}{\varepsilon(t)}\,\,\,\widehat{p}_k(t) 
\qquad\frac{d\widehat{p}_k(t)}{dt}=-\frac{k^2c^2}{\mu(t)}\,\,\,\widehat{q}_k(t). 
\label{operators_evolution}
\end{align}
These are two coupled differential equations that could be solved in an analytical closed-form for some particular time-modulations, but that can be straightforwardly solved numerically in a large number of cases. To this end, we first
discretize the time $t_{n}=n\Delta t$ from 0 to $t$ by N steps of $\Delta t=t/N$, so that we obtain the following input-output relation:

\begin{align}
	\left[\begin{array}{c}
		\widehat{q}_k\left(t\right)\\
		\widehat{p}_k\left(t\right)
	\end{array}\right]
	=\mathbf{M}\left(t\right)
	\left[\begin{array}{c}
		\widehat{q}_k\left(0\right)\\
		\widehat{p}_k\left(0\right)
	\end{array}\right],
	\label{eq:input_output_relation}
\end{align}

\noindent where
\begin{align}
\mathbf{M}\left(t\right)=
\left[\begin{array}{cc}
	M_{11}\left(t\right) & M_{12}\left(t\right)\\
	M_{21}\left(t\right) & M_{22}\left(t\right)
\end{array}\right]
=
\prod_{n=1}^{N}\mathbf{M}_\Delta\left(n\triangle t\right),
\end{align}

\noindent and 
\begin{align}
	\mathbf{M}_\Delta\left(n\triangle t\right)
	=\left[\begin{array}{cc}
		1 & \frac{\varDelta t}{\varepsilon(n\Delta t)}\\
		\frac{-\Delta t k^2c^2}{\mu(n\Delta t)} & 1
	\end{array}\right].
\end{align}

Equation\,(\ref{eq:input_output_relation}) describes the dynamics of the position and momentum operators as a function of the initial-time operators. Similarly, any quantum statistic at time $t$ can be written as a function of initial-time statistics. For example, the time-evolution of the energy expectation value for a system initially in its vacuum state can be explicitly written as (see Supplementary Material):

\begin{align}
	\nonumber\langle\widehat{\mathcal{H}}_{k}\left(t\right)\rangle&=\frac{\hslash\omega_{k}(0)}{2}\left\{ \frac{\mu\left(0\right)}{\mu(t)}\,\left|M_{11}\left(t\right)\right|^{2}+\frac{\varepsilon\left(0\right)}{\varepsilon\left(t\right)}\,\left|M_{22}\left(t\right)\right|^{2}\right.
	\\\
	&\left.+\frac{\mu\left(0\right)}{\varepsilon\left(t\right)\left(kc\right)^{2}}\,\left|M_{21}\left(t\right)\right|^{2}+\frac{\varepsilon\left(0\right)\left(kc\right)^{2}}{\mu(t)}\,\left|M_{12}\left(t\right)\right|^{2}\right\}
	\label{Expected_energy}
\end{align}

\indent We use this formalism to address photon production mechanism corresponding to temporal sequences of recent significance. Due to the inherently challenging task of addressing (theoretically or experimentally) a simultaneous controlled modulation of $\varepsilon(t)$ and $\mu(t)$, we focus on purely dielectric time-modulations. 

\paragraph*{Tapered photonic switching.-}
The most elemental unit of a temporal metamaterial is a temporal boundary, i.e., an abrupt (theoretically instantaneous) switch in optical properties at a localized instant. Early studies \cite{morgenthaler1958velocity,felsen1970wave,fante1971transmission,fante1973propagation,xiao2014reflection} unveiled that  temporal boundaries arise on a mixing of forward and backward waves, exhibiting frequency shifts with respect to the waves preceding the temporal boundary. Photon production from quantum vacuum amplification effects at temporal boundaries have also been investigated \cite{vazquez2022shaping,liberal2023quantum,horsley2023quantum,lyubarov2022amplified,mendoncca2000quantum}. Nevertheless, the transition times $\tau$ of any practical implementation of a temporal boundary are unavoidably finite, expecting to have an impact in both scattering and photon production effects. 

\begin{figure}[h!]
	 \centering
	\includegraphics[scale=0.47]{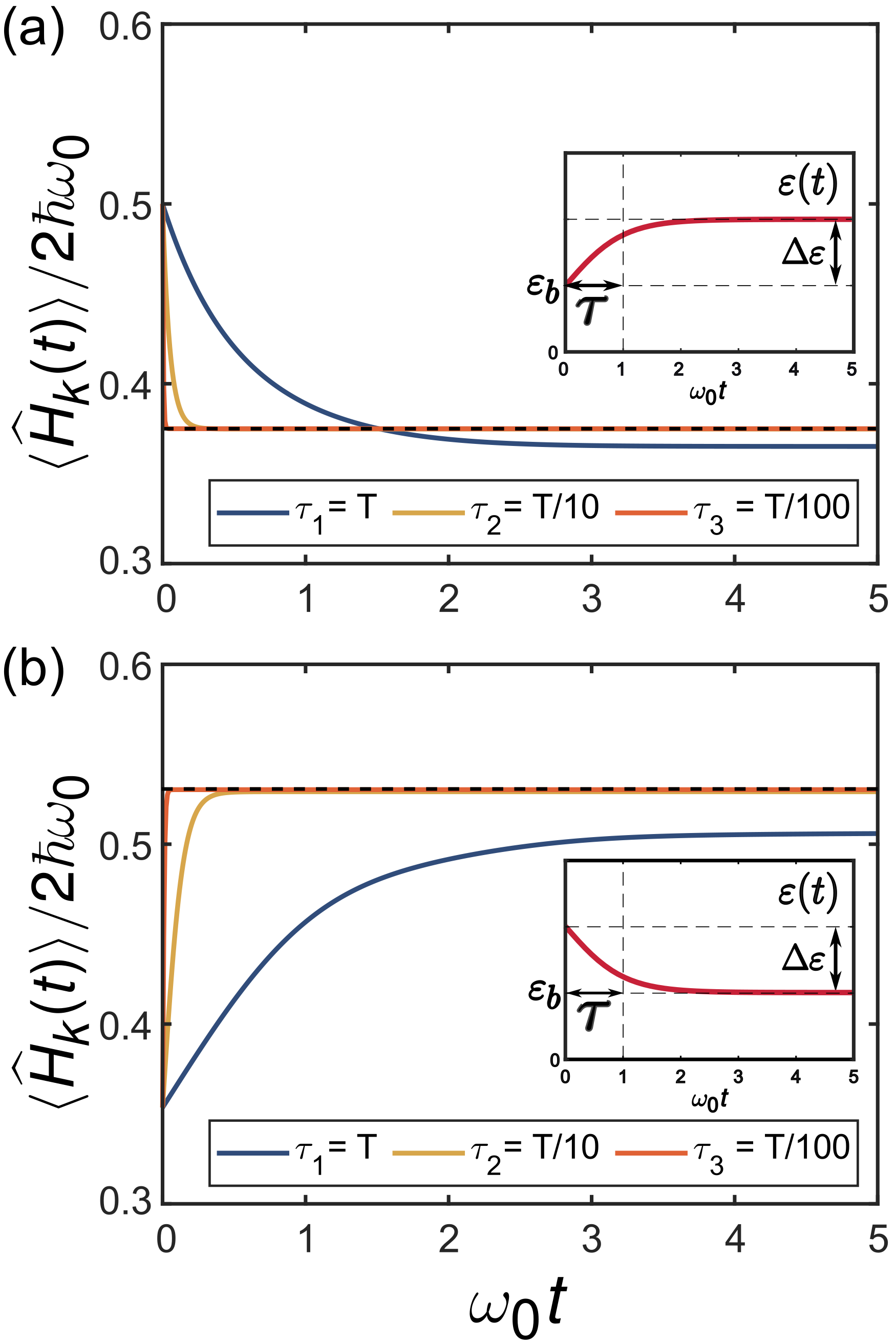}
	\caption{\label{Tapered_photonic_switch_and_dielectric_relaxation} Photon production with respect to vacuum energy for (a) photonic tapered switching, and (b) dielectric relaxation. Time-modulated $\varepsilon(t)$ ($\varepsilon_b=\Delta\varepsilon=1$) are depicted as red curves in their insets. Long characteristic time $\tau_1$ (in blue) mimics the behavior of adiabatic limit $\hbar\omega_k(t)$, leading to expected energies below the temporal boundary prediction in both cases. Faster modulations, $\tau_2$ (in yellow) and $\tau_3$ (in orange), tend quickly to the theoretical limit predicted for a temporal boundary (dashed horizontal line).}
\end{figure}

The classical response of a temporal boundary with a finite response time was recently addressed under the name of tapered photonic switching \cite{galiffi2022tapered}. Here, we use our theoretical framework to address quantum photon production in tapered photonic switching, elucidating the impact of a nonzero transition time. To this end,  we use a time-modulation of the permittivity following  $\varepsilon(t)=\varepsilon_b+\Delta\varepsilon\tanh(t/\tau)$ [\hyperref[Tapered_photonic_switch_and_dielectric_relaxation]{Fig. \ref*{Tapered_photonic_switch_and_dielectric_relaxation}a}]. 

Evaluating the expected energy with respect to the vacuum energy $2\hbar\omega_0/2$, the starting point matches it and two tendencies arise. Firstly, for long transition times $\tau$, the expected energy approaches the adiabatic regime $\hbar\omega_k(t)$ leading consequently to the absence of photon production via quantum vacuum amplification, as the system evolves slowly enough so that the $k$ modes simply adiabatically shift in frequency. In contrast, for ultra-fast modulations the results tend toward the result for a temporal boundary \cite{liberal2023quantum,vazquez2022shaping} $ 2\hbar\omega_0\left(\mathrm{sinh}^{2}\left(s{_{21}}\right)+1/2\right)$, where $s_{21}=\log[(\varepsilon(0)/\varepsilon(\infty))^{1/4}]$, highlighting the requirement of fast modulations for photon production via quantum vacuum amplification.
 
\paragraph*{Exponential relaxation.-}
The inverse process to tapered photonic switching is that of dielectric relaxation $\varepsilon(t)=\varepsilon_b+\Delta\varepsilon\exp(-t/\tau)$ [\hyperref[Tapered_photonic_switch_and_dielectric_relaxation]{Fig. \ref*{Tapered_photonic_switch_and_dielectric_relaxation}(b)}], which is of interest as it can model several material processes. In this case, the starting point is below the final vacuum energy since $\omega_k(0)<\omega_0$, but the adiabatic limit is still recovered for long transition times, and fast modulations lead to the predicted value of an abrupt temporal boundary to a lower dielectric permittivity.

The findings of these examples can be encapsulated under the intuitive notion of the swifter the modulation, the larger amount of photons are produced. Furthermore, for the studied parameters, a range of transition times  $T/4\leq\tau\leq T/10$ are sufficient to mimic a temporal boundary with great accuracy (see Supplementary Material for additional numerical examples for different configurations). Therefore, our study concludes that transition times smaller but comparable to the period are sufficient for implementing temporal boundaries in terms of quantum vacuum amplification effects.

\paragraph*{Finite-time pulsed modulations.-}
Next, we address the response of finite-time pulsed modulations. This example is of particular practical relevance, since many practical configurations would take advantage of discrete pulsed excitations, resulting in a time-localized increase (or decrease) of the dielectric permittivity of the whole medium. Interestingly, the very intuitive notion that photon production increases along with the abruptness of the modulation is challenged by finite-duration pulses.
 
\begin{figure}[t!]
	\centering
	\includegraphics[width=\columnwidth]{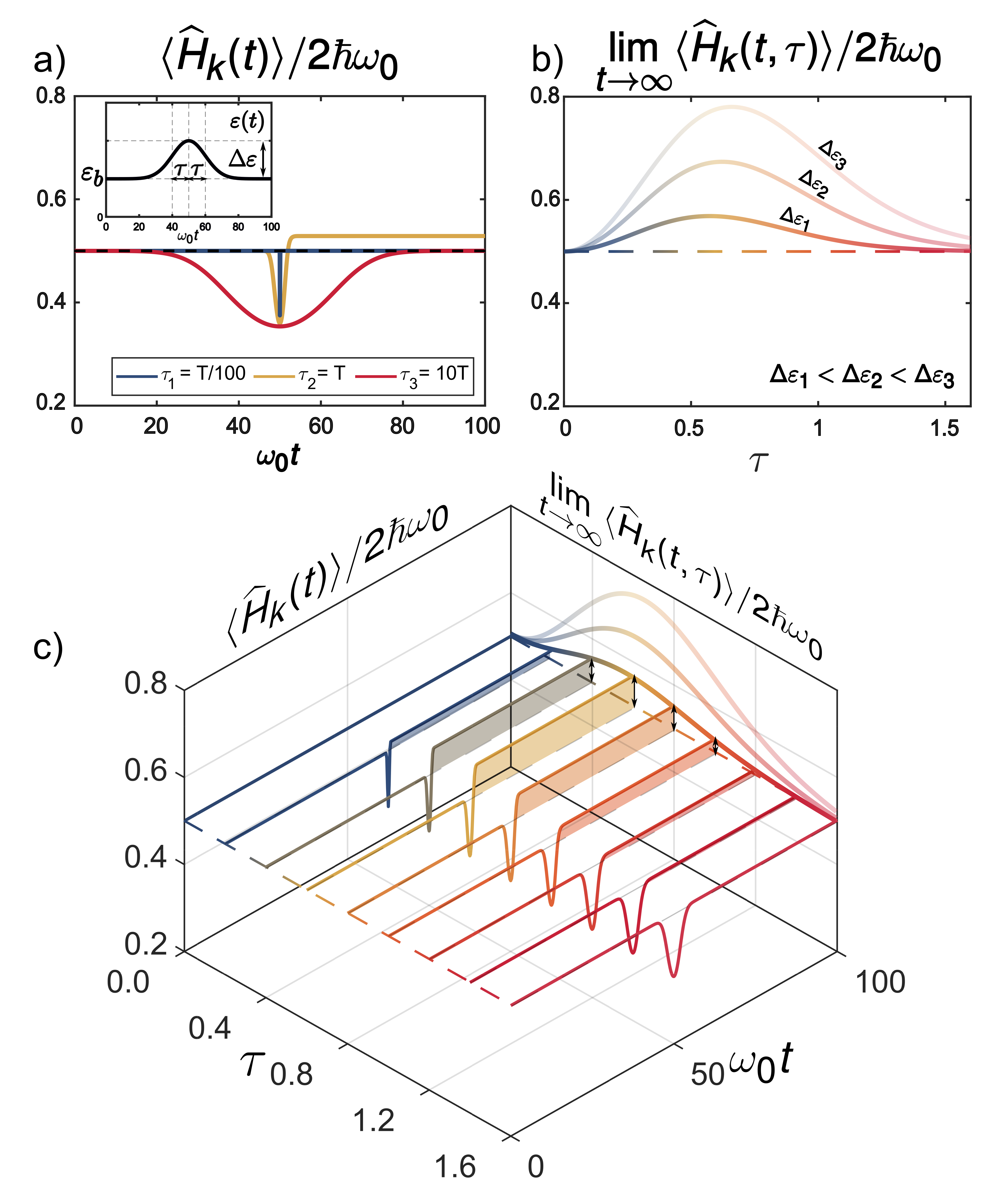}
	\caption{\label{Gaussian_pulse} Photons production for several Gaussian profiles. (a) Slow pulse $\tau_1$ (red) reverts the system energy to the initial state, short pulse $\tau_2$ (yellow) leads to a final excitation of energetic states, and ultra-short pulse $\tau_3$ (blue) approaches the limit of $\tau\rightarrow0$. Gaussian profile $\varepsilon(t)$ ($\varepsilon_b=\Delta\varepsilon=1$) is included as an inset. (b) Photon production distribution for long times after the pulse: the peak of production is achieved for a certain $\tau$ which shifts for increasing values as the pulse amplitude $\Delta\varepsilon$ also increases ($\Delta\varepsilon_1=1<\Delta\varepsilon_2=2<\Delta\varepsilon_3=3$). (c) Photons production in terms of $\omega_0t$ and $\tau$: a final excited energy state is observed for long times ($\varepsilon_b=1$ and $\Delta\varepsilon=1$), which finds a maximum around $\tau=0.57T$. }
\end{figure}

As the most representative example, we address a time-modulated dielectric function that obeys a Gaussian profile, $\varepsilon(t)=\varepsilon_b+\Delta\varepsilon\exp[-(t-t_0)^2/(2\tau^2)]$, where the initial state $\varepsilon_b$ is raised up to $\Delta\varepsilon$ by pulse, on a localized time $t_0$, with a tunable duration $\tau$. Numerical results for the time evolution of the energy of the system are depicted in \hyperref[Gaussian_pulse]{Fig. \ref*{Gaussian_pulse}}, concluding that the response of the systems to a Gaussian time-modulated profile is characterized by three different regimes. 

The first regime corresponds to a slow, wide width, Gaussian profile. The dielectric function is increased, and then decreased, through a slowly-modulated Gaussian process. As in the previous examples, the slow temporal modulation mimics the adiabatic limit, inasmuch the energy of the whole system simply follows the frequency shift, while reverting initial vacuum energy as the pulse passes [$\tau_1$ curve in \hyperref[Gaussian_pulse]{Fig. \ref*{Gaussian_pulse}(a)}], with no photon production.

The second one corresponds to pulses of vanishingly small duration ($\tau\rightarrow0$), intensely localized, approaching a delta-like distribution $\varepsilon(t)=\varepsilon_b+\Delta\varepsilon\cdot\delta(t-t_0)$ . Once again, the starting point of the expected energy is the vacuum energy, but unlike the cases of tapered photonic switching or exponential relaxation, these results reveal that a faster modulation no longer leads into the highest photon production for pulsed modulations, as a result of the Delta function properties. First, the continuity of $\mathbf{D}$ and $\textbf{B}$ can be derived from Maxwell–Faraday $\nabla\times\mathbf{E}=-\partial_t\textbf{B}$ and Ampère-Maxwell equations $\nabla\times\mathbf{H}=\partial_t[(\varepsilon_b+\Delta\varepsilon\cdot\delta(t-t_0))\textbf{E}]$. In addition, for Delta-like time-modulated barrier, we can use the shifting property $\int dt\thinspace f(t)\thinspace\delta(t-t_{0})\equiv f(t_{0})$ and the definition of the derivative of the delta distribution $\int dt\thinspace f(t)\thinspace\delta^{(n)}(t-t_{0})\equiv-\int dt\thinspace{\partial_{t}f(t)}\thinspace\delta^{(n-1)}(t-t_{0})$, to also demonstrate the continuity of $\textbf{E}$ and $\textbf{H}$, justifying classically the recovery of the initial energy state for a narrow dielectric pulsed modulation [$\tau_3$ curve in \hyperref[Gaussian_pulse]{Fig. \ref*{Gaussian_pulse}(a)}] and thus, no  photons are produced from the quantum vacuum for ultra-short pulses. Additional numerical examples showing the convergence to a delta-function response with equal-area pulses is also reported in the Supplementary Material. 

Between these two illustrative extreme regimes, short but finite-time pulses do lead to photon production. This effect can be appreciated in the fact that the final state results in an energy increase with respect to the vacuum state [$\tau_2$ curve in \hyperref[Gaussian_pulse]{Fig. \ref*{Gaussian_pulse}(a)}]. It can be then concluded from our results that there is an optimal pulse duration that maximizes photon production. Therefore, it is found that for pulses with a finite duration, the fastest and more abrupt modulation does not lead to the higher photon production, as shown for equal step and equal area pulses, in contrast to temporal boundaries. The position of this non-trivial maximum in terms of characteristic time $\tau$ is tracked down in \hyperref[Gaussian_pulse]{Fig. \ref*{Gaussian_pulse}(b)}, fitting on a distribution proportional to $\tau^4/(\exp(\omega_0\tau)-1)$ whose maximum shifts for increasing $\tau$ as $\Delta\varepsilon$ also increases.

\begin{figure}[h!]
	\centering
	\includegraphics[width=\columnwidth]{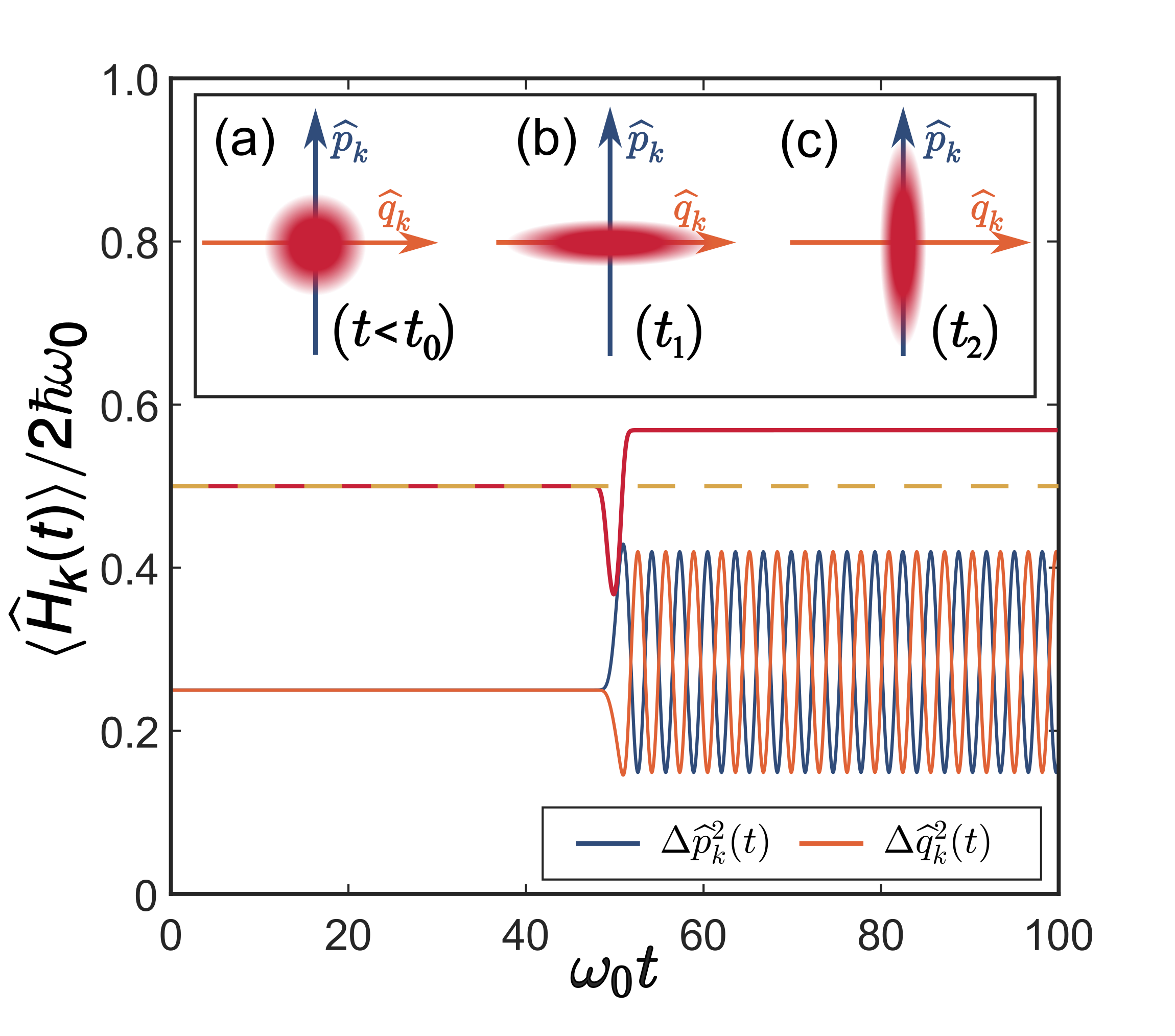}
	\caption{\label{Noise_driven_oscillator} Enhancement of expected energy for a Gaussian time-modulation profile,  ($\varepsilon_b=\Delta\varepsilon=1$), and $\tau=0.57T$ (red curve), with respect to the vacuum energy (yellow dashed line). Before the pulse (a), the equal valued variances describe a vacuum state. After the pulse, on (b) and (c): the variances of momentum operator (blue curve) and position operator (orange curve), associated with electric displacement field $\mathbf{D}$ and magnetic field $\textbf{B}$, respectively, evolve following an oscillatory counter-phase behavior. Adding up both numerically computed variances, the expected energy after the pulse reaches a steady state of $\langle \widehat{\mathcal{H}}_{k}\left(t,\tau\right)\rangle =0.5682\hbar\omega_{0}.$}
\end{figure}

These results are shown in conjunction on \hyperref[Gaussian_pulse]{Fig. \ref*{Gaussian_pulse}(c)}, where the photons production is illustrated in terms of $\omega_0t$ for several pulse widths $\tau$, thereby emphasizing the existence of a final state of maximum energy related with a non-trivial election of this latter parameter. Besides, contrasting with the previous profiles, since shorter pulses lead no longer to the highest expected energy, but a maximum for $\tau$ relying on $\Delta\varepsilon$/$\varepsilon_b$ ratio, it suggests that the design of time-modulation sequences is required in order to maximize photon production from quantum vacuum amplification.
\paragraph*{Noise driven oscillator.-} The analogy with the time-varying HO also provides an intuitive picture of the field dynamics in time-varying media after the interaction with a short pulse, which results into squeezed vacuum states. Exerting energy on the system, via increasing or decreasing the dielectric permittivity (i.e., the mass of the conjugated HO) with a Gaussian profile, one would be tempted to conclude that the HO has been set in motion, so that both position (magnetic field $\mathbf{B}$) and momentum (electric displacement field $\mathbf{D}$) oscillate in time.  However, the expected values of position and momentum are zero
$\langle \widehat{q}_k(t)\rangle=\langle\widehat{p}_k(t)\rangle=0$. 
In other words, expected fields dynamics are centered and so the global picture does not correspond to an oscillating mass in an usual harmonic oscillator. Besides, evaluating the respective variances show that they are nonzero, $\Delta\widehat{q}_k^2(t)\neq 0$ and $\Delta\widehat{p}_k^2(t)\neq 0$. On the contrary, the variances of $\widehat{q}_k(t)$ and $\widehat{p}_k(t)$ oscillate in time in a periodic fashion around the usual 1/4 value, as illustrated in \hyperref[Noise_driven_oscillator]{Fig. \ref*{Noise_driven_oscillator}(a)}. This is the usual behavior for a squeezed vacuum state \cite{loudon2000quantum}. Therefore, the state of the electromagnetic field after the interaction with the short pulse can be understood as that of a noise-driven oscillator, where the $\mathbf{D}$ and $\mathbf{B}$ fields have zero average amplitude, and their nonzero variances oscillate periodically with points of minimal and maximum uncertainty [see $t_1$ in  \hyperref[Noise_driven_oscillator]{Fig. \ref*{Noise_driven_oscillator}(b)} and $t_2$ \hyperref[Noise_driven_oscillator]{Fig. \ref*{Noise_driven_oscillator}(c)}]. 
\paragraph*{Conclusions.-}

We addressed photon production for temporal metamaterials with arbitrary time-modulated optical profiles. We have presented a general method based on the concept of conjugated harmonic oscillators that is not restricted to instantaneous and/or periodic modulations, and it provides additional physical insight. We used such method to clarify the impact of a finite transition times in temporal tapering. Intuitively, it is shown that the fastest the temporal tapering the larger the photon production from quantum vacuum amplification effects. At the same time, it is found that transition times smaller but on the same order of magnitude than the period already converge to an instantaneous temporal boundary, revealing what are the necessary time-scales to mimic systems modeled with instantaneous temporal boundaries. Our results also show that finite-time pulse modulations show a qualitatively different behavior: there is a non-trivial optimal width $\tau$ that maximize photon production, as justified by the existence of adiabatic and delta-function limits with zero photon production. Therefore, the highest photon production is not always obtained with the fastest modulation, and our results motivate further research in the design of temporal sequences that maximize quantum light production effects in ultra-fast temporal metamaterials.

This work was supported by ERC Starting Grant No. ERC-2020-STG948504-NZINATECH. J.E.V.-L. acknowledges support from Juan de la Cierva–Formación fellowship FJC2021-047776-I. I.L. further acknowledges support from Ramón y Cajal fellowship RYC2018-024123-I.
\newpage

\bibliography{apssamp}

\clearpage
\appendix
\onecolumngrid
\renewcommand\theequation{S\arabic{equation}}
\setcounter{equation}{0}
\renewcommand\thepage{S.\arabic{page}}
\setcounter{page}{1}
\setcounter{figure}{0}
\renewcommand{\thefigure}{S\arabic{figure}}

\begin{center}
	{\large \textbf{Supplemental Material:\\Quantum vacuum amplification in time-varying media with arbitrary temporal profiles}}
\end{center}
\begin{center}
	A. Ganfornina-Andrades, J. E. Vázquez-Lozano, I. Liberal\\
	{\small\textit{Department of Electrical, Electronic and Communications Engineering,\\
			Institute of Smart Cities (ISC), Public University of Navarre (UPNA), 31006 Pamplona, Spain.}}\\
\end{center}
\begin{quotation}
In this Supplemental Material, we provide further details in the derivation of the theoretical formulation of photons production for arbitrary time-modulated dielectric profiles. In Part I, we develop time-varying media dynamics directly from first principles, starting from Maxwell equations and proceeding through the Lagrangian, we derive the corresponding Hamiltonian of such system. In part II, we address the new notion of conjugated oscillators with time-varying parameters, promoting the dynamical variables into quantum operators. In Part III, we prove that the quantum Hamiltonian of a generalized parametric oscillator and the aforementioned Hamiltonian also satisfies SU(2) Lie Algebra, establishing the analogy between establishing the analogy between both time-dependent mass and spring force with time-modulated optical constitutive parameters. In Part IV, using Heisenberg equations of motion and numerical iterative methods, we explain two particular cases: photonic tapered switching and a Gaussian dielectric perturbation.
\end{quotation}
\part{Part I: Formalism in time-varying media.}
In this part, we first address field equations for 1D polarized electromagnetic wave (EM) in a time-varying media. Under the assumption of a free-charge space, the vector potential and wave equation are derived. Consequently, identifying the vector potential as the dynamical variable of such a system, the Lagrangian along with the conjugate momentum and the Hamiltonian are formulated. Finally, we check the wave equation directly by means of both Euler-Lagrange and Hamilton equation of motions.
\section{I. FIELD EQUATIONS}
First, we start by setting down the basic framework of the problem, regarding the field equations for a pure-time medium (i.e., a temporal metamaterial). Let us consider a linear, isotropic and homogeneous media whose optical properties, i.e., the dielectric permittivity $\varepsilon(t)$ and magnetic permeability $\mu(t)$, vary in time. To simplify the calculation, and without loss of generality, we address the problem for one polarization with the modes (k-vector) propagating along the z-axis, so that the fields are given by:

\begin{align}
\mathbf{E}\left(z,t\right)=\widehat{\mathbf{x}}\,E_{x}\left(z,t\right),
\label{S1}
\end{align}
\begin{align}
\mathbf{H}\left(z,t\right)=\widehat{\mathbf{y}}\,H_{y}\left(z,t\right).
\label{S2}
\end{align}

Accordingly, the curl of the fields are re-written in terms of their spatial derivatives,

\begin{align}
\nabla\times\mathbf{E}\left(z,t\right)=\widehat{\mathbf{y}}\,\partial_{z}E_{x}\left(z,t\right),
\label{SM3}
\end{align}
\begin{align}
	\nabla\times\mathbf{H}\left(z,t\right)=-\widehat{\mathbf{x}}\,\partial_{z}H_{y}\left(z,t\right).
	\label{SM4}
\end{align}
Thus, magnetic flux density $\boldsymbol{B}$ and the electric displacement $\boldsymbol{D}$ are:

\begin{align}
	\mathbf{D}\left(z,t\right)=\widehat{\mathbf{x}}D_{x}\left(z,t\right)=\widehat{\mathbf{x}}\left(\varepsilon_{0}\varepsilon\left(t\right)E_{x}\left(z,t\right)\right),
	\label{SM5}
\end{align}
\begin{align}
	\mathbf{B}\left(z,t\right)=\widehat{\mathbf{y}}B_{y}\left(z,t\right)=\widehat{\mathbf{y}}\left(\mu_{0}\mu\left(t\right)H_{y}\left(z,t\right)\right),
	\label{SM6}
\end{align}

and Maxwell curl equations are clearly identified:

\begin{align}
	\nabla\times\mathbf{E}\left(z,t\right)=-\partial_{t}\mathbf{B}\left(z,t\right)\rightarrow\partial_{z}E_{x}\left(z,t\right)=-\partial_{t}\left(\mu_{0}\mu\left(t\right)H_{y}\left(z,t\right)\right)
	\label{SM7}
\end{align}
\begin{align}
	\nabla\times\mathbf{H}\left(z,t\right)=\partial_{t}\mathbf{D}\left(z,t\right)\rightarrow-\partial_{z}H_{y}\left(z,t\right)=\partial_{t}\left(\varepsilon_{0}\varepsilon\left(t\right)E_{x}\left(z,t\right)\right)
	\label{SM8}
\end{align}

\section{II. VECTOR POTENTIAL AND WAVE EQUATION}
We derive the vector potential in the Coulomb gauge $\left(\nabla\cdot\boldsymbol{A}=0\right)$ writing the potentials in terms of instantaneous values of the fields in a free-charge region. Coulomb gauge admits a natural Hamiltonian formulation of the evolution equations of the EM fields interacting with a conserved current (which remains zero in our case of interest). Thus, the only component of vector potential is,

\begin{align}
	\mathbf{A}\left(z,t\right)=\widehat{\mathbf{x}}\,A_{x}\left(z,t\right),
	\label{SM9}
\end{align}

and the fields are then recast as follows,

\begin{align}
	\mathbf{E}\left(z,t\right)=-\partial_{t}\mathbf{A}\left(z,t\right)=-\widehat{\mathbf{x}}\,\partial_{t}A_{x}\left(z,t\right),
	\label{SM10}
\end{align}
\begin{align}
	\mathbf{B}\left(z,t\right)=\nabla\times\mathbf{A}\left(z,t\right)=\widehat{\mathbf{y}}\,\partial_{z}A_{x}\left(z,t\right).
	\label{SM11}
\end{align}

Field expression for $D_{x}\left(z,t\right)$ as a function of the vector potential can be straightforwardly derived from (\ref{SM5}) and (\ref{SM10}):

\begin{align}
	D_{x}\left(z,t\right)=\varepsilon_{0}\varepsilon\left(t\right)E_{x}\left(z,t\right),
	\label{SM12}
\end{align}
\begin{align}
	D_{x}\left(z,t\right)=-\left(\varepsilon_{0}\varepsilon\left(t\right)\right)\partial_{t}A_{x}\left(z,t\right).
	\label{SM13}
\end{align}

The same procedure applies to $B_{y}\left(z,t\right)$, using (\ref{SM6}) and (\ref{SM11}):

\begin{align}
	B_{y}\left(z,t\right)=\mu_{0}\mu\left(t\right)H_{y}\left(z,t\right),
	\label{SM14}
\end{align}
\begin{align}
	H_{y}\left(z,t\right)=\frac{1}{\mu_{0}\mu\left(t\right)}\left(\partial_{z}A_{x}\left(z,t\right)\right).
	\label{SM15}
\end{align}

Thus, the wave equation in terms of $A_{x}\left(z,t\right)$ is derived readily from Ampère-Maxwell equation:

\begin{align}
	\nabla\times\mathbf{H}\left(z,t\right)=\partial_{t}\mathbf{D}\left(z,t\right),
	\label{SM16}
\end{align}
\begin{equation*}
-\partial_{z}\left[\frac{1}{\mu_{0}\mu\left(t\right)}\left(\partial_{z}A_{x}\left(z,t\right)\right)\right]=-\partial_{t}\left[\left(\varepsilon_{0}\varepsilon\left(t\right)\right)\partial_{t}A_{x}\left(z,t\right)\right],
\end{equation*}
\begin{align}
	\boxed{\frac{1}{\mu_{0}\mu\left(t\right)}\partial_{z}^{2}A_{x}\left(z,t\right)=\partial_{t}\left\{ \left(\varepsilon_{0}\varepsilon\left(t\right)\right)\partial_{t}A_{x}\left(z,t\right)\right\}.}
	\label{SM17}
\end{align}
\section{III. LAGRANGIAN FORMALISM}
Regarding $A_{x}\left(z,t\right)$ as the dynamical variable of the system, the Lagrangian reads as follows:
\begin{equation*}
	L=\frac{1}{2}\int dV\,\mathcal{L}=\frac{1}{2}\int dV\,\left(\mathbf{E}\left(z,t\right)\cdot\mathbf{D}\left(z,t\right)-\mathbf{H}\left(z,t\right)\cdot\mathbf{B}\left(z,t\right)\right)
\end{equation*}
\begin{align}
	=\frac{1}{2}\int dV\,\left[\varepsilon_{0}\varepsilon\left(t\right)\left(\partial_{t}A_{x}\left(z,t\right)\right)^{2}-\frac{1}{\mu_{0}\mu\left(t\right)}\left(\partial_{z}A_{x}\left(z,t\right)\right)^{2}\right].
	\label{SM18}
\end{align}
\subsection{Checking the wave equation}
Lagrange equations should recover the wave equation (\ref{SM17})
\begin{align}
	\frac{d}{dt}\,\frac{\partial\mathcal{L}}{\partial\left(\partial_{t}A_{x}\left(z,t\right)\right)}=\frac{\partial\mathcal{L}}{\partial A_{x}\left(z,t\right)}-\partial_{z}\,\frac{\partial\mathcal{L}}{\partial\left(\partial A_{x}\left(z,t\right)\right)}.
	\label{SM19}
\end{align}
\begin{align}
	\frac{d}{dt}\,\left\{ \varepsilon_{0}\varepsilon\left(t\right)\partial_{t}A_{x}\left(z,t\right)\right\} =-\partial_{z}\,\left\{ -\frac{1}{\mu_{0}\mu\left(t\right)}\partial_{z}A_{x}\left(z,t\right)\right\} .
	\label{SM20}
\end{align}
\begin{align}
	\frac{1}{\mu_{0}\mu\left(t\right)}\partial_{z}^{2}\,A_{x}\left(z,t\right)=\partial_{t}\left\{ \varepsilon_{0}\varepsilon\left(t\right)\partial_{t}A_{x}\left(z,t\right)\right\} .
	\label{SM21}
\end{align}
The conjugate momentum is then derived:
\begin{align}
	\Pi_{x}\left(z,t\right)=\frac{\partial\mathcal{L}}{\partial\left(\partial_{t}A_{x}\left(z,t\right)\right)}=\left[\varepsilon_{0}\varepsilon\left(t\right)\partial_{t}A_{x}\left(z,t\right)\right]=-D_{x}\left(z,t\right).
	\label{SM22}
\end{align}
\section{IV. CLASSICAL TIME-VARYING HAMILTONIAN}
From the above, the classical Hamiltonian of a time-varying medium is:
\begin{equation*}
\mathcal{H}=\int dV\,\mathcal{H}=\int dV\,\Pi_{x}\left(z,t\right)\partial_{t}A_{x}\left(z,t\right)-\mathcal{L}=\int dV\,D_{x}\left(z,t\right)E_{x}\left(z,t\right)-\mathcal{L}
\end{equation*}
\begin{align}
=\frac{1}{2}\int dV\,\left(\mathbf{E}\left(z,t\right)\cdot\mathbf{D}\left(z,t\right)+\mathbf{H}\left(z,t\right)\cdot\mathbf{B}\left(z,t\right)\right)=\frac{1}{2}\int dV\,\left(D_{x}\left(z,t\right)E_{x}\left(z,t\right)+B_{y}\left(z,t\right)H_{y}\left(z,t\right)\right).
	\label{SM23}
\end{align}
\begin{align}
	\boxed{\mathcal{H}=\frac{1}{2}\int dV\,\left[\varepsilon_{0}\varepsilon\left(t\right)\left(\partial_{t}A_{x}\left(z,t\right)\right)^{2}+\frac{1}{\mu_{0}\mu\left(t\right)}\left(\partial_{z}A_{x}\left(z,t\right)\right)^{2}\right].}
	\label{SM24}
\end{align}
Despite the compactness of this notation, in order to correctly derive the equations of motion from the Hamiltonian and the wave equation, first we need to rewrite the Hamiltonian in terms of $\Pi_{x}\left(z,t\right)=-D_{x}\left(z,t\right)$ and $\partial_{z}A_{x}\left(z,t\right)=B_{y}\left(z,t\right)$:
\begin{equation*}
	\mathcal{H}=\frac{1}{2}\int dV\,\left(E_{x}\left(z,t\right)D_{x}\left(z,t\right)+H_{y}\left(z,t\right)B_{y}\left(z,t\right)\right)
\end{equation*}
\begin{equation*}
=\frac{1}{2}\frac{1}{\mu_{0}\varepsilon_{0}\varepsilon\left(t\right)\mu\left(t\right)}\int dV\,\left(\mu_{0}\mu\left(t\right)\,D_{x}^{2}\left(z,t\right)+\varepsilon_{0}\varepsilon\left(t\right)\,B_{y}^{2}\left(z,t\right)\right)
\end{equation*}
\begin{align}
	\mathcal{H}\left(t\right)=\frac{1}{2}\int dz\,\left(\frac{\mu_{0}c^{2}}{\varepsilon\left(t\right)}\Pi_{x}^{2}\left(z,t\right)+\frac{\varepsilon_{0}c^{2}}{\mu\left(t\right)}\,\left(\partial_{z}A_{x}\left(z,t\right)\right)^{2}\right).
	\label{SM25}
\end{align}
\subsection{Equations of motion}
Then we have the following set of equations of motion:
\begin{align}
	\partial_{t}A_{x}\left(z,t\right)=\frac{\partial\mathcal{H}}{\partial\Pi_{x}\left(z,t\right)}=\frac{\Pi_{x}\left(z,t\right)}{\varepsilon_{0}\varepsilon\left(t\right)},
	\label{SM26}
\end{align}
\begin{align}
	\partial_{t}\Pi_{x}\left(z,t\right)=-\frac{\partial\mathcal{H}}{\partial A_{x}\left(z,t\right)}+\partial_{z}\,\frac{\partial\mathcal{H}}{\partial\left(\partial_{z}A_{x}\left(z,t\right)\right)}=\frac{\partial_{z}^{2}A_{x}}{\mu_{0}\mu\left(t\right)}.
	\label{SM27}
\end{align}
Equation (\ref{SM26}) recovers the definition for the conjugate momentum (\ref{SM22}):

\begin{align}
	\Pi_{x}\left(z,t\right)=\varepsilon_{0}\varepsilon\left(t\right)\partial_{t}A_{x}\left(z,t\right).
	\label{SM28}
\end{align}
Furthermore, taking the time derivative on (\ref{SM28}), we find that:
\begin{equation*}
\partial_{t}\Pi_{x}\left(z,t\right)=\partial_{t}\left\{ \varepsilon_{0}\varepsilon\left(t\right)\partial_{t}A_{x}\left(z,t\right)\right\} =\frac{\partial_{z}^{2}A_{x}\left(z,t\right)}{\mu_{0}\mu\left(t\right)}.
\end{equation*}
Reorganizing the terms we recover again the wave equation (\ref{SM17}) for $A_{x}\left(z,t\right)$:
\begin{align*}
\partial_{t}\left\{ \varepsilon_{0}\varepsilon\left(t\right)\partial_{t}A_{x}\left(z,t\right)\right\} =\frac{1}{\mu_{0}\mu\left(t\right)}\,\partial_{z}^{2}A_{x}\left(z,t\right).
\end{align*}
\part{Part II: Quantization process}
In this part, in order to derive the quantum version of the Hamiltonian for time-varying media, we promote the dynamical variables into quantum operators, using the corresponding commutation rules and applying Fourier transforms over k-space. Then, due to the symmetry of the problem, the integral expressions are evaluated from $k=0$ up to $k\rightarrow\infty$ and the time-varying media Hamiltonian is finally derived in terms of position and momentum operators. Moreover, a description of this system is stated introducing the notion of conjugated parametric oscillators, i.e., two oscillators whose position and momentum operators move on on the complex plane, describing complex conjugate trajectories.
\section{V. TRANSITION TO QUANTUM SPACE}
The starting point of this derivations is the Hamiltonian of the system given in (\ref{SM25}), which is cast in terms of conjugate canonical variables. For the quantization, transition to quantum momentum space is encompassed by the promotion of the canonical variables $A_{x}\left(z,t\right)$ and $\Pi_{x}\left(z,t\right)$ into operators, which should satisfy the canonical commutation relation $\left[\widehat{A}_{x}\left(z,t\right),\widehat{\Pi}_{x}^{\dagger}\left(z',t\right)\right]=i\hslash\,\delta\left(z-z'\right)$:
\begin{align}
	\boxed{\widehat{\mathcal{H}}\left(t\right)=\frac{1}{2}\int dz\,\left\{ \frac{\mu_{0}c^{2}}{\varepsilon\left(t\right)}\widehat{\Pi}_{x}^{2}\left(z,t\right)+\frac{\varepsilon_{0}c^{2}}{\mu\left(t\right)}\,\left(\partial_{z}\widehat{A}_{x}\left(z,t\right)\right)^{2}\right\} .}
	\label{SM29}
\end{align}
Using the Fourier transform in k-space, those operators can be written as a combination of forward and backward modes:
\begin{align}
	\widehat{A}_{x}\left(z,t\right)=\frac{1}{\sqrt{2\pi}}\,\int dk\,\frac{1}{\sqrt{\varepsilon_{0}}}\,\widehat{q}_{k}\left(t\right)\,e^{ikz}+h.c,
	\label{SM30}
\end{align}
\begin{align}
	\partial_{z}\widehat{A}_{x}\left(z,t\right)=\frac{1}{\sqrt{2\pi}}\,\int dk\,ik\,\frac{1}{\sqrt{\varepsilon_{0}}}\,\widehat{q}_{k}\left(t\right)\,e^{ikz}+h.c,
	\label{SM31}
\end{align}
\begin{align}
	\widehat{\Pi}_{x}\left(z,t\right)=\frac{1}{\sqrt{2\pi}}\,\int dk\,\sqrt{\varepsilon_{0}}\,\widehat{p}_{k}\left(t\right)\,e^{ikz}+h.c.
	\label{SM32}
\end{align}

Moreover, since both quantum operators actually are physical observable, the reality condition shall be satisfied.

\begin{align}
	\widehat{q}_{-k}\left(t\right)=\widehat{q}_{k}^{\dagger}\left(t\right)
	\label{SM33}
\end{align}
\begin{align}
	\widehat{p}_{-k}\left(t\right)=\widehat{p}_{k}^{\dagger}\left(t\right)
	\label{SM34}
\end{align}

\subsection{Conjugated oscillators}
The commutation relations can be carried over to momentum space, by means of the Fourier transform of Dirac's delta $\delta\left(k-k'\right)=\frac{1}{2\pi}\int dz\,e^{i\left(k-k'\right)z}$, thus yielding to:

\begin{align}
	\left[\widehat{A}_{x}\left(z,t\right),\widehat{\Pi}_{x}^{\dagger}\left(z',t\right)\right]=i\hslash\,\delta\left(z-z'\right)=\frac{1}{2\pi}\,\int dk\,\int dk'\,\,e^{ikz}e^{-ik'z'}\left[\widehat{q}_{k}\left(t\right),\widehat{p}_{k'}^{\dagger}\left(t\right)\right],
	\label{SM35}
\end{align}
\begin{align}
	\left[\widehat{A}_{x}\left(z,t\right),\widehat{A}_{x}\left(z',t\right)\right]=0=\frac{1}{2\pi\varepsilon_{0}}\,\int dk\,\int dk'\,\,e^{ikz}e^{ik'z'}\left[\widehat{q}_{k}\left(t\right),\widehat{q}_{k'}\left(t\right)\right],
	\label{SM36}
\end{align}
\begin{align}
	\left[\widehat{\Pi}_{x}\left(z,t\right),\widehat{\Pi}_{x}\left(z',t\right)\right]=0=\frac{\varepsilon_{0}}{2\pi}\,\int dk\,\int dk'\,\,e^{-ikz}e^{-ik'z'}\left[\widehat{p}_{k}\left(t\right),\widehat{p}_{k'}\left(t\right)\right].
	\label{SM37}
\end{align}

So, taking the commutation relation stated between $\widehat{A}_{x}\left(z,t\right)$ and $\widehat{\Pi}_{x}^{\dagger}\left(z',t\right)$, the following commutation relations between $\widehat{q}_{k}\left(t\right)$ and $\widehat{p}_{k}\left(t\right)$ are derived:

\begin{align}
	\left[\widehat{A}_{x}\left(z,t\right),\widehat{\Pi}_{x}^{\dagger}\left(z',t\right)\right]=i\hslash\,\delta\left(z-z'\right)\rightarrow\left[\widehat{q}_{k}\left(t\right),\widehat{p}_{k'}^{\dagger}\left(t\right)\right]=i\hslash\,\delta\left(k-k'\right),
	\label{SM38}
\end{align}
\begin{align}
	\left[\widehat{q}_{k}\left(t\right),\widehat{p}_{k'}\left(t\right)\right]=\left[\widehat{q}_{k}\left(t\right),\widehat{p}_{-k'}^{\dagger}\left(t\right)\right]=0,
	\label{SM39}
\end{align}

\begin{align}
	\left[\widehat{A}_{x}\left(z,t\right),\widehat{A}_{x}\left(z',t\right)\right]=0\rightarrow\left[\widehat{q}_{k}\left(t\right),\widehat{q}_{k'}\left(t\right)\right]=0,
	\label{SM40}
\end{align}
\begin{align}
	\left[\widehat{q}_{k}\left(t\right),\widehat{q}_{k'}^{\dagger}\left(t\right)\right]=\left[\widehat{q}_{k}\left(t\right),\widehat{q}_{-k'}\left(t\right)\right]=0,
	\label{SM41}
\end{align}

\begin{align}
	\left[\widehat{\Pi}_{x}\left(z,t\right),\widehat{\Pi}_{x}\left(z',t\right)\right]=0\rightarrow\left[\widehat{p}_{k}\left(t\right),\widehat{p}_{k'}\left(t\right)\right]=0,
	\label{SM42}
\end{align}
\begin{align}
	\left[\widehat{p}_{k}\left(t\right),\widehat{p}_{k'}^{\dagger}\left(t\right)\right]=\left[\widehat{p}_{k}\left(t\right),\widehat{p}_{-k'}\left(t\right)\right]=0.
	\label{SM43}
\end{align}

Equation (\ref{SM39}) states the mutual interference between position $\widehat{q}_{k}\left(t\right)$ of the forward wave and conjugate transpose momentum $\widehat{p}_{-k}^{\dagger}\left(t\right)$ of the backward wave. In other words, the magnetic field $B_{y}\left(z,t\right)$ in (\ref{SM31}) of the forward (+k) time-scattered wave and the electric displacement vector of $D_{x}\left(z,t\right)$ in (\ref{SM32}) of the backward (-k) time-scattered wave, evaluated at the same time, cannot be measured without mutual interference.

The Fourier transform on k-space of the Hamiltonian (\ref{SM29}) allows us to connect the conjugated momentum $\widehat{\Pi}_{x}\left(z,t\right)$ with $\widehat{p}_{k}\left(t\right)$:

\begin{align}
	\begin{split}
		&\nonumber\int dz\,\widehat{\Pi}_{x}^{2}\left(z,t\right)= \frac{\varepsilon_{0}}{2\pi}\int dz\,\int dk\,\widehat{p}_{k}\left(t\right)\,e^{ikz}\int dk'\,\widehat{p}_{k'}\left(t\right)e^{ik'z} \\
		&= \int dk\,\widehat{p}_{k}\left(t\right)\int dk'\,\widehat{p}_{k'}\left(t\right)\frac{\varepsilon_{0}}{2\pi}\int dz\,e^{i(k+k')z}=\varepsilon_{0}\int dk\,\widehat{p}_{k}\left(t\right)\int dk'\,\widehat{p}_{k'}\left(t\right)\delta\left(k+k'\right)
	\end{split} \\
	&= \varepsilon_{0}\int dk\,\widehat{p}_{k}\left(t\right)\,\widehat{p}_{-k}\left(t\right)=\varepsilon_{0}\int dk\,\widehat{p}_{k}\left(t\right)\,\widehat{p}_{k}^{\dagger}\left(t\right),
	\label{SM44}
\end{align}

and the operator associated with the dynamical variable with $\widehat{q}_{k}\left(t\right)$, as follows:

\begin{align}
	\begin{split}
		&\nonumber\int dz\,\left(\partial_{z}\widehat{A}_{x}\left(z,t\right)\right)^{2}= \frac{1}{2\pi\varepsilon_{0}}\int dz\,\int dk\,ik\,\widehat{q}_{k}\left(t\right)\,e^{ikz}\int dk'\,ik'\,\widehat{q}_{k'}\left(t\right)\,e^{ik'z} \\
		&= -\int dk\,k\,\widehat{q}_{k}\left(t\right)\int dk'\,k'\,\widehat{q}_{k'}\left(t\right)\frac{1}{2\pi\varepsilon_{0}}\int dz\,e^{i(k+k')z}=-\frac{1}{\varepsilon_{0}}\int dk\,k\,\widehat{q}_{k}\left(t\right)\int dk'\,k'\,\widehat{q}_{k'}\left(t\right)\delta\left(k+k'\right)
	\end{split} \\
	&= \frac{1}{\varepsilon_{0}}\int dk\,k^{2}\,\widehat{q}_{k}\left(t\right)\,\widehat{q}_{-k}\left(t\right)=\frac{1}{\varepsilon_{0}}\int dk\,k^{2}\,\widehat{q}_{k}\left(t\right)\,\widehat{q}_{k}^{\dagger}\left(t\right).
	\label{SM45}
\end{align}

From these expressions, the Hamiltonian (\ref{SM29}) can be finally written in modal form:
\begin{align}
	\widehat{\mathcal{H}}\left(t\right)=\frac{1}{2}\,\int_{-\infty}^{\infty}dk\,\left(\frac{1}{\varepsilon\left(t\right)}\,\widehat{p}_{k}\left(t\right)\,\widehat{p}_{k}^{\dagger}\left(t\right)+\frac{k^{2}c^{2}}{\mu\left(t\right)}\,\widehat{q}_{k}\left(t\right)\,\widehat{q}_{k}^{\dagger}\left(t\right)\right).
	\label{SM46}
\end{align}
Integrating over $+k$ domain,

\begin{align}
	\widehat{\mathcal{H}}\left(t\right)=\frac{1}{2}\,\int_{0}^{\infty}dk\,\left(\frac{1}{\varepsilon\left(t\right)}\,\left(\widehat{p}_{k}\left(t\right)\,\widehat{p}_{k}^{\dagger}\left(t\right)+\widehat{p}_{-k}\left(t\right)\,\widehat{p}_{-k}^{\dagger}\left(t\right)\right)+\frac{k^{2}c^{2}}{\mu\left(t\right)}\,\left(\widehat{q}_{k}\left(t\right)\,\widehat{q}_{k}^{\dagger}\left(t\right)+\widehat{q}_{-k}\left(t\right)\,\widehat{q}_{-k}^{\dagger}\left(t\right)\right)\right),
	\label{SM47}
\end{align}
and applying the reality conditions (\ref{SM33}) and (\ref{SM34}), it finally follows that:
\begin{align}
	\boxed{\widehat{\mathcal{H}}\left(t\right)=\frac{1}{2}\,\int_{0}^{\infty}dk\,\left(\frac{1}{\varepsilon\left(t\right)}\,\left(\widehat{p}_{k}\left(t\right)\,\widehat{p}_{k}^{\dagger}\left(t\right)+\widehat{p}_{k}^{\dagger}\left(t\right)\,\widehat{p}_{k}\left(t\right)\right)+\frac{k^{2}c^{2}}{\mu\left(t\right)}\,\left(\widehat{q}_{k}\left(t\right)\,\widehat{q}_{k}^{\dagger}\left(t\right)+\widehat{q}_{k}^{\dagger}\left(t\right)\,\widehat{q}_{k}\left(t\right)\right)\right)=\int_{0}^{\infty}dk\,\widehat{\mathcal{H}}_{k}\left(t\right)}
	\label{SM48}
\end{align}
It is worth noticing that the integrand $\widehat{\mathcal{H}}_{k}\left(t\right)$ in (\ref{SM48}) masks the underlying behavior of a time-varying oscillator, as it intertwines the quantum dynamics of forward and backward waves inherent to the temporal modulation. Indeed, as it is shown in  \hyperref[HO_Conjugated_oscillator_picture]{Fig. \ref*{HO_Conjugated_oscillator_picture}} in the main text, time-varying media is depicted by two conjugated oscillators, i.e., the operators linked to both the forward $(+k)$ and backward $(-k)$ modes, evolve on the complex plane along conjugated trajectories.
\part{Part III: Generalized parametric oscillator}
The purpose of this part is to prove that the Hamiltonian \ref{SM48} fulfill SU(2) Lie Algebra, enabling the application of the time-evolution operator disentangling technique proposed by Cheng (for a step-by-step derivation of the required equations, see Appendix A). Furthermore, we address the evolution of annihilation quantum operator through the direct application of the derived time-evolution operator. Finally, we explicitly show the inapplicability of this method for scenarios involving time-varying media. This justify the alternative approach described in Part IV. Nevertheless, despite the non-applicability of such a disentangling technique, an intuitive and practical connection is established between generalized parametric oscillator parameters and time-dependent optical properties. All these derivations are conducted and expressed in terms of creation and annihilation quantum operators (see Appendix B1 and B2), though for the main purpose of this work (computing photons production), using position and momentum operators is fairly enough.
\section{VI. BRIEF REVIEW OF PARAMETRIC OSCILLATOR}
We can see at glance that the Hamiltonian (\ref{SM48}) is quite similar to the generalized parametric oscillator with varying mass, i.e., an harmonic oscillator whose parameters: mass, spring force, or both, are time-dependent. Such simplicity and elegance have served as the foundation for a myriad of applications in several branches. In the quantum context, right after promoting dynamical variables and momenta into quantum operators, the above-mentioned formalism is encapsulated within the Hamiltonian:
\begin{align}
	\widehat{\mathcal{H}}(t)=\frac{\widehat{p}^{2}\left(t\right)}{2m(t)}+\frac{1}{2}m(t)\omega^{2}(t)\widehat{q}^{2}\left(t\right).
	\label{SM49}
\end{align}
Despite one could be tempted to make a term-by-term comparative, in order to establish a parallelism between the physical magnitudes in the quantum electrodynamic case (\ref{SM48}) and the quantum mechanic case (\ref{SM48}), it is useful to firstly verify whether (\ref{SM48}) can be rewritten in the following form:
\begin{align}
	\widehat{\mathcal{H}}(t)=a_{1}(t)\widehat{J}_{+}+a_{2}(t)\widehat{J}_{0}+a_{3}(t)\widehat{J}_{-},
	\label{SM50}
\end{align}
where $\widehat{J}_{+}=\widehat{q}^{2}\left(t\right)/2\hbar$, $\widehat{J}_{0}=i\left(\widehat{p}\left(t\right)\widehat{q}\left(t\right)+\widehat{q}\left(t\right)\widehat{p}\left(t\right)\right)/4\hbar$, and $\widehat{J}_{-}=\widehat{p}^{2}\left(t\right)/2\hbar$ stand for the group generators satisfying the SU(2) Lie algebra:

\begin{align}
	\left[\widehat{J}_{+},\widehat{J}_{-}\right]=2\widehat{J}_{0},
	\label{SM51}
\end{align}
\begin{align}
	\left[\widehat{J}_{0},\widehat{J}_{\pm}\right]=\pm\widehat{J}_{\pm}.
	\label{SM52}
\end{align}

By identifying terms in the Hamiltonian (\ref{SM49}), it follows that

\begin{align}
	a_{1}(t)=\hbar m(t)\omega^{2}(t),
	\label{SM53}
\end{align}
\begin{align}
	a_{2}(t)=0,
	\label{SM54}
\end{align}
\begin{align}
	a_{3}(t)=\frac{\hbar}{m(t)}.
	\label{SM55}
\end{align}
Note that $\widehat{J}_{+}$, $\widehat{J}_{0}$, and $\widehat{J}_{-}$ are defined as indicated for a hermitian Hamiltonian (\ref{SM49}), and they can be applied in the method introduced by Cheng, as a the disentangling technique for exponential operators which are not necessarily unitary. The evolution operator method has long been used to solve problems in quantum mechanics, but finding an explicit solution for the Schrödinger equation:
\begin{align}
		\widehat{\mathcal{H}}(t)\left(t\right)\left|\psi(t)\right\rangle =i\hbar\partial_{t}\left|\psi(t)\right\rangle ,
	\label{SM56}
\end{align}
it is a fairly involved task. Besides, if a certain Hamiltonian admits to be written as in (\ref{SM50}) (i.e., satisfying SU(2) Lie Algebra), the evolution operator can be expressed in the following form:
\begin{align}
	\widehat{U}\left(t,0\right)=\exp\left(c_{1}(t)\widehat{J}_{+}\right)\exp\left(c_{2}(t)\widehat{J}_{0}\right)\exp\left(c_{3}(t)\widehat{J}_{-}\right),
	\label{SM57}
\end{align}
where $c_{i}(t)$ (with i=\{1,2,3\}) functions are to be determined. Since the time-evolution operator proposed (\ref{SM57}) is written as a product of exponential operators, it has a direct differentiation with respect to time. These procedures (see Appendix A) lead into a set of differential equations:
\begin{align}
	\partial_{t}c_{1}=\frac{a_{1}(t)}{i\hbar}+\frac{a_{2}(t)}{i\hbar}c_{1}(t)-\frac{a_{3}(t)}{i\hbar}c_{1}^{2}(t),
	\label{SM58}
\end{align}
\begin{align}
	\partial_{t}c_{2}=\frac{a_{2}(t)}{i\hbar}-2\frac{a_{3}(t)}{i\hbar}c_{1}(t),
	\label{SM59}
\end{align}
\begin{align}
	\partial_{t}c_{3}=\frac{a_{3}(t)}{i\hbar}\exp(c_{2}(t)),
	\label{SM60}
\end{align}

with the following initial conditions

\begin{align}
	c_{1}(0)=c_{2}(0)=c_{3}(0)=0.
	\label{SM61}
\end{align}

It must be noticed that, once solved (\ref{SM58}), which is just the Riccati equation, (\ref{SM59})-(\ref{SM60}) are readily evaluated. Hence, differential equation (\ref{SM58}) enables an already studied transformation for Riccati equation, i.e., $c_{1}(t)=i\hbar\partial_{t}u(t)/\left(a_{3}(t)u(t)\right)$, leading into a second order-differential equation:
\begin{align}
	\partial_{t}^{2}u(t)+\gamma_{0}(t)\partial_{t}u(t)+\omega^{2}(t)u(t)=0.
	\label{SM62}
\end{align}
where,
\begin{align}
	\gamma_{0}(t)=\frac{\partial_{t}m(t)}{m(t)}.
	\label{SM63}
\end{align}
Equation (\ref{SM62}) is the differential equation for a parametric oscillator whose coefficients, related with mass and spring constant, are time-dependent. As a matter of proof, this equation resembles the equation of motion for a damped oscillator for constant mass and oscillation frequency. Thus, once solved, it provides the dynamics of dictated by (\ref{SM58})-(\ref{SM61}), i.e., the evolution of each exponential operator in (\ref{SM57}):
\begin{align}
	c_{1}(t)=im(t)\frac{\partial_{t}u(t)}{u(t)},
	\label{SM64}
\end{align}
\begin{align}
	c_{2}(t)=-2\log\frac{u(t)}{u(0)},
	\label{SM65}
\end{align}
\begin{align}
	c_{3}(t)=-iu^{2}(0)\int_{0}^{t}\frac{dt'}{m(t')u^{2}(t')}.
	\label{SM66}
\end{align}
This set of equations (\ref{SM64})-(\ref{SM66}), inserted on (\ref{SM57}), and then applied to any operator at $t=0$, will allow for computing the quantum dynamics of their corresponding expected values and variances (as we develop in IX Bogoliubov transformations).

\section{VII. ENSURING SU(2) LIE ALGEBRA}
We identify the corresponding equations for group operators $\widehat{J}_{+}$ and $\widehat{J}_{-}$ from (\ref{SM48}):
\begin{align}
	\widehat{J}_{+}=\frac{1}{2\hbar}\left(\widehat{q}_{k}\left(t\right)\,\widehat{q}_{k}^{\dagger}\left(t\right)+\widehat{q}_{k}^{\dagger}\left(t\right)\,\widehat{q}_{k}\left(t\right)\right),
	\label{SM67}
\end{align}
\begin{align}
	\widehat{J}_{-}=\frac{1}{2\hbar}\left(\widehat{p}_{k}\left(t\right)\,\widehat{p}_{k}^{\dagger}\left(t\right)+\widehat{p}_{k}^{\dagger}\left(t\right)\,\widehat{p}_{k}\left(t\right)\right).
	\label{SM68}
\end{align}

\subsection{7.1 Computing $\widehat{J}_{0}$}
In order to evaluate commutation relations described in (\ref{SM51}) and (\ref{SM52}), $\widehat{J}_{0}$ must be first evaluated using the above definitions given in (\ref{SM67}) and (\ref{SM68}):

\begin{align}
\nonumber\left[\widehat{J}_{+},\widehat{J}_{-}\right]=\frac{1}{4\hslash^{2}}\,\left[\widehat{q}_{k}\left(t\right)\,\widehat{q}_{k}^{\dagger}\left(t\right)+\widehat{q}_{k}^{\dagger}\left(t\right)\,\widehat{q}_{k}\left(t\right),\widehat{p}_{k}\left(t\right)\,\widehat{p}_{k}^{\dagger}\left(t\right)+\widehat{p}_{k}^{\dagger}\left(t\right)\,\widehat{p}_{k}\left(t\right)\right]&\\\nonumber
={\color{blue}\left[\widehat{q}_{k}\left(t\right)\,\widehat{q}_{k}^{\dagger}\left(t\right),\widehat{p}_{k}\left(t\right)\,\widehat{p}_{k}^{\dagger}\left(t\right)\right]}+{\color{red}\left[\widehat{q}_{k}\left(t\right)\,\widehat{q}_{k}^{\dagger}\left(t\right),\widehat{p}_{k}^{\dagger}\left(t\right)\,\widehat{p}_{k}\left(t\right)\right]}&\\\nonumber
+{\color{magenta}\left[\widehat{q}_{k}^{\dagger}\left(t\right)\,\widehat{q}_{k}\left(t\right),\widehat{p}_{k}\left(t\right)\,\widehat{p}_{k}^{\dagger}\left(t\right)\right]}+{\color{teal}\left[\widehat{q}_{k}^{\dagger}\left(t\right)\,\widehat{q}_{k}\left(t\right),\widehat{p}_{k}^{\dagger}\left(t\right)\,\widehat{p}_{k}\left(t\right)\right]};
\end{align}

\begin{align}
\nonumber{\color{blue}\left[\widehat{q}_{k}\left(t\right)\,\widehat{q}_{k}^{\dagger}\left(t\right),\widehat{p}_{k}\left(t\right)\,\widehat{p}_{k}^{\dagger}\left(t\right)\right]}=\widehat{q}_{k}\left(t\right)\left[\widehat{q}_{k}^{\dagger}\left(t\right),\widehat{p}_{k}\left(t\right)\right]\widehat{p}_{k}^{\dagger}\left(t\right)+\cancelto{0}{\left[\widehat{q}_{k}\left(t\right),\widehat{p}_{k}\left(t\right)\right]\widehat{q}_{k}^{\dagger}\left(t\right)\widehat{p}_{k}^{\dagger}\left(t\right)}&\\+\widehat{p}_{k}\left(t\right)\widehat{q}_{k}\left(t\right)\cancelto{0}{\left[\widehat{q}_{k}^{\dagger}\left(t\right),\widehat{p}_{k}^{\dagger}\left(t\right)\right]}+\widehat{p}_{k}\left(t\right)\left[\widehat{q}_{k}\left(t\right),\widehat{p}_{k}^{\dagger}\left(t\right)\right]\widehat{q}_{k}^{\dagger}\left(t\right)=i\hslash(\widehat{q}_{k}\left(t\right)\widehat{p}_{k}^{\dagger}\left(t\right)+\widehat{p}_{k}\left(t\right)\widehat{q}_{k}^{\dagger}\left(t\right)),
	\label{SM69}
\end{align}

\begin{align}
\nonumber{\color{red}\left[\widehat{q}_{k}\left(t\right)\,\widehat{q}_{k}^{\dagger}\left(t\right),\widehat{p}_{k}^{\dagger}\left(t\right)\,\widehat{p}_{k}\left(t\right)\right]}=\widehat{q}_{k}\left(t\right)\cancelto{0}{\left[\widehat{q}_{k}^{\dagger}\left(t\right),\widehat{p}_{k}^{\dagger}\left(t\right)\right]}\widehat{p}_{k}\left(t\right)+\left[\widehat{q}_{k}\left(t\right),\widehat{p}_{k}^{\dagger}\left(t\right)\right]\widehat{q}_{k}^{\dagger}\left(t\right)\widehat{p}_{k}\left(t\right)&\\+\widehat{p}_{k}^{\dagger}\left(t\right)\widehat{q}_{k}\left(t\right)\left[\widehat{q}_{k}^{\dagger}\left(t\right),\widehat{p}_{k}\left(t\right)\right]+\widehat{p}_{k}^{\dagger}\left(t\right)\cancelto{0}{\left[\widehat{q}_{k}\left(t\right),\widehat{p}_{k}\left(t\right)\right]}\widehat{q}_{k}^{\dagger}\left(t\right)=i\hslash(\widehat{q}_{k}^{\dagger}\left(t\right)\widehat{p}_{k}\left(t\right)+\widehat{p}_{k}^{\dagger}\left(t\right)\widehat{q}_{k}\left(t\right)),
	\label{SM70}
\end{align}

\begin{align}
\nonumber{\color{magenta}\left[\widehat{q}_{k}^{\dagger}\left(t\right)\,\widehat{q}_{k}\left(t\right),\widehat{p}_{k}\left(t\right)\,\widehat{p}_{k}^{\dagger}\left(t\right)\right]}=\widehat{q}_{k}^{\dagger}\left(t\right)\cancelto{0}{\left[\widehat{q}_{k}\left(t\right),\widehat{p}_{k}\left(t\right)\right]}\widehat{p}_{k}^{\dagger}\left(t\right)+\left[\widehat{q}_{k}^{\dagger}\left(t\right),\widehat{p}_{k}\left(t\right)\right]\widehat{q}_{k}\left(t\right)\widehat{p}_{k}^{\dagger}\left(t\right)&\\+\widehat{p}_{k}\left(t\right)\widehat{q}_{k}^{\dagger}\left(t\right)\left[\widehat{q}_{k}\left(t\right),\widehat{p}_{k}^{\dagger}\left(t\right)\right]+\widehat{p}_{k}\left(t\right)\cancelto{0}{\left[\widehat{q}_{k}^{\dagger}\left(t\right),\widehat{p}_{k}^{\dagger}\left(t\right)\right]}\widehat{q}_{k}\left(t\right)=i\hslash\left(\widehat{q}_{k}\left(t\right)\widehat{p}_{k}^{\dagger}\left(t\right)+\widehat{p}_{k}\left(t\right)\widehat{q}_{k}^{\dagger}\left(t\right)\right),
	\label{SM71}
\end{align}

\begin{align}
\nonumber{\color{teal}\left[\widehat{q}_{k}^{\dagger}\left(t\right)\,\widehat{q}_{k}\left(t\right),\widehat{p}_{k}^{\dagger}\left(t\right)\,\widehat{p}_{k}\left(t\right)\right]}=\widehat{q}_{k}^{\dagger}\left(t\right)\left[\widehat{q}_{k}\left(t\right),\widehat{p}_{k}^{\dagger}\left(t\right)\right]\widehat{p}_{k}\left(t\right)+\cancelto{0}{\left[\widehat{q}_{k}^{\dagger}\left(t\right),\widehat{p}_{k}^{\dagger}\left(t\right)\right]}\widehat{q}_{k}\left(t\right)\widehat{p}_{k}\left(t\right)&\\+\widehat{p}_{k}^{\dagger}\left(t\right)\widehat{q}_{k}^{\dagger}\left(t\right)\cancelto{0}{\left[\widehat{q}_{k}\left(t\right),\widehat{p}_{k}\left(t\right)\right]}+\widehat{p}_{k}^{\dagger}\left(t\right)\left[\widehat{q}_{k}^{\dagger}\left(t\right),\widehat{p}_{k}\left(t\right)\right]\widehat{q}_{k}\left(t\right)=i\hslash\left(\widehat{q}_{k}^{\dagger}\left(t\right)\widehat{p}_{k}\left(t\right)+\widehat{p}_{k}^{\dagger}\left(t\right)\widehat{q}_{k}\left(t\right)\right),
	\label{SM72}
\end{align}

\begin{align}
	\left[\widehat{J}_{+},\widehat{J}_{-}\right]=\frac{i}{2\hslash}\,\left\{ \left(\widehat{p}_{k}\left(t\right)\widehat{q}_{k}^{\dagger}\left(t\right)+\widehat{q}_{k}\left(t\right)\widehat{p}_{k}^{\dagger}\left(t\right)\right)+\left(\widehat{q}_{k}^{\dagger}\left(t\right)\widehat{p}_{k}\left(t\right)+\widehat{p}_{k}^{\dagger}\left(t\right)\widehat{q}_{k}\left(t\right)\right)\right\} =2\widehat{J}_{0}.
	\label{SM73}
\end{align}
Therefore,
\begin{align}
	\boxed{\widehat{J}_{0}=\frac{i}{4\hslash}\,\left\{ \left(\widehat{p}_{k}\left(t\right)\widehat{q}_{k}^{\dagger}\left(t\right)+\widehat{q}_{k}\left(t\right)\widehat{p}_{k}^{\dagger}\left(t\right)\right)+\left(\widehat{q}_{k}^{\dagger}\left(t\right)\widehat{p}_{k}\left(t\right)+\widehat{p}_{k}^{\dagger}\left(t\right)\widehat{q}_{k}\left(t\right)\right)\right\} .}
	\label{SM74}
\end{align}
\subsection{7.2 Verifying $\left[\widehat{J}_{0},\widehat{J}_{+}\right]=+\widehat{J}_{+}$}
We check the plus sign outcome of commutation relation (\ref{SM52}):

\begin{align}
	\nonumber\left[\widehat{J}_{0},\widehat{J}_{+}\right]=\frac{i}{8\hslash^{2}}\left[\left(\widehat{p}_{k}\left(t\right)\widehat{q}_{k}^{\dagger}\left(t\right)+\widehat{q}_{k}\left(t\right)\widehat{p}_{k}^{\dagger}\left(t\right)\right),\left(\widehat{q}_{k}\left(t\right)\,\widehat{q}_{k}^{\dagger}\left(t\right)+\widehat{q}_{k}^{\dagger}\left(t\right)\,\widehat{q}_{k}\left(t\right)\right)\right]&\\+\left[\left(\widehat{q}_{k}^{\dagger}\left(t\right)\widehat{p}_{k}\left(t\right)+\widehat{p}_{k}^{\dagger}\left(t\right)\widehat{q}_{k}\left(t\right)\right),\left(\widehat{q}_{k}\left(t\right)\,\widehat{q}_{k}^{\dagger}\left(t\right)+\widehat{q}_{k}^{\dagger}\left(t\right)\,\widehat{q}_{k}\left(t\right)\right)\right].
	\label{SM75}
\end{align}
We address the first commutator of (\ref{SM75}):
\begin{align*}
\nonumber{\color{blue}\left[\widehat{p}_{k}\left(t\right)\widehat{q}_{k}^{\dagger}\left(t\right),\widehat{q}_{k}\left(t\right)\widehat{q}_{k}^{\dagger}\left(t\right)\right]}+{\color{red}\left[\widehat{p}_{k}\left(t\right)\widehat{q}_{k}^{\dagger}\left(t\right),\widehat{q}_{k}^{\dagger}\left(t\right)\widehat{q}_{k}\left(t\right)\right]}&\\+{\color{magenta}\left[\widehat{q}_{k}\left(t\right)\widehat{p}_{k}^{\dagger}\left(t\right),\widehat{q}_{k}\left(t\right)\widehat{q}_{k}^{\dagger}\left(t\right)\right]}+{\color{teal}\left[\widehat{q}_{k}\left(t\right)\widehat{p}_{k}^{\dagger}\left(t\right),\widehat{q}_{k}^{\dagger}\left(t\right)\widehat{q}_{k}\left(t\right)\right];}
\end{align*}

\begin{align}
\nonumber{\color{blue}\left[\widehat{p}_{k}\left(t\right)\widehat{q}_{k}^{\dagger}\left(t\right),\widehat{q}_{k}\left(t\right)\widehat{q}_{k}^{\dagger}\left(t\right)\right]}=\widehat{p}_{k}\left(t\right)\cancelto{0}{\left[\widehat{q}_{k}^{\dagger}\left(t\right),\widehat{q}_{k}\left(t\right)\right]}\widehat{q}_{k}^{\dagger}\left(t\right)+\cancelto{0}{\left[\widehat{p}_{k}\left(t\right),\widehat{q}_{k}\left(t\right)\right]}\widehat{q}_{k}^{\dagger}\left(t\right)\widehat{q}_{k}^{\dagger}\left(t\right)&\\+\widehat{q}_{k}\left(t\right)\widehat{p}_{k}\left(t\right)\cancelto{0}{\left[\widehat{q}_{k}^{\dagger}\left(t\right),\widehat{q}_{k}^{\dagger}\left(t\right)\right]}+\widehat{q}_{k}\left(t\right)\left[\widehat{p}_{k}\left(t\right),\widehat{q}_{k}^{\dagger}\left(t\right)\right]\widehat{q}_{k}^{\dagger}\left(t\right)=-i\hslash\left(\widehat{q}_{k}\left(t\right)\widehat{q}_{k}^{\dagger}\left(t\right)\right),
	\label{SM76}
\end{align}

\begin{align}
\nonumber{\color{red}\left[\widehat{p}_{k}\left(t\right)\widehat{q}_{k}^{\dagger}\left(t\right),\widehat{q}_{k}^{\dagger}\left(t\right)\widehat{q}_{k}\left(t\right)\right]}=\widehat{p}_{k}\left(t\right)\cancelto{0}{\left[\widehat{q}_{k}^{\dagger}\left(t\right),\widehat{q}_{k}^{\dagger}\left(t\right)\right]}\widehat{q}_{k}\left(t\right)+\left[\widehat{p}_{k}\left(t\right),\widehat{q}_{k}^{\dagger}\left(t\right)\right]\widehat{q}_{k}^{\dagger}\left(t\right)\widehat{q}_{k}\left(t\right)&\\+\widehat{q}_{k}^{\dagger}\left(t\right)\widehat{p}_{k}\left(t\right)\cancelto{0}{\left[\widehat{q}_{k}^{\dagger}\left(t\right),\widehat{q}_{k}\left(t\right)\right]}+\widehat{q}_{k}^{\dagger}\left(t\right)\cancelto{0}{\left[\widehat{p}_{k}\left(t\right),\widehat{q}_{k}\left(t\right)\right]}\widehat{q}_{k}^{\dagger}\left(t\right)=-i\hslash\left(\widehat{q}_{k}^{\dagger}\left(t\right)\widehat{q}_{k}\left(t\right)\right),
	\label{SM77}
\end{align}

\begin{align}
\nonumber{\color{magenta}\left[\widehat{q}_{k}\left(t\right)\widehat{p}_{k}^{\dagger}\left(t\right),\widehat{q}_{k}\left(t\right)\widehat{q}_{k}^{\dagger}\left(t\right)\right]}=\widehat{q}_{k}\left(t\right)\left[\widehat{p}_{k}^{\dagger}\left(t\right),\widehat{q}_{k}\left(t\right)\right]\widehat{q}_{k}^{\dagger}\left(t\right)+\cancelto{0}{\left[\widehat{q}_{k}\left(t\right),\widehat{q}_{k}\left(t\right)\right]}\widehat{p}_{k}^{\dagger}\left(t\right)\widehat{q}_{k}^{\dagger}\left(t\right)&\\+\widehat{q}_{k}\left(t\right)\widehat{q}_{k}\left(t\right)\cancelto{0}{\left[\widehat{p}_{k}^{\dagger}\left(t\right),\widehat{q}_{k}^{\dagger}\left(t\right)\right]}+\widehat{q}_{k}\left(t\right)\cancelto{0}{\left[\widehat{q}_{k}\left(t\right),\widehat{q}_{k}^{\dagger}\left(t\right)\right]}\widehat{p}_{k}^{\dagger}\left(t\right)=-i\hslash\left(\widehat{q}_{k}\left(t\right)\widehat{q}_{k}^{\dagger}\left(t\right)\right),
	\label{SM78}
\end{align}

\begin{align}
\nonumber{\color{teal}\left[\widehat{q}_{k}\left(t\right)\widehat{p}_{k}^{\dagger}\left(t\right),\widehat{q}_{k}^{\dagger}\left(t\right)\widehat{q}_{k}\left(t\right)\right]}=\widehat{q}_{k}\left(t\right)\cancelto{0}{\left[\widehat{p}_{k}^{\dagger}\left(t\right),\widehat{q}_{k}^{\dagger}\left(t\right)\right]}\widehat{q}_{k}\left(t\right)+\cancelto{0}{\left[\widehat{q}_{k}\left(t\right),\widehat{q}_{k}^{\dagger}\left(t\right)\right]}\widehat{p}_{k}^{\dagger}\left(t\right)\widehat{q}_{k}\left(t\right)&\\+\widehat{q}_{k}^{\dagger}\left(t\right)\widehat{q}_{k}\left(t\right)\left[\widehat{p}_{k}^{\dagger}\left(t\right),\widehat{q}_{k}\left(t\right)\right]+\widehat{q}_{k}^{\dagger}\left(t\right)\cancelto{0}{\left[\widehat{q}_{k}\left(t\right),\widehat{q}_{k}\left(t\right)\right]}\widehat{p}_{k}^{\dagger}\left(t\right)=-i\hslash\left(\widehat{q}_{k}^{\dagger}\left(t\right)\widehat{q}_{k}\left(t\right)\right).
\end{align}
And the the same for the second commutator of (\ref{SM75}):

\begin{align}
\nonumber{\color{blue}\left[\widehat{q}_{k}^{\dagger}\left(t\right)\widehat{p}_{k}\left(t\right),\widehat{q}_{k}\left(t\right)\widehat{q}_{k}^{\dagger}\left(t\right)\right]}=\widehat{q}_{k}^{\dagger}\left(t\right)\cancelto{0}{\left[\widehat{p}_{k}\left(t\right),\widehat{q}_{k}\left(t\right)\right]}\widehat{q}_{k}^{\dagger}\left(t\right)+\cancelto{0}{\left[\widehat{q}_{k}^{\dagger}\left(t\right),\widehat{q}_{k}\left(t\right)\right]}\widehat{p}_{k}\left(t\right)\widehat{q}_{k}^{\dagger}\left(t\right)&\\+\widehat{q}_{k}\left(t\right)\widehat{q}_{k}^{\dagger}\left(t\right)\left[\widehat{p}_{k}\left(t\right),\widehat{q}_{k}^{\dagger}\left(t\right)\right]+\widehat{q}_{k}\left(t\right)\cancelto{0}{\left[\widehat{q}_{k}^{\dagger}\left(t\right),\widehat{q}_{k}^{\dagger}\left(t\right)\right]}\widehat{p}_{k}\left(t\right)=-i\hslash\left(\widehat{q}_{k}\left(t\right)\widehat{q}_{k}^{\dagger}\left(t\right)\right),
	\label{SM80}
\end{align}

\begin{align}
	\nonumber{\color{red}\left[\widehat{q}_{k}^{\dagger}\left(t\right)\widehat{p}_{k}\left(t\right),\widehat{q}_{k}^{\dagger}\left(t\right)\widehat{q}_{k}\left(t\right)\right]}=\widehat{q}_{k}^{\dagger}\left(t\right)\left[\widehat{p}_{k}\left(t\right),\widehat{q}_{k}^{\dagger}\left(t\right)\right]\widehat{q}_{k}\left(t\right)+\cancelto{0}{\left[\widehat{q}_{k}^{\dagger}\left(t\right),\widehat{q}_{k}^{\dagger}\left(t\right)\right]}\widehat{p}_{k}\left(t\right)\widehat{q}_{k}\left(t\right)&\\+\widehat{q}_{k}^{\dagger}\left(t\right)\widehat{q}_{k}^{\dagger}\left(t\right)\cancelto{0}{\left[\widehat{p}_{k}\left(t\right),\widehat{q}_{k}\left(t\right)\right]}+\widehat{q}_{k}^{\dagger}\left(t\right)\cancelto{0}{\left[\widehat{q}_{k}^{\dagger}\left(t\right),\widehat{q}_{k}\left(t\right)\right]}\widehat{p}_{k}\left(t\right)=-i\hslash\left(\widehat{q}_{k}^{\dagger}\left(t\right)\widehat{q}_{k}\left(t\right)\right),
	\label{SM81}
\end{align}

\begin{align}
\nonumber{\color{magenta}\left[\widehat{p}_{k}^{\dagger}\left(t\right)\widehat{q}_{k}\left(t\right),\widehat{q}_{k}\left(t\right)\widehat{q}_{k}^{\dagger}\left(t\right)\right]}=\widehat{p}_{k}^{\dagger}\left(t\right)\cancelto{0}{\left[\widehat{q}_{k}\left(t\right),\widehat{q}_{k}\left(t\right)\right]}\widehat{q}_{k}^{\dagger}\left(t\right)+\left[\widehat{p}_{k}^{\dagger}\left(t\right),\widehat{q}_{k}\left(t\right)\right]\widehat{q}_{k}\left(t\right)\widehat{q}_{k}^{\dagger}\left(t\right)&\\+\widehat{q}_{k}\left(t\right)\widehat{p}_{k}^{\dagger}\left(t\right)\cancelto{0}{\left[\widehat{q}_{k}\left(t\right),\widehat{q}_{k}^{\dagger}\left(t\right)\right]}+\widehat{q}_{k}\left(t\right)\cancelto{0}{\left[\widehat{p}_{k}^{\dagger}\left(t\right),\widehat{q}_{k}^{\dagger}\left(t\right)\right]}\widehat{q}_{k}\left(t\right)=-i\hslash\left(\widehat{q}_{k}\left(t\right)\widehat{q}_{k}^{\dagger}\left(t\right)\right),
	\label{SM82}
\end{align}

\begin{align}
\nonumber{\color{teal}\left[\widehat{p}_{k}^{\dagger}\left(t\right)\widehat{q}_{k}\left(t\right),\widehat{q}_{k}^{\dagger}\left(t\right)\widehat{q}_{k}\left(t\right)\right]}=\widehat{p}_{k}^{\dagger}\left(t\right)\cancelto{0}{\left[\widehat{q}_{k}\left(t\right),\widehat{q}_{k}^{\dagger}\left(t\right)\right]}\widehat{q}_{k}\left(t\right)+\cancelto{0}{\left[\widehat{p}_{k}^{\dagger}\left(t\right),\widehat{q}_{k}^{\dagger}\left(t\right)\right]}\widehat{q}_{k}\left(t\right)\widehat{q}_{k}\left(t\right)&\\+\widehat{q}_{k}^{\dagger}\left(t\right)\widehat{p}_{k}^{\dagger}\left(t\right)\cancelto{0}{\left[\widehat{q}_{k}\left(t\right),\widehat{q}_{k}\left(t\right)\right]}+\widehat{q}_{k}^{\dagger}\left(t\right)\left[\widehat{p}_{k}^{\dagger}\left(t\right),\widehat{q}_{k}\left(t\right)\right]\widehat{q}_{k}\left(t\right)=-i\hslash\left(\widehat{q}_{k}^{\dagger}\left(t\right)\widehat{q}_{k}\left(t\right)\right).
	\label{SM83}
\end{align}

Therefore,

\begin{align}
	\boxed{\left[\widehat{J}_{0},\widehat{J}_{+}\right]=\frac{1}{2\hslash}\left\{ \widehat{q}_{k}^{\dagger}\left(t\right)\widehat{q}_{k}\left(t\right)+\widehat{q}_{k}\left(t\right)\widehat{q}_{k}^{\dagger}\left(t\right)\right\} =\widehat{J}_{+}.}
	\label{SM84}
\end{align}

\subsection{7.3 Verifying $\left[\widehat{J}_{0},\widehat{J}_{-}\right]=-\widehat{J}_{-}$}
We now check the minus sign outcome of commutation relation (\ref{SM52}) following the same procedure as above:
\begin{align}
\nonumber\left[\widehat{J}_{0},\widehat{J}_{-}\right]=\frac{i}{8\hslash^{2}}\left[\left(\widehat{p}_{k}\left(t\right)\widehat{q}_{k}^{\dagger}\left(t\right)+\widehat{q}_{k}\left(t\right)\widehat{p}_{k}^{\dagger}\left(t\right)\right),\left(\widehat{p}_{k}\left(t\right)\widehat{p}_{k}^{\dagger}\left(t\right)+\widehat{p}_{k}^{\dagger}\left(t\right)\widehat{p}_{k}\left(t\right)\right)\right]&\\+\left[\left(\widehat{q}_{k}^{\dagger}\left(t\right)\widehat{p}_{k}\left(t\right)+\widehat{p}_{k}^{\dagger}\left(t\right)\widehat{q}_{k}\left(t\right)\right),\left(\widehat{p}_{k}\left(t\right)\widehat{p}_{k}^{\dagger}\left(t\right)+\widehat{p}_{k}^{\dagger}\left(t\right)\widehat{p}_{k}\left(t\right)\right)\right].
	\label{SM85}
\end{align}

The first commutator in (\ref{SM85}) leads to,
\begin{align*}
{\color{blue}\left[\widehat{p}_{k}\left(t\right)\widehat{q}_{k}^{\dagger}\left(t\right),\widehat{p}_{k}\left(t\right)\widehat{p}_{k}^{\dagger}\left(t\right)\right]}+{\color{red}\left[\widehat{p}_{k}\left(t\right)\widehat{q}_{k}^{\dagger}\left(t\right),\widehat{p}_{k}^{\dagger}\left(t\right)\widehat{p}_{k}\left(t\right)\right]}&\\+{\color{magenta}\left[\widehat{q}_{k}\left(t\right)\widehat{p}_{k}^{\dagger}\left(t\right),\widehat{p}_{k}\left(t\right)\widehat{p}_{k}^{\dagger}\left(t\right)\right]}+{\color{teal}\left[\widehat{q}_{k}\left(t\right)\widehat{p}_{k}^{\dagger}\left(t\right),\widehat{p}_{k}^{\dagger}\left(t\right)\widehat{p}_{k}\left(t\right)\right]};
\end{align*}

\begin{align}
\nonumber{\color{blue}\left[\widehat{p}_{k}\left(t\right)\widehat{q}_{k}^{\dagger}\left(t\right),\widehat{p}_{k}\left(t\right)\widehat{p}_{k}^{\dagger}\left(t\right)\right]}=\widehat{p}_{k}\left(t\right)\left[\widehat{q}_{k}^{\dagger}\left(t\right),\widehat{p}_{k}\left(t\right)\right]\widehat{p}_{k}^{\dagger}\left(t\right)+\cancelto{0}{\left[\widehat{p}_{k}\left(t\right),\widehat{p}_{k}\left(t\right)\right]}\widehat{q}_{k}^{\dagger}\left(t\right)\widehat{p}_{k}^{\dagger}\left(t\right)&\\+\widehat{p}_{k}\left(t\right)\widehat{p}_{k}\left(t\right)\cancelto{0}{\left[\widehat{q}_{k}^{\dagger}\left(t\right),\widehat{p}_{k}^{\dagger}\left(t\right)\right]}+\widehat{p}_{k}\left(t\right)\cancelto{0}{\left[\widehat{p}_{k}\left(t\right),\widehat{p}_{k}^{\dagger}\left(t\right)\right]}\widehat{q}_{k}^{\dagger}\left(t\right)=i\hslash\left(\widehat{p}_{k}\left(t\right)\widehat{p}_{k}^{\dagger}\left(t\right)\right),
	\label{SM86}
\end{align}

\begin{align}
\nonumber{\color{red}\left[\widehat{p}_{k}\left(t\right)\widehat{q}_{k}^{\dagger}\left(t\right),\widehat{p}_{k}^{\dagger}\left(t\right)\widehat{p}_{k}\left(t\right)\right]}=\widehat{p}_{k}\left(t\right)\cancelto{0}{\left[\widehat{q}_{k}^{\dagger}\left(t\right),\widehat{p}_{k}^{\dagger}\left(t\right)\right]}\widehat{p}_{k}\left(t\right)+\cancelto{0}{\left[\widehat{p}_{k}\left(t\right),\widehat{p}_{k}^{\dagger}\left(t\right)\right]}\widehat{q}_{k}^{\dagger}\left(t\right)\widehat{p}_{k}\left(t\right)&\\+\widehat{p}_{k}^{\dagger}\left(t\right)\widehat{p}_{k}\left(t\right)\left[\widehat{q}_{k}^{\dagger}\left(t\right),\widehat{p}_{k}\left(t\right)\right]+\widehat{p}_{k}^{\dagger}\left(t\right)\cancelto{0}{\left[\widehat{p}_{k}\left(t\right),\widehat{p}_{k}^{\dagger}\left(t\right)\right]}\widehat{q}_{k}^{\dagger}\left(t\right)=i\hslash\left(\widehat{p}_{k}^{\dagger}\left(t\right)\widehat{p}_{k}\left(t\right)\right),
	\label{SM87}
\end{align}

\begin{align}
\nonumber{\color{magenta}\left[\widehat{q}_{k}\left(t\right)\widehat{p}_{k}^{\dagger}\left(t\right),\widehat{p}_{k}\left(t\right)\widehat{p}_{k}^{\dagger}\left(t\right)\right]}=\widehat{q}_{k}\left(t\right)\cancelto{0}{\left[\widehat{p}_{k}^{\dagger}\left(t\right),\widehat{p}_{k}\left(t\right)\right]}\widehat{p}_{k}^{\dagger}\left(t\right)+\cancelto{0}{\left[\widehat{q}_{k}\left(t\right),\widehat{p}_{k}\left(t\right)\right]}\widehat{p}_{k}^{\dagger}\left(t\right)\widehat{p}_{k}^{\dagger}\left(t\right)&\\+\widehat{p}_{k}\left(t\right)\widehat{q}_{k}\left(t\right)\cancelto{0}{\left[\widehat{p}_{k}^{\dagger}\left(t\right),\widehat{p}_{k}^{\dagger}\left(t\right)\right]}+\widehat{p}_{k}\left(t\right)\left[\widehat{q}_{k}\left(t\right),\widehat{q}_{k}^{\dagger}\left(t\right)\right]\widehat{p}_{k}^{\dagger}\left(t\right)=i\hslash\left(\widehat{p}_{k}\left(t\right)\widehat{p}_{k}^{\dagger}\left(t\right)\right),
	\label{SM88}
\end{align}

\begin{align}
\nonumber{\color{teal}\left[\widehat{q}_{k}\left(t\right)\widehat{p}_{k}^{\dagger}\left(t\right),\widehat{p}_{k}^{\dagger}\left(t\right)\widehat{p}_{k}\left(t\right)\right]}=\widehat{q}_{k}\left(t\right)\cancelto{0}{\left[\widehat{p}_{k}^{\dagger}\left(t\right),\widehat{p}_{k}^{\dagger}\left(t\right)\right]}\widehat{p}_{k}\left(t\right)+\left[\widehat{q}_{k}\left(t\right),\widehat{p}_{k}^{\dagger}\left(t\right)\right]\widehat{p}_{k}^{\dagger}\left(t\right)\widehat{p}_{k}\left(t\right)&\\+\widehat{p}_{k}^{\dagger}\left(t\right)\widehat{q}_{k}\left(t\right)\cancelto{0}{\left[\widehat{p}_{k}^{\dagger}\left(t\right),\widehat{p}_{k}\left(t\right)\right]}+\widehat{p}_{k}^{\dagger}\left(t\right)\cancelto{0}{\left[\widehat{q}_{k}\left(t\right),\widehat{p}_{k}\left(t\right)\right]}\widehat{p}_{k}^{\dagger}\left(t\right)=i\hslash\left(\widehat{p}_{k}^{\dagger}\left(t\right)\widehat{p}_{k}\left(t\right)\right).
	\label{SM89}
\end{align}

And similarly for the second commutator in (\ref{SM85}):

\begin{align*}
{\color{blue}\left[\widehat{q}_{k}^{\dagger}\left(t\right)\widehat{p}_{k}\left(t\right),\widehat{p}_{k}\left(t\right)\widehat{p}_{k}^{\dagger}\left(t\right)\right]}+{\color{red}\left[\widehat{q}_{k}^{\dagger}\left(t\right)\widehat{p}_{k}\left(t\right),\widehat{p}_{k}^{\dagger}\left(t\right)\widehat{p}_{k}\left(t\right)\right]}&\\+{\color{magenta}\left[\widehat{p}_{k}^{\dagger}\left(t\right)\widehat{q}_{k}\left(t\right),\widehat{p}_{k}\left(t\right)\widehat{p}_{k}^{\dagger}\left(t\right)\right]}+{\color{teal}\left[\widehat{p}_{k}^{\dagger}\left(t\right)\widehat{q}_{k}\left(t\right),\widehat{p}_{k}^{\dagger}\left(t\right)\widehat{p}_{k}\left(t\right)\right];}
\end{align*}

\begin{align}
\nonumber{\color{blue}\left[\widehat{q}_{k}^{\dagger}\left(t\right)\widehat{p}_{k}\left(t\right),\widehat{p}_{k}\left(t\right)\widehat{p}_{k}^{\dagger}\left(t\right)\right]}=\widehat{q}_{k}^{\dagger}\left(t\right)\cancelto{0}{\left[\widehat{p}_{k}\left(t\right),\widehat{p}_{k}\left(t\right)\right]}\widehat{p}_{k}^{\dagger}\left(t\right)+\left[\widehat{q}_{k}^{\dagger}\left(t\right),\widehat{p}_{k}\left(t\right)\right]\widehat{p}_{k}\left(t\right)\widehat{p}_{k}^{\dagger}\left(t\right)&\\+\widehat{p}_{k}\left(t\right)\widehat{q}_{k}^{\dagger}\left(t\right)\cancelto{0}{\left[\widehat{p}_{k}\left(t\right),\widehat{p}_{k}^{\dagger}\left(t\right)\right]}+\widehat{p}_{k}\left(t\right)\cancelto{0}{\left[\widehat{q}_{k}^{\dagger}\left(t\right),\widehat{p}_{k}^{\dagger}\left(t\right)\right]}\widehat{p}_{k}\left(t\right)=i\hslash\left(\widehat{p}_{k}\left(t\right)\widehat{p}_{k}^{\dagger}\left(t\right)\right),
	\label{SM90}
\end{align}

\begin{align}
\nonumber{\color{red}\left[\widehat{q}_{k}^{\dagger}\left(t\right)\widehat{p}_{k}\left(t\right),\widehat{p}_{k}^{\dagger}\left(t\right)\widehat{p}_{k}\left(t\right)\right]}=\widehat{q}_{k}^{\dagger}\left(t\right)\cancelto{0}{\left[\widehat{p}_{k}\left(t\right),\widehat{p}_{k}^{\dagger}\left(t\right)\right]}\widehat{p}_{k}\left(t\right)+\cancelto{0}{\left[\widehat{q}_{k}^{\dagger}\left(t\right),\widehat{p}_{k}^{\dagger}\left(t\right)\right]}\widehat{p}_{k}\left(t\right)\widehat{p}_{k}\left(t\right)&\\+\widehat{p}_{k}^{\dagger}\left(t\right)\widehat{q}_{k}^{\dagger}\left(t\right)\cancelto{0}{\left[\widehat{p}_{k}\left(t\right),\widehat{p}_{k}\left(t\right)\right]}+\widehat{p}_{k}^{\dagger}\left(t\right)\left[\widehat{q}_{k}^{\dagger}\left(t\right),\widehat{p}_{k}\left(t\right)\right]\widehat{p}_{k}\left(t\right)=i\hslash\left(\widehat{p}_{k}^{\dagger}\left(t\right)\widehat{p}_{k}\left(t\right)\right),
	\label{SM91}
\end{align}

\begin{align}
\nonumber{\color{magenta}\left[\widehat{p}_{k}^{\dagger}\left(t\right)\widehat{q}_{k}\left(t\right),\widehat{p}_{k}\left(t\right)\widehat{p}_{k}^{\dagger}\left(t\right)\right]}=\widehat{p}_{k}^{\dagger}\left(t\right)\cancelto{0}{\left[\widehat{q}_{k}\left(t\right),\widehat{p}_{k}\left(t\right)\right]}\widehat{p}_{k}^{\dagger}\left(t\right)+\cancelto{0}{\left[\widehat{p}_{k}^{\dagger}\left(t\right),\widehat{p}_{k}\left(t\right)\right]}\widehat{q}_{k}\left(t\right)\widehat{p}_{k}^{\dagger}\left(t\right)&\\+\widehat{p}_{k}\left(t\right)\widehat{p}_{k}^{\dagger}\left(t\right)\left[\widehat{q}_{k}\left(t\right),\widehat{p}_{k}^{\dagger}\left(t\right)\right]+\widehat{p}_{k}\left(t\right)\cancelto{0}{\left[\widehat{p}_{k}^{\dagger}\left(t\right),\widehat{p}_{k}^{\dagger}\left(t\right)\right]}\widehat{q}_{k}\left(t\right)=i\hslash\left(\widehat{p}_{k}\left(t\right)\widehat{p}_{k}^{\dagger}\left(t\right)\right),
	\label{SM92}
\end{align}

\begin{align}
\nonumber{\color{teal}\left[\widehat{p}_{k}^{\dagger}\left(t\right)\widehat{q}_{k}\left(t\right),\widehat{p}_{k}^{\dagger}\left(t\right)\widehat{p}_{k}\left(t\right)\right]}=\widehat{p}_{k}^{\dagger}\left(t\right)\left[\widehat{q}_{k}\left(t\right),\widehat{p}_{k}^{\dagger}\left(t\right)\right]\widehat{p}_{k}\left(t\right)+\cancelto{0}{\left[\widehat{p}_{k}^{\dagger}\left(t\right),\widehat{p}_{k}^{\dagger}\left(t\right)\right]}\widehat{q}_{k}\left(t\right)\widehat{p}_{k}\left(t\right)&\\+\widehat{p}_{k}^{\dagger}\left(t\right)\widehat{p}_{k}^{\dagger}\left(t\right)\cancelto{0}{\left[\widehat{q}_{k}\left(t\right),\widehat{p}_{k}\left(t\right)\right]}+\widehat{p}_{k}^{\dagger}\left(t\right)\cancelto{0}{\left[\widehat{p}_{k}^{\dagger}\left(t\right),\widehat{p}_{k}\left(t\right)\right]}\widehat{q}_{k}\left(t\right)=i\hslash\left(\widehat{p}_{k}^{\dagger}\left(t\right)\widehat{p}_{k}\left(t\right)\right).
	\label{SM93}
\end{align}

Again, as expected:

\begin{align}
	\boxed{\left[\widehat{J}_{0},\widehat{J}_{-}\right]=-\frac{1}{2\hslash}\left\{ \widehat{p}_{k}^{\dagger}\left(t\right)\widehat{p}_{k}\left(t\right)+\widehat{p}_{k}\left(t\right)\widehat{p}_{k}^{\dagger}\left(t\right)\right\} =-\widehat{J}_{-}.}
	\label{SM94}
\end{align}

\section{VIII. MASS AND FREQUENCY IN TIME-VARYING MEDIA}
Thus far it has been proved that the Hamiltonian (\ref{SM48}) can be written in terms of (\ref{SM50}), allowing the use of disentangling technique proposed by Cheng. Likewise, the time-dependent prefactors $a_{1}(t)$, $a_{2}(t)$, and $a_{3}(t)$ have been clearly identified in (\ref{SM53})-(\ref{SM55}). Therefore, on account of the form of the Hamiltonian given in (\ref{SM48}), it can be seen that the time-varying dielectric permittivity $\varepsilon(t)$ portrays the role of time-dependent mass $m(t)$:

\begin{align}
	\boxed{a_{3}\left(t\right)=\hslash\,\frac{1}{\varepsilon\left(t\right)}=\frac{\hbar}{m(t)}\Rightarrow m(t)=\varepsilon\left(t\right),}
	\label{SM95}
\end{align}
and that the time-dependent frequency $\omega(t)$ is more intricately related with the modulation of both the permittivity and the permeability:
\begin{align}
	\boxed{a_{1}\left(t\right)=\hslash\,\frac{k^{2}c^{2}}{\mu\left(t\right)}\Rightarrow\hslash\,\frac{k^{2}c^{2}}{\mu\left(t\right)}=\hbar\,m(t)\omega^{2}(t)\Rightarrow\omega^{2}(t)=\frac{k^{2}c^{2}}{\varepsilon\left(t\right)\mu\left(t\right)}.}
	\label{SM96}
\end{align}
Furthermore, the absence of crossed terms (as presented in $\widehat{J}_{0}$) leads unequivocally to,
\begin{align}
	a_{2}\left(t\right)=0.
	\label{SM97}
\end{align}
In virtue of these relationships, here below we examine two potential cases of interest.

\subsection{Dielectric temporal metamaterial}
Considering a non-magnetized ($\mu(t)=1$) temporal metamaterial solely characterized by its dielectric permittivity $\varepsilon(t)$, the Hamiltonian (\ref{SM48}) presents the following form:
\begin{align}
	\widehat{\mathcal{H}}\left(t\right)=\frac{1}{2}\,\int_{0}^{\infty}dk\,\left(\frac{1}{\varepsilon\left(t\right)}\,\left(\widehat{p}_{k}\left(t\right)\,\widehat{p}_{k}^{\dagger}\left(t\right)+\widehat{p}_{k}^{\dagger}\left(t\right)\,\widehat{p}_{k}\left(t\right)\right)+k^{2}c^{2}\,\left(\widehat{q}_{k}\left(t\right)\,\widehat{q}_{k}^{\dagger}\left(t\right)+\widehat{q}_{k}^{\dagger}\left(t\right)\,\widehat{q}_{k}\left(t\right)\right)\right)=\int_{0}^{\infty}dk\,\widehat{\mathcal{H}}_{k}\left(t\right).
	\label{SM98}
\end{align}
Using again (\ref{SM53})-(\ref{SM55}), the notion of $\varepsilon(t)$ as the time-dependent mass of the system prevails,

\begin{align}
	a_{3}(t)=\hslash\,\frac{1}{\varepsilon\left(t\right)}=\frac{\hbar}{m(t)}\Rightarrow m(t)=\varepsilon\left(t\right),
	\label{SM99}
\end{align}
and time-dependent frequency $\omega(t)$ is only related with the modulation of $\varepsilon(t)$,
\begin{align}
	a_{1}(t)=\hslash\,\varepsilon\left(t\right)\frac{k^{2}c^{2}}{\varepsilon\left(t\right)}=\hbar m(t)\omega^{2}(t)\Rightarrow\omega^{2}(t)=\frac{k^{2}c^{2}}{\varepsilon\left(t\right)}.
	\label{SM100}
\end{align}
Again, owing to the absence of crossed terms,
\begin{align}
	a_{2}(t)=0.
	\label{SM101}
\end{align}
Inserting (\ref{SM99})-(\ref{SM101}) into (\ref{SM58})-(\ref{SM60}), along with the initial condition given in (\ref{SM61}), and using Riccati transformation, we recover the second order-differential equation (\ref{SM62}) with time-dependent coefficients:
\begin{align}
	\boxed{\partial_{t}^{2}u(t)+\frac{\partial_{t}\varepsilon\left(t\right)}{\varepsilon\left(t\right)}\partial_{t}u(t)+\frac{k^{2}c^{2}}{\varepsilon\left(t\right)}u(t)=0,}
	\label{SM102}
\end{align}
 
\subsection{Magnetic temporal metamaterial}
Similarly, a non-dielectric ($\varepsilon(t)=1$) temporal metamaterial solely characterized by its magnetic permeability $\mu(t)$, the Hamiltonian (\ref{SM48}) presents the following form:
\begin{align}
	\widehat{\mathcal{H}}\left(t\right)=\frac{1}{2}\,\int_{0}^{\infty}dk\,\left(\left(\widehat{p}_{k}\left(t\right)\,\widehat{p}_{k}^{\dagger}\left(t\right)+\widehat{p}_{k}^{\dagger}\left(t\right)\,\widehat{p}_{k}\left(t\right)\right)+\frac{k^{2}c^{2}}{\mu\left(t\right)}\,\left(\widehat{q}_{k}\left(t\right)\,\widehat{q}_{k}^{\dagger}\left(t\right)+\widehat{q}_{k}^{\dagger}\left(t\right)\,\widehat{q}_{k}\left(t\right)\right)\right)=\int_{0}^{\infty}dk\,\widehat{\mathcal{H}}_{k}\left(t\right).
	\label{SM103}
\end{align}
Using again (\ref{SM54})-(\ref{SM55}), the magnetic time-modulated scenario corresponds to a system of time-independent (constant) mass:
\begin{align}
	a_{3}(t)=\hslash=\frac{\hbar}{m(t)}\Rightarrow m(t)=1,
	\label{SM104}
\end{align}
and the time-dependent frequency $\omega(t)$ will be now only related with the modulation of $\mu(t)$,
\begin{align}
	a_{1}(t)=\hslash\,\varepsilon\left(t\right)\frac{k^{2}c^{2}}{\mu\left(t\right)}=\hbar m(t)\omega^{2}(t)\Rightarrow\omega^{2}(t)=\frac{k^{2}c^{2}}{\mu\left(t\right)}.
	\label{SM105}
\end{align}
And once again, 
\begin{align}
	a_{2}(t)=0.
	\label{SM106}
\end{align}
Inserting (\ref{SM104})-(\ref{SM106}) into (\ref{SM58})-(\ref{SM60}), with the initial condition given in (\ref{SM61}), the Riccati transformation allows us to recover the second order-differential equation (\ref{SM62}). In contrast with the previous case, now the damping factor is $\gamma_{0}(t)=0$ and the frequency $\omega(t)$ is the only time-dependent parameter:
\begin{align}
	\boxed{\partial_{t}^{2}u(t)+\frac{k^{2}c^{2}}{\mu\left(t\right)}u(t)=0.}
	\label{SM107}
\end{align}

\section{IX. BOGOLIUBOV TRANSFORMATIONS}
Using the time-evolution operator (\ref{SM57}), we can study the evolution of a coherent state and discuss its squeezing and coherence properties. Assuming an initial coherent state at $t=0$, the wave function at any later time will be given by,
\begin{align}
	\left|\psi(t)\right\rangle =\widehat{U}\left(t,0\right)\left|\alpha\right\rangle .
	\label{SM108}
\end{align}
Performing the usual quantization procedure at $t=0$, and identifying $m(0)=\varepsilon(0)$, and $\omega_{k}(0)$, creation and annihilation operators and their inverse relations (see Appendix B1 and B2) can be written as:
\begin{align}
	\widehat{a}_{k}(0)=\frac{1}{\sqrt{2\hbar\varepsilon(0)\omega_{k}(0)}}\left(\varepsilon(0)\omega_{k}(0)\widehat{q}_{k}(0)+i\widehat{p}_{k}(0)\right),
	\label{SM109}
\end{align}
\begin{align}
	\widehat{a}_{k}^{\dagger}(0)=\frac{1}{\sqrt{2\hbar\varepsilon(0)\omega_{k}(0)}}\left(\varepsilon(0)\omega_{k}(0)\widehat{q}_{k}(0)-i\widehat{p}_{k}(0)\right),
	\label{SM110}
\end{align}
\begin{align}
	\widehat{q}_{k}(0)=\frac{1}{2}\sqrt{\frac{2\hbar}{\varepsilon(0)\omega_{k}(0)}}\left(\widehat{a}_{k}(0)+\widehat{a}_{-k}^{\dagger}(0)\right)=\widehat{q}_{-k}^{\dagger}(0),
	\label{SM111}
\end{align}
\begin{align}
	\widehat{p}_{k}(0)=\frac{1}{2i}\sqrt{2\hbar\varepsilon(0)\omega_{k}(0)}\left(\widehat{a}_{k}(0)-\widehat{a}_{-k}^{\dagger}(0)\right)=\widehat{p}_{-k}^{\dagger}(0).
	\label{SM112}
\end{align}

In the Heisenberg picture, all state vectors are considered to remain constant at their initial values, whereas operators evolve in time according to:

\begin{align}
	\widehat{O}(t)\equiv\widehat{U}^{\dagger}(t,0)\widehat{O}(0)\widehat{U}(t,0).
	\label{SM113}
\end{align}

Thus, regarding the annihilation operator at $t=0$ and applying the rule of transformation into the Heisenberg picture, it follows that:

\begin{align}
	\hat{a}_{k}(t)=\sqrt{\frac{\varepsilon(0)\omega_{k}(0)}{2\hbar}}\hat{U}^{\dagger}(t,0)\hat{q}_{k}(0)\hat{U}(t,0)+\frac{i}{\sqrt{2\hbar\varepsilon(0)\omega_{k}(0)}}\hat{U}^{\dagger}(t,0)\hat{p}_{k}(0)\hat{U}(t,0)
	\label{SM114}
\end{align}

\begin{align}
\nonumber\hat{a}_{k}(t)=\sqrt{\frac{\varepsilon(0)\omega_{k}(0)}{2\hbar}}{\color{gray}{\color{blue}\exp\left(-c_{3}(t)\widehat{J}_{-}\right)\exp\left(-c_{2}(t)\widehat{J}_{0}\right)\exp\left(-c_{1}(t)\widehat{J}_{+}\right)}}\hat{q}_{k}(0){\color{green}{\color{purple}\exp\left(c_{1}(t)\widehat{J}_{+}\right)\exp\left(c_{2}(t)\widehat{J}_{0}\right)\exp\left(c_{3}(t)\widehat{J}_{-}\right)}}&\\+\frac{i}{\sqrt{2\hbar\varepsilon(0)\omega_{k}(0)}}{\color{gray}{\color{gray}{\color{blue}\exp\left(-c_{3}(t)\widehat{J}_{-}\right)\exp\left(-c_{2}(t)\widehat{J}_{0}\right)\exp\left(-c_{1}(t)\widehat{J}_{+}\right)}}}\hat{p}_{k}(0){\color{purple}{\color{green}{\color{purple}\exp\left(c_{1}(t)\widehat{J}_{+}\right)\exp\left(c_{2}(t)\widehat{J}_{0}\right)\exp\left(c_{3}(t)\widehat{J}_{-}\right)}}}.
	\label{SM115}
\end{align}
Notice that we have assumed that $c_{2}(t)$ along with $c_{1}(t)c_{3}(t)$ are real. Noteworthly, and without loss of generality, we can take both $c_{1}(t)$ and $c_{3}(t)$ as pure imaginary functions.

\subsection{Non-applicability of Cheng method}
Unfortunately, regarding time-modulated properties, the time-dependent functions $c_{i}(t)$ (with $i=\{1,2,3\}$) given in (\ref{SM64})-(\ref{SM66}) to be then inserted into the disentangled exponentials of time-evolution operator (\ref{SM62}) have the following form:
\begin{align}
	c_{1}(t)=i\varepsilon(t)\frac{\partial_{t}u(t)}{u(t)},
	\label{SM116}
\end{align}
\begin{align}
	c_{2}(t)=-2\log\frac{u(t)}{u(0)},
	\label{SM117}
\end{align}
\begin{align}
	c_{3}(t)=-iu^{2}(0)\int_{0}^{t}\frac{dt'}{\varepsilon(t')u^{2}(t')}.
	\label{SM118}
\end{align}

Both $c_{1}(t)$ and $c_{3}(t)$ bring about a rapid divergence issue in cases where both mass and frequency (and accordingly, $\varepsilon(t)$ and $\mu(t)$ in our case) are time-modulated functions exhibiting a significant non-linear behavior. The solution for second-order time coefficients differential equation (\ref{SM62}) leads to oscillatory solutions of $u(t)$ for a wide variety of scenarios wherein optical properties are modulated in time. Furthermore, using numerical solvers directly on Riccati equation (\ref{SM58}), and consequently obtaining (\ref{SM59}) and (\ref{SM60}), yield non-convergence problems for long times. For these reasons, we finally opt for addressing the problem under minimal assumptions, directly by elaborating from Heisenberg equations of motion.

\part{Part IV: Numerical methods}
In this part, we propose iterative matrix numerical methods as an alternative to address squeezed vacuum states and photons production. First, we write down the input-output relation for the generalized parametric oscillator. Despite the non-applicability of Cheng method, as far as the Hamiltonian (\ref{SM48}) can be written as (\ref{SM49}), the equivalences $m(t)=\varepsilon(t)$ and $\omega^{2}(t)=k^{2}c^{2}/\varepsilon(t)\mu(t)$ for the quantum electrodynamic case hold on. Then, we particularize the general method for dielectric temporal metamaterials. Finally, we examine in detail two purely dielectric time-modulated cases, namely (i) tapered photonic switching, which has been recently studied under the lenses of classical physics and (ii) Gaussian pulse modulation, which it is a common scenario in a variety of applications. In the quantum context, the former opens a discussion about the proximity of ultra-fast modulations and temporal boundaries. The later exhibits an interesting behavior for long times, reporting a maximum amount of squeezed photons for for a specific non-trivial dispersion of the Gaussian profile.

\section{X. IMPLEMENTATION OF THE THEORETICAL METHOD FOR PHOTON PRODUCTION IN ARBITRARY TIME-VARYING MEDIA}
\subsection{Generalized parametric oscillator}
The Hamiltonian for a harmonic oscillator with time-varying mass and frequency,
\begin{align}
	\widehat{\mathcal{H}}\left(t\right)=\frac{1}{2}\,\frac{1}{m\left(t\right)}\,\widehat{p}^{2}\left(t\right)+m\left(t\right)\omega^{2}\left(t\right)\,\widehat{q}^{2}\left(t\right),
	\label{SM119}
\end{align}
along with the well-behaved commutation relations for the time-varying position and momentum operator,
\begin{align}
	\left[\widehat{q}\left(t\right),\widehat{p}\left(t\right)\right]=i\hslash,
	\label{SM120}
\end{align}
allow us to directly obtain the equations of motion:
\begin{align}
	\frac{d\widehat{q}\left(t\right)}{dt}=\frac{1}{i\hslash}\,\left[\widehat{q}\left(t\right),\widehat{H}\left(t\right)\right]=\frac{1}{m\left(t\right)}\,\widehat{p}\left(t\right),
	\label{SM121}
\end{align}
\begin{align}
	\frac{d\widehat{p}\left(t\right)}{dt}=\frac{1}{i\hslash}\,\left[\widehat{p}\left(t\right),\widehat{H}\left(t\right)\right]=-m\left(t\right)\omega^{2}\left(t\right)\,\widehat{q}\left(t\right).
	\label{SM122}
\end{align}
This can be expressed in a discretized fashion as:
\begin{align}
	\frac{\widehat{q}\left(t+\varDelta t\right)-\widehat{q}\left(t\right)}{\varDelta t}=\frac{1}{m\left(t\right)}\,\widehat{p}\left(t\right),
	\label{SM123}
\end{align}
\begin{align}
	\frac{\widehat{p}\left(t+\varDelta t\right)-\widehat{p}\left(t\right)}{\varDelta t}=-m\left(t\right)\omega^{2}\left(t\right)\,\widehat{q}\left(t\right),
	\label{SM124}
\end{align}
thereby fulfilling the following matrix form,
\begin{align}
	\left[\begin{array}{c}
		\widehat{q}\left(t+\varDelta t\right)\\
		\widehat{p}\left(t+\varDelta t\right)
	\end{array}\right]=\mathbf{M}\left(t\right)\,\left[\begin{array}{c}
		\widehat{q}\left(t\right)\\
		\widehat{p}\left(t\right)
	\end{array}\right],
	\label{SM125}
\end{align}
where,

\begin{align}
	\mathbf{M}\left(t\right)=\left[\begin{array}{cc}
		1 & \frac{\varDelta t}{m\left(t\right)}\\
		-\varDelta t\,m\left(t\right)\omega^{2}\left(t\right) & 1
	\end{array}\right].
	\label{SM126}
\end{align}

Then, taking a discretized interval of time $t_{n}=n\triangle t$, from $0$ to $t$ by $N$ steps of $\triangle t=t/N$, leads to
\begin{align}
	\left[\begin{array}{c}
		\widehat{q}\left(t\right)\\
		\widehat{p}\left(t\right)
	\end{array}\right]=\left(\prod_{n=1}^{N}\mathbf{M}\left(n\triangle t\right)\right)\,\left[\begin{array}{c}
		\widehat{q}\left(0\right)\\
		\widehat{p}\left(0\right)
	\end{array}\right].
	\label{SM127}
\end{align}

These two coupled equations can be numerically solved to obtain an input-output relation of the form:

\begin{align}
	\left[\begin{array}{c}
		\widehat{q}\left(t\right)\\
		\widehat{p}\left(t\right)
	\end{array}\right]=\mathbf{M}\left(t\right)\,\left[\begin{array}{c}
		\widehat{q}\left(0\right)\\
		\widehat{p}\left(0\right)
	\end{array}\right]=\left[\begin{array}{cc}
		M_{11}\left(t\right) & M_{12}\left(t\right)\\
		M_{21}\left(t\right) & M_{22}\left(t\right)
	\end{array}\right]\,\left[\begin{array}{c}
		\widehat{q}\left(0\right)\\
		\widehat{p}\left(0\right)
	\end{array}\right],
	\label{SM128}
\end{align}

with,

\begin{align}
	\widehat{q}\left(t\right)=M_{11}\left(t\right)\widehat{q}\left(0\right)+M_{12}\left(t\right)\widehat{p}\left(0\right),
	\label{SM129}
\end{align}
\begin{align}
	\widehat{p}\left(t\right)=M_{21}\left(t\right)\widehat{q}\left(0\right)+M_{22}\left(t\right)\widehat{p}\left(0\right).
	\label{SM130}
\end{align}

Thus, quantum correlations of the position and momentum operators only depend on the correlations at the initial-time. Let us assume that the initial-time correlations are those of a harmonic oscillator in the ground state:
\begin{align}
	\left\langle \widehat{q}\left(0\right)\right\rangle =0,
	\label{SM131}
\end{align}
\begin{align}
	\left\langle \widehat{p}\left(0\right)\right\rangle =0,
	\label{SM132}
\end{align}
\begin{align}
	\left\langle \widehat{q}^{2}\left(0\right)\right\rangle =\frac{\hslash}{2}\,\frac{1}{m\left(0\right)\omega\left(0\right)},
	\label{SM133}
\end{align}
\begin{align}
	\left\langle \widehat{p}^{2}\left(0\right)\right\rangle =\frac{\hslash}{2}\,m\left(0\right)\omega\left(0\right),
	\label{SM134}
\end{align}
\begin{align}
	\left\langle \widehat{q}\left(0\right)\widehat{p}\left(0\right)\right\rangle =i\,\frac{\hslash}{2},
	\label{SM135}
\end{align}
\begin{align}
	\left\langle \widehat{p}\left(0\right)\widehat{q}\left(0\right)\right\rangle =-i\,\frac{\hslash}{2}.
	\label{SM136}
\end{align}

From this, it follows that:
\begin{align}
	\left\langle \widehat{q}\left(t\right)\right\rangle =M_{11}\left(t\right)\left\langle \widehat{q}\left(0\right)\right\rangle +M_{12}\left(t\right)\left\langle \widehat{p}\left(0\right)\right\rangle =0,
	\label{SM137}
\end{align}
\begin{align}
	\left\langle \widehat{p}\left(t\right)\right\rangle =M_{21}\left(t\right)\left\langle \widehat{q}\left(0\right)\right\rangle +M_{22}\left(t\right)\left\langle \widehat{p}\left(0\right)\right\rangle =0,
	\label{SM138}
\end{align}
\begin{align}
\left\langle \widehat{q}^{2}\left(t\right)\right\rangle=\left\langle \widehat{q}^{\dagger}\left(t\right)\widehat{q}\left(t\right)\right\rangle =\frac{\hslash}{2}\,\left\{ \frac{\left|M_{11}\left(t\right)\right|^{2}}{m\left(0\right)\omega\left(0\right)}+\left|M_{12}\left(t\right)\right|^{2}m\left(0\right)\omega\left(0\right)\right\},
	\label{SM139}
\end{align}
\begin{align}
	\left\langle \widehat{p}^{2}\left(t\right)\right\rangle=\left\langle \widehat{p}^{\dagger}\left(t\right)\widehat{p}\left(t\right)\right\rangle=\frac{\hslash}{2}\,\left\{\frac{\left|M_{21}\left(t\right)\right|^{2}}{m\left(0\right)\omega\left(0\right)}+\left|M_{22}\left(t\right)\right|^{2}m\left(0\right)\omega\left(0\right)\right\} ,
	\label{SM140}
\end{align}

\begin{align}
	\widehat{\mathcal{H}}\left(t\right)=\frac{1}{2}\,\frac{1}{m\left(t\right)}\,\left\langle \widehat{p}^{2}\left(t\right)\right\rangle +m\left(t\right)\omega^{2}\left(t\right)\,\left\langle \widehat{q}^{2}\left(t\right)\right\rangle .
	\label{SM141}
\end{align}
\subsection{Time-varying media case}
As it has been pointed out before, the equivalences $m(t)=\varepsilon(t)$ and $\omega^{2}(t)=k^{2}c^{2}/\varepsilon(t)\mu(t)$ for the quantum electrodynamic case hold on. Thus, equations of motion read as:
\begin{align}
	\frac{d\widehat{q}_{k}\left(t\right)}{dt}=\frac{1}{i\hslash}\,\left[\widehat{q_{k}}\left(t\right),\widehat{H_{k}}\left(t\right)\right]=\frac{1}{\varepsilon\left(t\right)}\,\widehat{p}_{k}\left(t\right),
	\label{SM142}
\end{align}
\begin{align}
	\frac{d\widehat{p}_{k}\left(t\right)}{dt}=\frac{1}{i\hslash}\,\left[\widehat{p}_{k}\left(t\right),\widehat{H}_{k}\left(t\right)\right]=-\frac{k^{2}c^{2}}{\mu(t)}\,\widehat{q}_{k}\left(t\right).
	\label{SM143}
\end{align}

Once again, (\ref{SM142})-(\ref{SM143}) can be written in matrix form

\begin{align}
	\left[\begin{array}{c}
		\widehat{q}_{k}\left(t+\varDelta t\right)\\
		\widehat{p}_{k}\left(t+\varDelta t\right)
	\end{array}\right]=\mathbf{M}_{k}\left(t\right)\,\left[\begin{array}{c}
		\widehat{q}_{k}\left(t\right)\\
		\widehat{p}_{k}\left(t\right)
	\end{array}\right],
	\label{SM144}
\end{align}

where,
\begin{align}
	\mathbf{M}_{k}\left(t\right)=\left[\begin{array}{cc}
		1 & \frac{\varDelta t}{\varepsilon\left(t\right)}\\
		-\varDelta t\,\frac{k^{2}c^{2}}{\mu(t)} & 1
	\end{array}\right].
	\label{SM145}
\end{align}

Then, assuming sufficiently small time steps, the input-output relation becomes,
\begin{align}
	\left[\begin{array}{c}
		\widehat{q}_{k}\left(t\right)\\
		\widehat{p}_{k}\left(t\right)
	\end{array}\right]=\left(\prod_{n=1}^{N}\mathbf{M}_{k}\left(n\triangle t\right)\right)\,\left[\begin{array}{c}
		\widehat{q}_{k}\left(0\right)\\
		\widehat{p}_{k}\left(0\right)
	\end{array}\right].
	\label{SM146}
\end{align}

In this case, we have no longer position, $\widehat{q}^{2}\left(t\right)$, and momentum, $\widehat{p}^{2}\left(t\right)$, associated with a generalized parametric oscillator, but two complex-conjugated oscillators related via reality conditions \ref{SM33} and \ref{SM34}. Thus, we need to check the variances of the forward and backward squeezed photons due to the temporal modulation. To do that, we start from,

\begin{align}
	\widehat{q}_{k}\left(t\right)=\sqrt{\frac{2\hslash}{\varepsilon\left(0\right)\omega_{k}\left(0\right)}}\,\frac{1}{2}\,\left(\widehat{a}_{-k}^{\dagger}\left(t\right)+\widehat{a}_{k}\left(t\right)\right),
	\label{SM147}
\end{align}
\begin{align}
	\widehat{q}_{k}^{\dagger}\left(t\right)=\sqrt{\frac{2\hslash}{\varepsilon\left(0\right)\omega_{k}\left(0\right)}}\,\frac{1}{2}\,\left(\widehat{a}_{-k}\left(t\right)+\widehat{a}_{k}^{\dagger}\left(t\right)\right),
	\label{SM148}
\end{align}
\begin{align}
	\widehat{p}_{k}\left(t\right)=\sqrt{2\hslash\varepsilon\left(0\right)\omega_{k}\left(0\right)}\,\frac{i}{2}\left(\widehat{a}_{-k}^{\dagger}\left(t\right)-\widehat{a}_{k}\left(t\right)\right),
	\label{SM149}
\end{align}
\begin{align}
	\widehat{p}_{k}^{\dagger}\left(t\right)=\sqrt{2\hslash\varepsilon\left(0\right)\omega_{k}\left(0\right)}\,\frac{i}{2}\left(\widehat{a}_{-k}\left(t\right)-\widehat{a}_{k}^{\dagger}\left(t\right)\right),
	\label{SM150}
\end{align}

which, evaluated in $t=0$, result in,
\begin{align}
	\widehat{q}_{k}\left(0\right)=\sqrt{\frac{2\hslash}{\varepsilon\left(0\right)\omega_{k}\left(0\right)}}\,\frac{1}{2}\,\left(\widehat{a}_{-k}^{\dagger}\left(0\right)+\widehat{a}_{k}\left(0\right)\right),
	\label{SM151}
\end{align}
\begin{align}
	\widehat{q}_{k}^{\dagger}\left(0\right)=\sqrt{\frac{2\hslash}{\varepsilon\left(0\right)\omega_{k}\left(0\right)}}\,\frac{1}{2}\,\left(\widehat{a}_{-k}\left(0\right)+\widehat{a}_{k}^{\dagger}\left(0\right)\right),
	\label{SM152}
\end{align}
\begin{align}
	\widehat{p}_{k}\left(0\right)=\sqrt{2\hslash\varepsilon\left(0\right)\omega_{k}\left(0\right)}\,\frac{i}{2}\left(\widehat{a}_{-k}^{\dagger}\left(0\right)-\widehat{a}_{k}\left(0\right)\right),
	\label{SM153}
\end{align}
\begin{align}
	\widehat{p}_{k}^{\dagger}\left(0\right)=-\sqrt{2\hslash\varepsilon\left(0\right)\omega_{k}\left(0\right)}\,\frac{i}{2}\left(\widehat{a}_{-k}\left(0\right)-\widehat{a}_{k}^{\dagger}\left(0\right)\right),
	\label{SM154}
\end{align}

thereby recovering expressions given in (\ref{SM111}) and (\ref{SM112}). From this, one can directly compute the following products of quantum operators:

\begin{align}
	\widehat{q}_{k}\left(0\right)\widehat{q}_{k}^{\dagger}\left(0\right)=\frac{1}{\varepsilon\left(0\right)\omega_{k}\left(0\right)}\,\frac{\hslash}{2}\left\{ \widehat{a}_{-k}^{\dagger}\left(0\right)\widehat{a}_{-k}\left(0\right)+\widehat{a}_{-k}^{\dagger}\left(0\right)\widehat{a}_{k}^{\dagger}\left(0\right)+\widehat{a}_{k}\left(0\right)\widehat{a}_{-k}\left(0\right)+\widehat{a}_{k}\left(0\right)\widehat{a}_{k}^{\dagger}\left(0\right)\right\} ,
	\label{SM155}
\end{align}
\begin{align}
	\widehat{q}_{k}^{\dagger}\left(0\right)\widehat{q}_{k}\left(0\right)=\frac{1}{\varepsilon\left(0\right)\omega_{k}\left(0\right)}\,\frac{\hslash}{2}\left\{ \widehat{a}_{-k}\left(0\right)\widehat{a}_{-k}^{\dagger}\left(0\right)+\widehat{a}_{-k}\left(0\right)\widehat{a}_{k}\left(0\right)+\widehat{a}_{k}^{\dagger}\left(0\right)\widehat{a}_{-k}^{\dagger}\left(0\right)+\widehat{a}_{k}^{\dagger}\left(0\right)\widehat{a}_{k}\left(0\right)\right\} ,
	\label{SM156}
\end{align}
\begin{align}
	\widehat{p}_{k}\left(0\right)\widehat{p}_{k}^{\dagger}\left(0\right)=\varepsilon\left(0\right)\omega_{k}\left(0\right)\,\frac{\hslash}{2}\left\{ \widehat{a}_{-k}^{\dagger}\left(0\right)\widehat{a}_{-k}\left(0\right)-\widehat{a}_{-k}^{\dagger}\left(0\right)\widehat{a}_{k}^{\dagger}\left(0\right)-\widehat{a}_{k}\left(0\right)\widehat{a}_{-k}\left(0\right)+\widehat{a}_{k}\left(0\right)\widehat{a}_{k}^{\dagger}\left(0\right)\right\} ,
	\label{SM157}
\end{align}
\begin{align}
	\widehat{p}_{k}^{\dagger}\left(0\right)\widehat{p}_{k}\left(0\right)=\varepsilon\left(0\right)\omega_{k}\left(0\right)\,\frac{\hslash}{2}\left\{ \widehat{a}_{-k}\left(0\right)\widehat{a}_{-k}^{\dagger}\left(0\right)-\widehat{a}_{-k}\left(0\right)\widehat{a}_{k}\left(0\right)-\widehat{a}_{k}^{\dagger}\left(0\right)\widehat{a}_{-k}^{\dagger}\left(0\right)+\widehat{a}_{k}^{\dagger}\left(0\right)\widehat{a}_{k}\left(0\right)\right\} .
	\label{SM158}
\end{align}

Equations (\ref{SM155})–(\ref{SM158}) prove the equal contribution of the two conjugated oscillators at the initial time ($t=0$),

\begin{align}
	\widehat{q}_{k}^{\dagger}\left(0\right)\widehat{q}_{k}\left(0\right)=\left(\widehat{q}_{k}\left(0\right)\widehat{q}_{k}^{\dagger}\left(0\right)\right)^{\dagger},
	\label{SM159}
\end{align}

\begin{align}
	\widehat{p}_{k}^{\dagger}\left(0\right)\widehat{p}_{k}\left(0\right)=\left(\widehat{p}_{k}\left(0\right)\widehat{p}_{k}^{\dagger}\left(0\right)\right)^{\dagger}.
	\label{SM160}
\end{align}

Evaluating crossed terms at $t=0$,

\begin{align}
	\widehat{q}_{k}\left(0\right)\widehat{p}_{k}\left(0\right)=i\frac{\hslash}{2}\left\{ \widehat{a}_{-k}^{\dagger}\left(0\right)\widehat{a}_{-k}^{\dagger}\left(0\right)-\widehat{a}_{-k}^{\dagger}\left(0\right)\widehat{a}_{k}\left(0\right)+\widehat{a}_{k}\left(0\right)\widehat{a}_{-k}^{\dagger}\left(0\right)-\widehat{a}_{k}\left(0\right)\widehat{a}_{k}\left(0\right)\right\} ,
	\label{SM161}
\end{align}
\begin{align}
	\widehat{q}_{k}\left(0\right)\widehat{p}_{k}^{\dagger}\left(0\right)=-i\frac{\hslash}{2}\left\{ \widehat{a}_{-k}^{\dagger}\left(0\right)\widehat{a}_{-k}\left(0\right)-\widehat{a}_{-k}^{\dagger}\left(0\right)\widehat{a}_{k}^{\dagger}\left(0\right)+\widehat{a}_{k}\left(0\right)\widehat{a}_{-k}\left(0\right)-\widehat{a}_{k}\left(0\right)\widehat{a}_{k}^{\dagger}\left(0\right)\right\} ,
	\label{SM162}
\end{align}
\begin{align}
	\widehat{q}_{k}^{\dagger}\left(0\right)\widehat{p}_{k}\left(0\right)=i\frac{\hslash}{2}\left\{ \widehat{a}_{-k}\left(0\right)\widehat{a}_{-k}^{\dagger}\left(0\right)-\widehat{a}_{-k}\left(0\right)\widehat{a}_{k}\left(0\right)+\widehat{a}_{k}^{\dagger}\left(0\right)\widehat{a}_{-k}^{\dagger}\left(0\right)-\widehat{a}_{k}^{\dagger}\left(0\right)\widehat{a}_{k}\left(0\right)\right\} ,
	\label{SM163}
\end{align}
\begin{align}
	\widehat{q}_{k}^{\dagger}\left(0\right)\widehat{p}_{k}^{\dagger}\left(0\right)=-i\frac{\hslash}{2}\left\{ \widehat{a}_{-k}\left(0\right)\widehat{a}_{-k}\left(0\right)-\widehat{a}_{-k}\left(0\right)\widehat{a}_{k}^{\dagger}\left(0\right)+\widehat{a}_{k}^{\dagger}\left(0\right)\widehat{a}_{-k}\left(0\right)-\widehat{a}_{k}^{\dagger}\left(0\right)\widehat{a}_{k}^{\dagger}\left(0\right)\right\} ,
	\label{SM164}
\end{align}

\begin{align}
	\widehat{p}_{k}\left(0\right)\widehat{q}_{k}\left(0\right)=i\frac{\hslash}{2}\left\{ \widehat{a}_{-k}^{\dagger}\left(0\right)\widehat{a}_{-k}^{\dagger}\left(0\right)+\widehat{a}_{-k}^{\dagger}\left(0\right)\widehat{a}_{k}\left(0\right)-\widehat{a}_{k}\left(0\right)\widehat{a}_{-k}^{\dagger}\left(0\right)-\widehat{a}_{k}\left(0\right)\widehat{a}_{k}\left(0\right)\right\},
	\label{SM165}
\end{align}
\begin{align}
	\widehat{p}_{k}\left(0\right)\widehat{q}_{k}^{\dagger}\left(0\right)=i\frac{\hslash}{2}\left\{ \widehat{a}_{-k}^{\dagger}\left(0\right)\widehat{a}_{-k}\left(0\right)+\widehat{a}_{-k}^{\dagger}\left(0\right)\widehat{a}_{k}^{\dagger}\left(0\right)-\widehat{a}_{k}\left(0\right)\widehat{a}_{-k}\left(0\right)-\widehat{a}_{k}\left(0\right)\widehat{a}_{k}^{\dagger}\left(0\right)\right\},
	\label{SM166}
\end{align}
\begin{align}
	\widehat{p}_{k}^{\dagger}\left(0\right)\widehat{q}_{k}\left(0\right)=-i\frac{\hslash}{2}\left\{ \widehat{a}_{-k}\left(0\right)\widehat{a}_{-k}^{\dagger}\left(0\right)+\widehat{a}_{-k}\left(0\right)\widehat{a}_{k}\left(0\right)-\widehat{a}_{k}^{\dagger}\left(0\right)\widehat{a}_{-k}^{\dagger}\left(0\right)-\widehat{a}_{k}^{\dagger}\left(0\right)\widehat{a}_{k}\left(0\right)\right\},
	\label{SM167}
\end{align}
\begin{align}
	\widehat{p}_{k}^{\dagger}\left(0\right)\widehat{q}_{k}^{\dagger}\left(0\right)=-i\frac{\hslash}{2}\left\{ \widehat{a}_{-k}\left(0\right)\widehat{a}_{-k}\left(0\right)+\widehat{a}_{-k}\left(0\right)\widehat{a}_{k}^{\dagger}\left(0\right)-\widehat{a}_{k}^{\dagger}\left(0\right)\widehat{a}_{-k}\left(0\right)-\widehat{a}_{k}^{\dagger}\left(0\right)\widehat{a}_{k}^{\dagger}\left(0\right)\right\},
	\label{SM168}
\end{align}

leads to,

\begin{align}
	\widehat{p}_{k}^{\dagger}\left(0\right)\widehat{q}_{k}\left(0\right)=\left(\widehat{q}_{k}^{\dagger}\left(0\right)\widehat{p}_{k}\left(0\right)\right)^{\dagger},
	\label{SM169}
\end{align}
\begin{align}
	\widehat{q}_{k}^{\dagger}\left(0\right)\widehat{p}_{k}\left(0\right)=\left(\widehat{p}_{k}^{\dagger}\left(0\right)\widehat{q}_{k}\left(0\right)\right)^{\dagger},
	\label{SM170}
\end{align}
\begin{align}
	\widehat{p}_{k}^{\dagger}\left(0\right)\widehat{q}_{k}^{\dagger}\left(0\right)=\left(\widehat{q}_{k}\left(0\right)\widehat{p}_{k}\left(0\right)\right)^{\dagger},
	\label{SM171}
\end{align}
\begin{align}
	\widehat{q}_{k}^{\dagger}\left(0\right)\widehat{p}_{k}^{\dagger}\left(0\right)=\left(\widehat{p}_{k}\left(0\right)\widehat{q}_{k}\left(0\right)\right)^{\dagger}.
	\label{SM172}
\end{align}

Hence, assuming that the initial-time correlations are those of a harmonic oscillator in the ground state, it follows that,

\begin{align}
	\langle \widehat{q}_{k}\left(0\right)\rangle =\sqrt{\frac{2\hslash}{\varepsilon\left(0\right)\omega_{k}\left(0\right)}}\,\frac{1}{2}\langle \cancelto{0}{\widehat{a}_{-k}^{\dagger}\left(0\right)}+\cancelto{0}{\widehat{a}_{k}\left(0\right)}\rangle =0,
	\label{SM173}
\end{align}
\begin{align}
	\langle \widehat{q}_{k}^{\dagger}\left(0\right)\rangle =\sqrt{\frac{2\hslash}{\varepsilon\left(0\right)\omega_{k}\left(0\right)}}\,\frac{1}{2}\langle \cancelto{0}{\widehat{a}_{-k}\left(0\right)}+\cancelto{0}{\widehat{a}_{k}^{\dagger}\left(0\right)}\rangle =0,
	\label{SM174}
\end{align}
\begin{align}
	\langle \widehat{p}_{k}\left(0\right)\rangle =\sqrt{2\hslash\varepsilon\left(0\right)\omega_{k}\left(0\right)}\,\frac{i}{2}\langle \cancelto{0}{\widehat{a}_{-k}^{\dagger}\left(0\right)}-\cancelto{0}{\widehat{a}_{k}\left(0\right)}\rangle =0,
	\label{SM175}
\end{align}
\begin{align}
	\langle \widehat{p}_{k}^{\dagger}\left(0\right)\rangle =-\sqrt{2\hslash\varepsilon\left(0\right)\omega_{k}\left(0\right)}\,\frac{i}{2}\langle \cancelto{0}{\widehat{a}_{-k}\left(0\right)}-\cancelto{0}{\widehat{a}_{k}^{\dagger}\left(0\right)}\rangle =0.
	\label{SM176}
\end{align}

Likewise,

\begin{gather}
	\begin{split}
		\nonumber\langle \widehat{q}_{k}\left(0\right)\widehat{q}_{k}^{\dagger}\left(0\right)\rangle &=\frac{1}{\varepsilon\left(0\right)\omega_{k}\left(0\right)}\,\frac{\hslash}{2}\langle \cancelto{0}{\widehat{a}_{-k}^{\dagger}\left(0\right)\widehat{a}_{-k}\left(0\right)}+\cancelto{0}{\widehat{a}_{-k}^{\dagger}\left(0\right)\widehat{a}_{k}^{\dagger}\left(0\right)}+\cancelto{0}{\widehat{a}_{k}\left(0\right)\widehat{a}_{-k}\left(0\right)}+\widehat{a}_{k}\left(0\right)\widehat{a}_{k}^{\dagger}\left(0\right)\rangle
	\end{split} \\
	\begin{split}
		&=\frac{1}{\varepsilon\left(0\right)\omega_{k}\left(0\right)}\,\frac{\hslash}{2}\langle 1-\widehat{a}_{k}^{\dagger}\left(0\right)\widehat{a}_{k}\left(0\right)\rangle=\frac{1}{\varepsilon\left(0\right)\omega_{k}\left(0\right)}\,\frac{\hslash}{2},
	\end{split}
	\label{SM177}
\end{gather}
\begin{gather}
	\begin{split}
		\nonumber\langle \widehat{q}_{k}^{\dagger}\left(0\right)\widehat{q}_{k}\left(0\right)\rangle &=\frac{1}{\varepsilon\left(0\right)\omega_{k}\left(0\right)}\,\frac{\hslash}{2}\langle \widehat{a}_{-k}\left(0\right)\widehat{a}_{-k}^{\dagger}\left(0\right)+\cancelto{0}{\widehat{a}_{-k}\left(0\right)\widehat{a}_{k}\left(0\right)}+\cancelto{0}{\widehat{a}_{k}^{\dagger}\left(0\right)\widehat{a}_{-k}^{\dagger}\left(0\right)}+\cancelto{0}{\widehat{a}_{k}^{\dagger}\left(0\right)\widehat{a}_{k}\left(0\right)}\rangle
	\end{split} \\
	\begin{split}
		&=\frac{1}{\varepsilon\left(0\right)\omega_{k}\left(0\right)}\,\frac{\hslash}{2}\langle 1-\widehat{a}_{-k}^{\dagger}\left(0\right)\widehat{a}_{-k}\left(0\right)\rangle =\frac{1}{\varepsilon\left(0\right)\omega_{k}\left(0\right)}\,\frac{\hslash}{2},
	\end{split}
	\label{SM178}
\end{gather}

\begin{gather}
	\begin{split}
		\nonumber\langle \widehat{p}_{k}\left(0\right)\widehat{p}_{k}^{\dagger}\left(0\right)\rangle &=\varepsilon\left(0\right)\omega_{k}\left(0\right)\,\frac{\hslash}{2}\langle \cancelto{0}{\widehat{a}_{-k}^{\dagger}\left(0\right)\widehat{a}_{-k}\left(0\right)}-\cancelto{0}{\widehat{a}_{-k}^{\dagger}\left(0\right)\widehat{a}_{k}^{\dagger}\left(0\right)}-\cancelto{0}{\widehat{a}_{k}\left(0\right)\widehat{a}_{-k}\left(0\right)}+\widehat{a}_{k}\left(0\right)\widehat{a}_{k}^{\dagger}\left(0\right)\rangle
	\end{split} \\
	\begin{split}
		&=\varepsilon\left(0\right)\omega_{k}\left(0\right)\,\frac{\hslash}{2}\langle 1-\widehat{a}_{k}^{\dagger}\left(0\right)\widehat{a}_{k}\left(0\right)\rangle =\varepsilon\left(0\right)\omega_{k}\left(0\right)\,\frac{\hslash}{2},
	\end{split}
	\label{SM179}
\end{gather}
\begin{gather}
	\begin{split}
		\nonumber\langle \widehat{p}_{k}^{\dagger}\left(0\right)\widehat{p}_{k}\left(0\right)\rangle &=\varepsilon\left(0\right)\omega_{k}\left(0\right)\,\frac{\hslash}{2}\langle \widehat{a}_{-k}\left(0\right)\widehat{a}_{-k}^{\dagger}\left(0\right)-\cancelto{0}{\widehat{a}_{-k}\left(0\right)\widehat{a}_{k}\left(0\right)}-\cancelto{0}{\widehat{a}_{k}^{\dagger}\left(0\right)\widehat{a}_{-k}^{\dagger}\left(0\right)}+\cancelto{0}{\widehat{a}_{k}^{\dagger}\left(0\right)\widehat{a}_{k}\left(0\right)}\rangle
	\end{split} \\
	\begin{split}
		&=\varepsilon\left(0\right)\omega_{k}\left(0\right)\,\frac{\hslash}{2}\langle 1-\widehat{a}_{-k}^{\dagger}\left(0\right)\widehat{a}_{-k}\left(0\right)\rangle =\varepsilon\left(0\right)\omega_{k}\left(0\right)\,\frac{\hslash}{2},
	\end{split}
	\label{SM180}
\end{gather}
\begin{gather}
	\begin{split}
		\nonumber\langle \widehat{q}_{k}\left(0\right)\widehat{p}_{k}\left(0\right)\rangle &=i\frac{\hslash}{2}\langle \cancelto{0}{\widehat{a}_{-k}^{\dagger}\left(0\right)\widehat{a}_{-k}^{\dagger}\left(0\right)}-\cancelto{0}{\widehat{a}_{-k}^{\dagger}\left(0\right)\widehat{a}_{k}\left(0\right)}+\widehat{a}_{k}\left(0\right)\widehat{a}_{-k}^{\dagger}\left(0\right)-\cancelto{0}{\widehat{a}_{k}\left(0\right)\widehat{a}_{k}\left(0\right)}\rangle
	\end{split} \\
	\begin{split}
		&=i\frac{\hslash}{2}\langle 1-\widehat{a}_{-k}^{\dagger}\left(0\right)\widehat{a}_{k}\left(0\right)\rangle =i\frac{\hslash}{2},
	\end{split}
	\label{SM181}
\end{gather}

\begin{gather}
	\begin{split}
		\nonumber\langle \widehat{p}_{k}\left(0\right)\widehat{q}_{k}\left(0\right)\rangle &=i\frac{\hslash}{2}\langle \cancelto{0}{\widehat{a}_{-k}^{\dagger}\left(0\right)\widehat{a}_{-k}^{\dagger}\left(0\right)}+\cancelto{0}{\widehat{a}_{-k}^{\dagger}\left(0\right)\widehat{a}_{k}\left(0\right)}-\widehat{a}_{k}\left(0\right)\widehat{a}_{-k}^{\dagger}\left(0\right)-\cancelto{0}{\widehat{a}_{k}\left(0\right)\widehat{a}_{k}\left(0\right)}\rangle
	\end{split} \\
	\begin{split}
		&=-i\frac{\hslash}{2}\langle 1-\widehat{a}_{-k}^{\dagger}\left(0\right)\widehat{a}_{k}\left(0\right)\rangle =-i\frac{\hslash}{2},
	\end{split}
	\label{SM182}
\end{gather}

\begin{gather}
	\begin{split}
		\nonumber\langle \widehat{q}_{k}\left(0\right)\widehat{p}_{k}^{\dagger}\left(0\right)\rangle &=-i\frac{\hslash}{2}\langle \cancelto{0}{\widehat{a}_{-k}^{\dagger}\left(0\right)\widehat{a}_{-k}\left(0\right)}-\cancelto{0}{\widehat{a}_{-k}^{\dagger}\left(0\right)\widehat{a}_{k}^{\dagger}\left(0\right)}+\cancelto{0}{\widehat{a}_{k}\left(0\right)\widehat{a}_{-k}\left(0\right)}-\widehat{a}_{k}\left(0\right)\widehat{a}_{k}^{\dagger}\left(0\right)\rangle
	\end{split} \\
	\begin{split}
		&=i\frac{\hslash}{2}\langle 1-\widehat{a}_{k}^{\dagger}\left(0\right)\widehat{a}_{k}\left(0\right)\rangle =i\frac{\hslash}{2},
	\end{split}
	\label{SM183}
\end{gather}

\begin{gather}
	\begin{split}
		\nonumber\langle \widehat{p}_{k}^{\dagger}\left(0\right)\widehat{q}_{k}\left(0\right)\rangle &=-i\frac{\hslash}{2}\langle \widehat{a}_{-k}\left(0\right)\widehat{a}_{-k}^{\dagger}\left(0\right)+\cancelto{0}{\widehat{a}_{-k}\left(0\right)\widehat{a}_{k}\left(0\right)}-\cancelto{0}{\widehat{a}_{k}^{\dagger}\left(0\right)\widehat{a}_{-k}^{\dagger}\left(0\right)}-\cancelto{0}{\widehat{a}_{k}^{\dagger}\left(0\right)\widehat{a}_{k}\left(0\right)}\rangle
	\end{split} \\
	\begin{split}
		&=-i\frac{\hslash}{2}\langle 1-\widehat{a}_{-k}^{\dagger}\left(0\right)\widehat{a}_{-k}\left(0\right)\rangle =-i\frac{\hslash}{2},
	\end{split}
	\label{SM184}
\end{gather}

\begin{gather}
	\begin{split}
		\nonumber\langle \widehat{q}_{k}^{\dagger}\left(0\right)\widehat{p}_{k}\left(0\right)\rangle &=i\frac{\hslash}{2}\langle \widehat{a}_{-k}\left(0\right)\widehat{a}_{-k}^{\dagger}\left(0\right)-\cancelto{0}{\widehat{a}_{-k}\left(0\right)\widehat{a}_{k}\left(0\right)}+\cancelto{0}{\widehat{a}_{k}^{\dagger}\left(0\right)\widehat{a}_{-k}^{\dagger}\left(0\right)}-\cancelto{0}{\widehat{a}_{k}^{\dagger}\left(0\right)\widehat{a}_{k}\left(0\right)}\rangle
	\end{split} \\
	\begin{split}
		&=i\frac{\hslash}{2}\langle 1-\widehat{a}_{-k}^{\dagger}\left(0\right)\widehat{a}_{-k}\left(0\right)\rangle =i\frac{\hslash}{2},
	\end{split}
	\label{SM185}
\end{gather}

\begin{gather}
	\begin{split}
		\nonumber\langle \widehat{p}_{k}\left(0\right)\widehat{q}_{k}^{\dagger}\left(0\right)\rangle &=i\frac{\hslash}{2}\langle \cancelto{0}{\widehat{a}_{-k}^{\dagger}\left(0\right)\widehat{a}_{-k}\left(0\right)}+\cancelto{0}{\widehat{a}_{-k}^{\dagger}\left(0\right)\widehat{a}_{k}^{\dagger}\left(0\right)}-\cancelto{0}{\widehat{a}_{k}\left(0\right)\widehat{a}_{-k}\left(0\right)}-\widehat{a}_{k}\left(0\right)\widehat{a}_{k}^{\dagger}\left(0\right)\rangle
	\end{split} \\
	\begin{split}
		&=-i\frac{\hslash}{2}\langle 1-\widehat{a}_{k}^{\dagger}\left(0\right)\widehat{a}_{k}\left(0\right)\rangle =-i\frac{\hslash}{2}.
	\end{split}
	\label{SM186}
\end{gather}

Then, we can compute:

\begin{align}
	\left\langle \widehat{q}_{k}\left(t\right)\right\rangle =M_{11}\left(t\right)\left\langle \widehat{q}_{k}\left(0\right)\right\rangle +M_{12}\left(t\right)\left\langle \widehat{p}_{k}\left(0\right)\right\rangle =0,
	\label{SM187}
\end{align}
\begin{align}
	\left\langle \widehat{q}_{k}^{\dagger}\left(t\right)\right\rangle =M_{11}\left(t\right)\left\langle \widehat{q}_{k}^{\dagger}\left(0\right)\right\rangle +M_{12}\left(t\right)\left\langle \widehat{p}_{k}^{\dagger}\left(0\right)\right\rangle =0,
	\label{SM188}
\end{align}
\begin{align}
	\left\langle \widehat{p}_{k}\left(t\right)\right\rangle =M_{21}\left(t\right)\left\langle \widehat{q}_{k}\left(0\right)\right\rangle +M_{22}\left(t\right)\left\langle \widehat{p}_{k}\left(0\right)\right\rangle =0,
	\label{SM189}
\end{align}
\begin{align}
	\left\langle \widehat{p}_{k}^{\dagger}\left(t\right)\right\rangle =M_{21}\left(t\right)\left\langle \widehat{q}_{k}^{\dagger}\left(0\right)\right\rangle +M_{22}\left(t\right)\left\langle \widehat{p}_{k}^{\dagger}\left(0\right)\right\rangle =0.
	\label{SM190}
\end{align}

Expected values (\ref{SM187})-(\ref{SM190}) stand for squeezed vacuum states whose position and momentum operators are centered on the origin, whereas their respective variances provide with a whole picture showing that statistical noise vary on time:

\begin{gather}
\begin{split}
\nonumber\langle \widehat{q}_{k}\left(t\right)\widehat{q}_{k}^{\dagger}\left(t\right)\rangle =\left|M_{11}\left(t\right)\right|^{2}\left\langle \widehat{q}_{k}\left(0\right)\widehat{q}_{k}^{\dagger}\left(0\right)\right\rangle +\left|M_{12}\left(t\right)\right|^{2}\left\langle \widehat{p}_{k}\left(0\right)\widehat{p}_{k}^{\dagger}\left(0\right)\right\rangle
\end{split}\\
\begin{split}
=\frac{\hslash}{2}\,\left\{ \frac{\left|M_{11}\left(t\right)\right|^{2}}{\varepsilon\left(0\right)\omega_{k}\left(0\right)}+\left|M_{12}\left(t\right)\right|^{2}\varepsilon\left(0\right)\omega_{k}\left(0\right)\right\},
\end{split}
\label{S191}
\end{gather}

\begin{gather}
	\begin{split}
\nonumber\langle \widehat{q}_{k}^{\dagger}\left(t\right)\widehat{q}_{k}\left(t\right)\rangle =\left|M_{11}\left(t\right)\right|^{2}\left\langle \widehat{q}_{k}^{\dagger}\left(0\right)\widehat{q}_{k}\left(0\right)\right\rangle +\left|M_{12}\left(t\right)\right|^{2}\left\langle \widehat{p}_{k}^{\dagger}\left(0\right)\widehat{p}_{k}\left(0\right)\right\rangle
	\end{split}\\
	\begin{split}
=\frac{\hslash}{2}\,\left\{ \frac{\left|M_{11}\left(t\right)\right|^{2}}{\varepsilon\left(0\right)\omega_{k}\left(0\right)}+\left|M_{12}\left(t\right)\right|^{2}\varepsilon\left(0\right)\omega_{k}\left(0\right)\right\},		
	\end{split}
	\label{S192}
\end{gather}

\begin{gather}
	\begin{split}
\nonumber\langle \widehat{p}_{k}\left(t\right)\widehat{p}_{k}^{\dagger}\left(t\right)\rangle =\left|M_{21}\left(t\right)\right|^{2}\left\langle \widehat{q}_{k}\left(0\right)\widehat{q}_{k}^{\dagger}\left(0\right)\right\rangle +\left|M_{22}\left(t\right)\right|^{2}\left\langle \widehat{p}_{k}\left(0\right)\widehat{p}_{k}^{\dagger}\left(0\right)\right\rangle
	\end{split}\\
	\begin{split}
=\frac{\hslash}{2}\,\left\{ \frac{\left|M_{21}\left(t\right)\right|^{2}}{\varepsilon\left(0\right)\omega_{k}\left(0\right)}+\left|M_{22}\left(t\right)\right|^{2}\varepsilon\left(0\right)\omega_{k}\left(0\right)\right\},		
	\end{split}
	\label{S193}
\end{gather}

\begin{gather}
	\begin{split}
\nonumber\langle \widehat{p}_{k}^{\dagger}\left(t\right)\widehat{p}_{k}\left(t\right)\rangle =\left|M_{21}\left(t\right)\right|^{2}\left\langle \widehat{q}_{k}^{\dagger}\left(0\right)\widehat{q}_{k}\left(0\right)\right\rangle +\left|M_{22}\left(t\right)\right|^{2}\left\langle \widehat{p}_{k}^{\dagger}\left(0\right)\widehat{p}_{k}\left(0\right)\right\rangle
	\end{split}\\
	\begin{split}
=\frac{\hslash}{2}\,\left\{ \frac{\left|M_{21}\left(t\right)\right|^{2}}{\varepsilon\left(0\right)\omega_{k}\left(0\right)}+\left|M_{22}\left(t\right)\right|^{2}\varepsilon\left(0\right)\omega_{k}\left(0\right)\right\}.		
	\end{split}
	\label{S194}
\end{gather}

From the above we finally have the expected energy of a 1D polarization electromagnetic wave propagating on a temporal metamaterial:

\begin{align}
	\boxed{\langle \widehat{\mathcal{H}}_{k}\left(t\right)\rangle =\frac{1}{2}\,\frac{1}{\varepsilon\left(t\right)}\,\langle \widehat{p}_{k}\left(t\right)\widehat{p}_{k}^{\dagger}\left(t\right)+\widehat{p}_{k}^{\dagger}\left(t\right)\widehat{p}_{k}\left(t\right)\rangle +\frac{1}{2}\,\varepsilon\left(t\right)\omega_{k}^{2}\left(t\right)\,\langle \widehat{q}_{k}\left(t\right)\widehat{q}_{k}^{\dagger}\left(t\right)+\widehat{q}_{k}^{\dagger}\left(t\right)\widehat{q}_{k}\left(t\right)\rangle}.
	\label{SM195}
\end{align}

\subsubsection{\textbf{Dielectric time metamaterial}}
wing to the theoretical and also the experimental difficulties in simultaneously modulating $\varepsilon(t)$ and $\mu(t)$, in this section we restrict the analysis to dielectric temporal-modulation in two specific cases: a tapered photonic switch and a pulse with a Gaussian profile. In such a case, the expected energy of the system, given in (\ref{SM195}), adopts the following form:
\begin{align}
	\langle \widehat{\mathcal{H}}_{k}\left(t\right)\rangle =\frac{1}{2}\,\frac{1}{\varepsilon\left(t\right)}\,\langle \widehat{p}_{k}\left(t\right)\widehat{p}_{k}^{\dagger}\left(t\right)+\widehat{p}_{k}^{\dagger}\left(t\right)\widehat{p}_{k}\left(t\right)\rangle +\frac{1}{2}\,k^{2}c^{2}\,\langle \widehat{q}_{k}\left(t\right)\widehat{q}_{k}^{\dagger}\left(t\right)+\widehat{q}_{k}^{\dagger}\left(t\right)\widehat{q}_{k}\left(t\right)\rangle .
	\label{SM196}
\end{align}

In addition, according to the Heisenberg equation of motions, the transfer matrix encompassing the input-output relation (\ref{SM145}) reads as:

\begin{align}
	\mathbf{M}_{k}\left(t\right)=\left[\begin{array}{cc}
		1 & \frac{\varDelta t}{\varepsilon\left(t\right)}\\
		-\varDelta t\,k^{2}c^{2} & 1
	\end{array}\right].
	\label{SM197}
\end{align}

\begin{figure*}[h!]
	\includegraphics[width=\columnwidth]{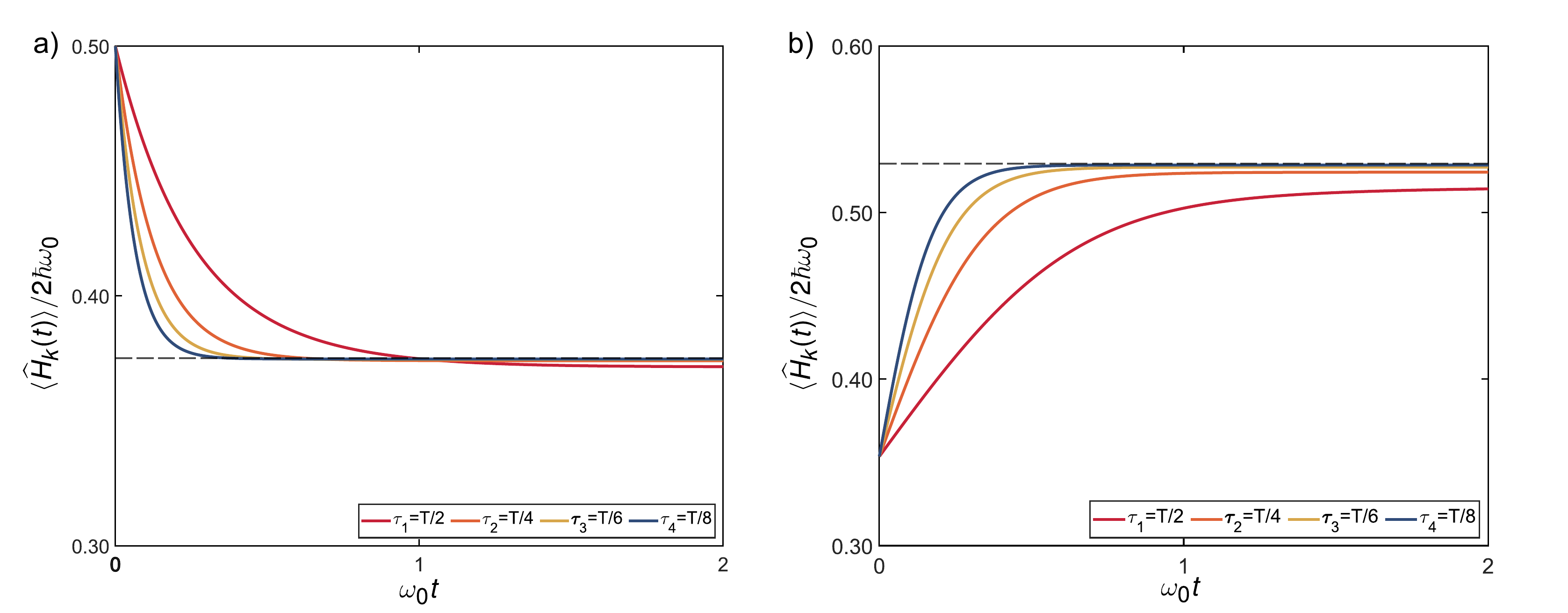}
	\caption{Graphical representation of a range of transition times for tapered photonic switching a) increasing the value of the dielectric permittivity ($\varepsilon\left(t\right)=1+\tanh(t/\tau)$), and b) decreasing the value of the dielectric permittivity ($\varepsilon\left(t\right)=1+\exp(-t/\tau)$). In the former, expected energy drops below reference energy of vacuum $\hbar\omega_{0}$, encompassing the adiabatic limit for slow transitions (red curve) and quickly approaching the predicted behavior of a temporal boundary, $\langle \widehat{H}_{k}\rangle =0.375\hbar\omega_{0}$ (black dashed line), with good accuracy for $T/4\leq\tau\leq T/8$ (orange, yellow, and blue curves, respectively). In the later case, decreasing the dielectric permittivity of the system yields to an expected energy above the aforementioned energy level of vacuum. Despite following closely the adiabatic regime for slow time transitions (red curve), reaching the predicted expected value for a temporal boundary, $\langle \widehat{H}_{k}\rangle =0.530\hbar\omega_{0}$ (black dashed line), slightly above vacuum energy, requires shorter time periods in this case in order to mimic a temporal boundary. According to the results, the temporal boundary behavior is reached faster in the process of dielectric permittivity increasing.}
	\label{fig1sm}
\end{figure*}

\paragraph{Tapered photonic switching}
Regarding the tapered photonic switch, we examine two parallel cases related with the increase and decrease of an initial background permittivity, respectively described by:
\begin{align}
	\varepsilon\left(t\right)=\varepsilon_{b}+\Delta\varepsilon\tanh(t/\tau),
	\label{SM198}
\end{align}
\begin{align}
	\varepsilon\left(t\right)=\varepsilon_{b}+\Delta\varepsilon\exp(-t/\tau),
	\label{SM199}
\end{align}

where $\varepsilon_{b}$ is the initial background dielectric permittivity of the system, $\Delta\varepsilon$ is the amplitude of the modulation (which leads into a final steady state), and $\tau$ stands for the swiftness of the modulation. When this latter parameter is slightly above a transition time $\tau=T$, the expected energy reaches the adiabatic limit, $\langle \widehat{\mathcal{H}}_{k}\left(t\right)\rangle =\hbar\omega_{k}(t)$. In contrast, for ultra-fast modulations, the energy tends toward the theoretical prediction tied to a temporal boundary $\langle \widehat{H}_{k}\rangle =\hslash\omega\,\left(\mathrm{sinh}^{2}\left(s_{21}\right)+1/2\right)$, where $s_{21}=\log[(\varepsilon(0)/\varepsilon(t\rightarrow\infty))^{1/4}]$ is the squeezing parameter. Remarkably, according to the set of parameters we have studied here [see \hyperref[fig1sm]{Fig. \ref*{fig1sm}}], transition times of $T/4\leq\tau\leq T/10$ are sufficient to accurately describe the behavior of an abrupt (theoretically instantaneous) temporal boundary.

\subsubsection{\textbf{Gaussian pulse profile}}
Regarding widely examined cases where dielectrics are modulated by ultra-short laser pulses, resulting in a time-localized increase (or decrease) of the dielectric permittivity of the whole medium, the very intuitive notion concerning photon production, namely, the swifter the modulation the bigger amount of photons are produced, is challenged. Here below we consider a system whose dielectric permittivity is modulated by means of a Gaussian pulse:
\begin{align}
	\varepsilon\left(t\right)=\varepsilon_{b}+\Delta\varepsilon\exp\left(-\frac{\left(t-t_{0}\right)^{2}}{2\tau^{2}}\right),
	\label{SM200}
\end{align}
where $\varepsilon_{b}$ is the dielectric permittivity of the system before and after the pulse, $\Delta\varepsilon$ is the amplitude of the pulse localized at $=t_{0}$, and $\tau$ is the time-width dispersion. According to the value of the modulation-swiftness, $\tau$, this Gaussian modulation profile exhibits three potential cases of interest.

\subparagraph*{Slow pulse (encompassing the adiabatic limit)}
The first one corresponds to a broad temporal-width Gaussian profile, $(\tau\gg T)$, reverting the energy of the whole system into the initial state. This slow modulation profile mimics the adiabatic limit, namely, reaching the initial energy, $\left\langle \widehat{\mathcal{H}}_{k}\left(t\right)\right\rangle =\hbar\omega_{k}(t)$, as the permittivity recover the original value prior the perturbation.
\subparagraph*{Short-width pulse (optimal pulse-width)}
When applying a Gaussian time-modulation profile, we computed several excitation values of the energy for long times. Those results got a maximum for a certain dispersion $\tau$ of the Gaussian pulse [\hyperref[Gaussian_pulse]{Fig. \ref*{Gaussian_pulse}} in the main text]. Here we deduce the behavior for such an optimal dispersion of the system. First, the expected energy (\ref{SM196}) is rewritten as a function of two variables, namely time, $t$, and pulse-width, $\tau$:
\begin{align}
	\left\langle \widehat{\mathcal{H}}_{k}\left(t,\tau\right)\right\rangle =\frac{1}{2}\,\frac{1}{\varepsilon\left(t,\tau\right)}\,\left\langle \widehat{p}_{k}\left(t,\tau\right)\widehat{p}_{k}^{\dagger}\left(t,\tau\right)+\widehat{p}_{k}^{\dagger}\left(t,\tau\right)\widehat{p}_{k}\left(t,\tau\right)\right\rangle +\frac{1}{2}\,k^{2}c^{2}\,\left\langle \widehat{q}_{k}\left(t,\tau\right)\widehat{q}_{k}^{\dagger}\left(t,\tau\right)+\widehat{q}_{k}^{\dagger}\left(t,\tau\right)\widehat{q}_{k}\left(t,\tau\right)\right\rangle ,
	\label{SM201}
\end{align}
where we have used the relation $\omega_{k}^{2}\left(t,\tau\right)=k^{2}c^{2}/\varepsilon\left(t,\tau\right)$. Regardless of the specific dispersion, it can be numerically demonstrated that there is a minimum of energy exactly at the instant that corresponds to the center of the Gaussian distribution (i.e., the dielectric permittivity reaches its higher value). However, we are interested in the optimal pulse-width dispersion for long times, i.e., so that

\begin{align}
	\partial_{\tau}\left\{ \lim_{t\rightarrow\infty}\left\langle \widehat{\mathcal{H}}_{k}\left(t,\tau\right)\right\rangle \right\} =0.
	\label{SM202}
\end{align}

Thus,
\begin{align}
	\lim_{t\rightarrow\infty}\partial_{\tau}\,\left\langle \widehat{p}_{k}\left(t,\tau\right)\widehat{p}_{k}^{\dagger}\left(t,\tau\right)+\widehat{p}_{k}^{\dagger}\left(t,\tau\right)\widehat{p}_{k}\left(t,\tau\right)\right\rangle =-\varepsilon_{b}k^{2}c^{2}\lim_{t\rightarrow\infty}\partial_{\tau}\,\left[\left\langle \widehat{q}_{k}\left(t,\tau\right)\widehat{q}_{k}^{\dagger}\left(t,\tau\right)+\widehat{q}_{k}^{\dagger}\left(t,\tau\right)\widehat{q}_{k}\left(t,\tau\right)\right\rangle \right].
	\label{SM203}
\end{align}

\begin{figure*}[h!]
	\includegraphics[width=\columnwidth]{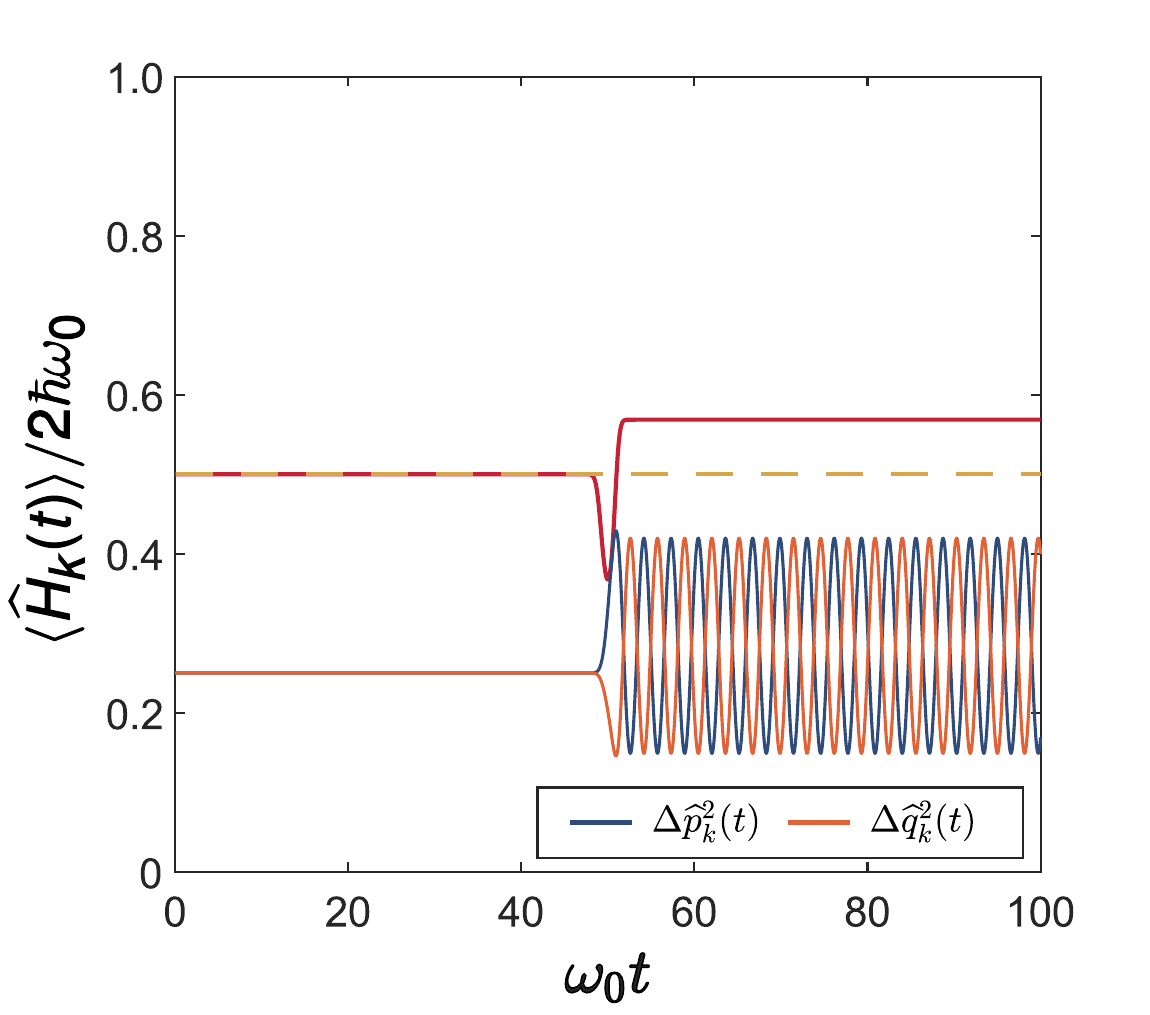}
	\caption{Enhancement of expected energy (red curve) for a Gaussian time-modulation profile, for $\varepsilon=\Delta\varepsilon=1$, centered at $t=t_{0}$, and $\tau=0.57T$, over the reference energy of vacuum (yellow dashed line). After the pulse, the variances of momentum operator $\left\langle \widehat{p}_{k}\left(t,\tau\right)\widehat{p}_{k}^{\dagger}\left(t,\tau\right)+\widehat{p}_{k}^{\dagger}\left(t,\tau\right)\widehat{p}_{k}\left(t,\tau\right)\right\rangle$  (blue curve) and position operator $\left\langle \widehat{q}_{k}\left(t,\tau\right)\widehat{q}_{k}^{\dagger}\left(t,\tau\right)+\widehat{q}_{k}^{\dagger}\left(t,\tau\right)\widehat{q}_{k}\left(t,\tau\right)\right\rangle$  (orange curve), associated with both conjugated parametric oscillators, evolve following an oscillatory counter-phase behavior. Adding up both numerically computed variances, the expected energy reaches a value of $\left\langle \widehat{\mathcal{H}}_{k}\left(t,\tau\right)\right\rangle =0.5682\hbar\omega_{0}.$}
	\label{fig2sm}
\end{figure*}

As the limits may not be defined for long times [see \hyperref[fig2sm]{Fig. \ref*{fig2sm}}], we can only assume from (\ref{SM203}) that derivatives take opposite signs. Evidently, (\ref{SM202}) exposes the existence of null-derivatives for both variances in (\ref{SM203}). This indicates that, in order to achieve the observed enhancement of expected energy of the system, those both conditions shall be satisfied at once. Therefore, when both position and momentum behave as oscillatory counter-phase functions for long times, a maximum amount of photon is produced.

\subparagraph*{Ultra-fast pulse (delta-like behavior)}
Ampère-Maxwell equation states that,
\begin{align}
	\nabla\times\mathbf{H}\equiv\frac{\partial\mathbf{D}}{\partial t}.
	\label{SM204}
\end{align}

Assuming a non-magnetized and pure-time medium:

\begin{align}
	\nabla\times\mathbf{B}=\varepsilon_{0}\mu_{0}\frac{\partial}{\partial t}\left(\varepsilon(t)\mathbf{E}\right)=\frac{1}{c^{2}}\left\{ \frac{\partial\varepsilon(t)}{\partial t}\mathbf{E}+\varepsilon(t)\frac{\partial\mathbf{E}}{\partial t}\right\} .
	\label{SM205}
\end{align}

Setting $\varepsilon(t)=\varepsilon_{b}+\Delta\varepsilon\thinspace\delta(t-t_{0})$,

\begin{align}
	\nabla\times\mathbf{B}=\frac{1}{c^{2}}\thinspace\left\{ \Delta\thinspace\frac{\partial\delta(t-t_{0})}{\partial t}\thinspace\mathbf{E}+\varepsilon_{b}\thinspace\frac{\partial\mathbf{E}}{\partial t}+\Delta\thinspace\delta(t-t_{0})\thinspace\frac{\partial\mathbf{E}}{\partial t}\right\} .
	\label{SM206}
\end{align}

And using the shifting property $\int dt\thinspace f(t)\thinspace\delta(t-t_{0})\equiv f(t_{0})$ and the equation that defines derivatives of the delta function $\int dt\thinspace f(t)\thinspace\delta^{(n)}(t-t_{0})\equiv-\int dt\thinspace{\partial_{t}f(t)}\thinspace\delta^{(n-1)}(t-t_{0})$, (see Appendix C), integration over (\ref{SM206}) leads to,

\begin{align}
	\frac{\Delta}{c^{2}}\thinspace\int_{-\infty}^{+\infty}dt\thinspace\frac{\partial\delta(t-t_{0})}{\partial t}\thinspace\mathbf{E}=-\frac{\Delta}{c^{2}}\thinspace\left.\frac{\partial\mathbf{E}}{\partial t}\right|_{t=t_{0}},
	\label{SM207}
\end{align}
\begin{align}
	\frac{\Delta}{c^{2}}\thinspace\int_{-\infty}^{+\infty}dt\thinspace\delta(t-t_{0})\thinspace\frac{\partial\mathbf{E}}{\partial t}=\frac{\Delta}{c^{2}}\thinspace\left.\frac{\partial\mathbf{E}}{\partial t}\right|_{t=t_{0}}.
	\label{SM208}
\end{align}
Thus, the continuity of electric field at both sides of the temporal delta barrier is proven.
\begin{align}
	\int_{t_{0}^{-}}^{t_{0}^{+}}dt\thinspace\nabla\times\mathbf{B}\left(\mathbf{r},t\right)=\frac{1}{c^{2}}\thinspace\int_{t_{0}^{-}}^{t_{0}^{+}}dt\thinspace\frac{\partial}{\partial t}\left(\varepsilon_{r}(t)\mathbf{E}\left(\mathbf{r},t\right)\right)=\frac{\varepsilon_{b}}{c^{2}}\thinspace\int_{t_{0}^{-}}^{t_{0}^{+}}dt\thinspace\frac{\partial\mathbf{E}\left(\mathbf{r},t\right)}{\partial t}=\frac{\varepsilon_{b}}{c^{2}}\thinspace\left\{ \mathbf{E}\left(\mathbf{r},t_{0}^{+}\right)-\mathbf{E}\left(\mathbf{r},t_{0}^{-}\right)\right\} .
	\label{SM209}
\end{align}

\begin{figure*}[h!]
	\includegraphics[width=\columnwidth]{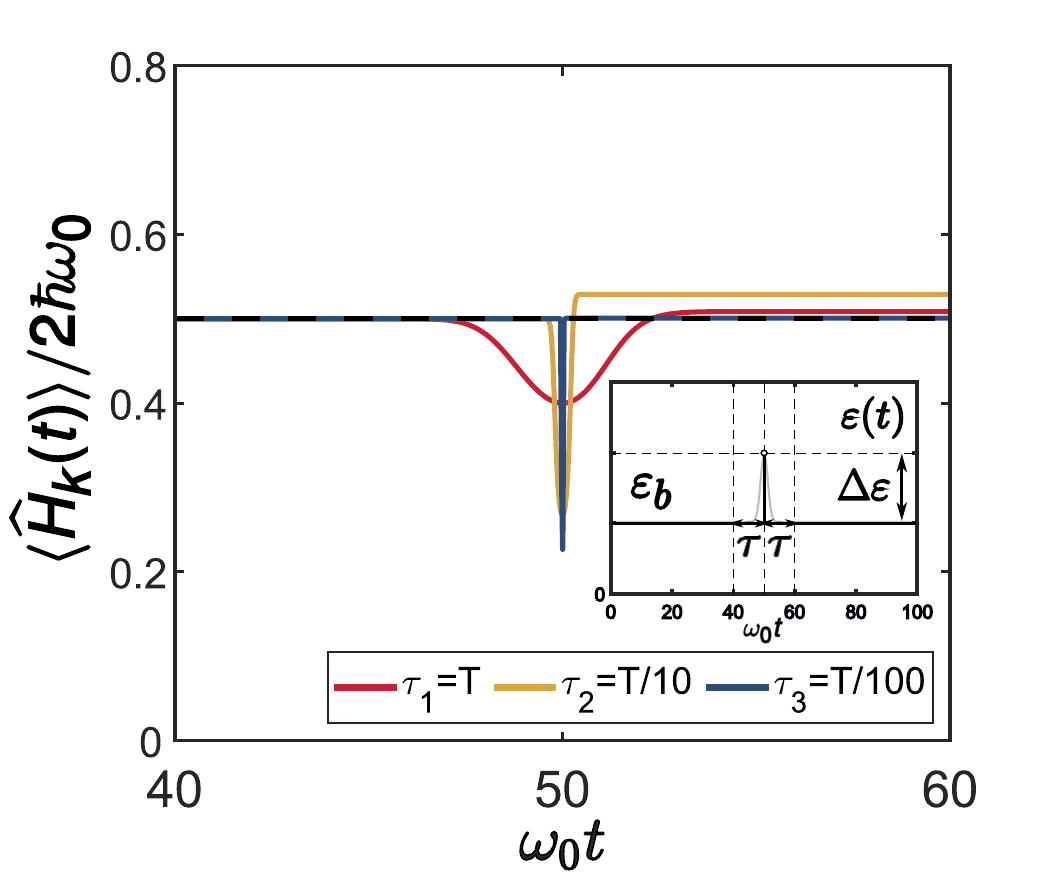}
	\caption{Expected energy with respect to vacuum energy reference for several Gaussian profiles approaching a delta-like response with equal-area pulses $\varepsilon(t)=\varepsilon_{b}+(\Delta\varepsilon/\omega_{0}\tau)\exp[-(t-t_{0})^{2}/(2\tau^{2})]$ (included as an inset, where $\varepsilon_{b}=\Delta\varepsilon=1$). Slow pulse $\tau_{1}$ (red) leads the system energy to a slightly excited state, whereas short pulse $\tau_{2}$ (yellow) leads to a noticeable final excitation of energetic states, and contrary to the previous cases, ultra-shorter pulses $\tau_{3}$ (blue) approaches the limit of a Delta-like barrier, reaching a limit expected energy of $\left\langle \widehat{\mathcal{H}}_{k}\left(t\right)\right\rangle =0.25\hbar\omega_{0}$ and recovering the initial energy almost immediately after the intense pulse, i.e., no photon production is achieved.}
	\label{fig3sm}
\end{figure*}

The continuity of the fields justify the behavior illustrated for ultra-narrow Gaussian profiles [see see \hyperref[fig3sm]{Fig. \ref*{fig3sm}}], where the maximum amount of squeezed photons is almost completely localized on the instant $t_{0}$, swiftly flowing into the initial expected energy since $\varepsilon(t_{0}^{-})=\varepsilon(t_{0}^{+})$.

\clearpage
\part{A. Set of differential equations: Cheng technique}
As we have pointed out before, $\widehat{J}_{+}$, $\widehat{J}_{0}$, and $\widehat{J}_{-}$, form the SU(2) Lie algebra and so, time-evolution operator can be expressed as follows:
\begin{align}
	\widehat{U}\left(t,0\right)=\exp\left(c_{1}(t)\widehat{J}_{+}\right)\exp\left(c_{2}(t)\widehat{J}_{0}\right)\exp\left(c_{3}(t)\widehat{J}_{-}\right)
	\label{S210}
\end{align}
Taking the time derivative and inserting it in $\widehat{\mathcal{H}}\left(t\right)\widehat{U}\left(t,0\right)=i\hbar\partial_{t}\widehat{U}\left(t,0\right)$, with $\widehat{U}\left(0,0\right)=1$, it follows that:
\begin{gather}
\begin{split}
\nonumber\partial_{t}{\color{green}{\color{teal}\left[\exp\left(c_{1}(t)\widehat{J}_{+}\right)\exp\left(c_{2}(t)\widehat{J}_{0}\right)\exp\left(c_{3}(t)\widehat{J}_{-}\right)\right]}}={\color{blue}\left\{ \partial_{t}\exp\left(c_{1}(t)\widehat{J}_{+}\right)\right\} \exp\left(c_{2}(t)\widehat{J}_{0}\right)\exp\left(c_{3}(t)\widehat{J}_{-}\right)}
\end{split}\\
\begin{split}
{\color{red}+\exp\left(c_{1}(t)\widehat{J}_{+}\right)\left\{ \partial_{t}\exp\left(c_{2}(t)\widehat{J}_{0}\right)\right\} \exp\left(c_{3}(t)\widehat{J}_{-}\right)}{\color{magenta}+\exp\left(c_{1}(t)\widehat{J}_{+}\right)\exp\left(c_{2}(t)\widehat{J}_{0}\right)\left\{ \partial_{t}\exp\left(c_{3}(t)\widehat{J}_{-}\right)\right\}}.
\end{split}
\label{S211}
\end{gather}
Where,
\begin{gather}
\begin{split}
\nonumber{\color{blue}\left\{ \partial_{t}\exp\left(c_{1}(t)\widehat{J}_{+}\right)\right\} \exp\left(c_{2}(t)\widehat{J}_{0}\right)\exp\left(c_{3}(t)\widehat{J}_{-}\right)}
\end{split}\\
\begin{split}
\partial_{t}c_{1}(t)\widehat{J}_{+}{\color{green}{\color{teal}\exp\left(c_{1}(t)\widehat{J}_{+}\right)\exp\left(c_{2}(t)\widehat{J}_{0}\right)\exp\left(c_{3}(t)\widehat{J}_{-}\right)}}=\partial_{t}c_{1}(t)\widehat{J}_{+}{\color{green}{\color{teal}\widehat{U}\left(t,0\right)}},
\end{split}
\label{S212}
\end{gather}

\begin{gather}
\begin{split}
\nonumber{\color{red}\exp\left(c_{1}(t)\widehat{J}_{+}\right)\left\{ \partial_{t}\exp\left(c_{2}(t)\widehat{J}_{0}\right)\right\} \exp\left(c_{3}(t)\widehat{J}_{-}\right)}=\partial_{t}c_{2}(t)\exp\left(c_{1}(t)\widehat{J}_{+}\right)\widehat{J}_{0}\exp\left(c_{2}(t)\widehat{J}_{0}\right)\exp\left(c_{3}(t)\widehat{J}_{-}\right)
\end{split}\\
\begin{split}
\nonumber=\partial_{t}c_{2}(t){\color{red}\exp\left(c_{1}(t)\widehat{J}_{+}\right)\widehat{J}_{0}\exp\left(-c_{1}(t)\widehat{J}_{+}\right)}{\color{green}{\color{teal}\exp\left(c_{1}(t)\widehat{J}_{+}\right)\exp\left(c_{2}(t)\widehat{J}_{0}\right)\exp\left(c_{3}(t)\widehat{J}_{-}\right)}}
\end{split}\\
\begin{split}
\nonumber=\partial_{t}c_{2}(t){\color{red}\left\{ \widehat{J}_{0}+c_{1}(t)\left[\widehat{J}_{+},\widehat{J}_{0}\right]+\frac{c_{1}^{2}(t)}{2!}\cancelto{0}{\left[\widehat{J}_{+},\left[\widehat{J}_{+},\widehat{J}_{0}\right]\right]}+\frac{c_{1}^{3}(t)}{3!}\cancelto{0}{\left[\widehat{J}_{+},\left[\widehat{J}_{+},\left[\widehat{J}_{+},\widehat{J}_{0}\right]\right]\right]}+\cancelto{0}{...}\right\} }{\color{green}{\color{teal}\widehat{U}\left(t,0\right)}}
\end{split}\\
\begin{split}
=\partial_{t}c_{2}(t){\color{red}\left\{ \widehat{J}_{0}-c_{1}(t)\left[\widehat{J}_{0},\widehat{J}_{+}\right]\right\} }{\color{green}{\color{teal}\widehat{U}\left(t,0\right)}}=\partial_{t}c_{2}(t)\widehat{J}_{0}{\color{green}{\color{teal}\widehat{U}\left(t,0\right)}}-\partial_{t}c_{2}(t)c_{1}(t)\widehat{J}_{+}{\color{green}{\color{teal}\widehat{U}\left(t,0\right)}},
\end{split}
\label{S213}
\end{gather}
\begin{gather}
\begin{split}
\nonumber{\color{magenta}\exp\left(c_{1}(t)\widehat{J}_{+}\right)\exp\left(c_{2}(t)\widehat{J}_{0}\right)\left\{ \partial_{t}\exp\left(c_{3}(t)\widehat{J}_{-}\right)\right\} }=\partial_{t}c_{3}(t)\exp\left(c_{1}(t)\widehat{J}_{+}\right)\exp\left(c_{2}(t)\widehat{J}_{0}\right)\widehat{J}_{-}\exp\left(c_{3}(t)\widehat{J}_{-}\right)
\end{split}\\
\begin{split}
\nonumber=\partial_{t}c_{3}(t)\exp\left(c_{1}(t)\widehat{J}_{+}\right){\color{magenta}\exp\left(c_{2}(t)\widehat{J}_{0}\right)\widehat{J}_{-}\exp\left(-c_{2}(t)\widehat{J}_{0}\right)}\exp\left(c_{2}(t)\widehat{J}_{0}\right)\exp\left(c_{3}(t)\widehat{J}_{-}\right)
\end{split}\\
\begin{split}
\nonumber=\partial_{t}c_{3}(t)\exp\left(c_{1}(t)\widehat{J}_{+}\right){\color{magenta}\widehat{J}_{-}\left\{ 1-c_{2}(t)+\frac{c_{2}^{2}(t)}{2!}-\frac{c_{2}^{3}(t)}{3!}+...\right\} }\exp\left(c_{2}(t)\widehat{J}_{0}\right)\exp\left(c_{3}(t)\widehat{J}_{-}\right)
\end{split}\\
\begin{split}
\nonumber=\partial_{t}c_{3}(t)\exp\left(c_{1}(t)\widehat{J}_{+}\right){\color{magenta}\widehat{J}_{-}\exp\left(-c_{2}(t)\right)}\exp\left(c_{2}(t)\widehat{J}_{0}\right)\exp\left(c_{3}(t)\widehat{J}_{-}\right)
\end{split}\\
\begin{split}
\nonumber=\partial_{t}c_{3}(t)\exp\left(-c_{2}(t)\right){\color{magenta}\exp\left(c_{1}(t)\widehat{J}_{+}\right){\color{magenta}\widehat{J}_{-}}\exp\left(-c_{1}(t)\widehat{J}_{+}\right)}{\color{green}{\color{teal}\exp\left(c_{1}(t)\widehat{J}_{+}\right)\exp\left(c_{2}(t)\widehat{J}_{0}\right)\exp\left(c_{3}(t)\widehat{J}_{-}\right)}}
\end{split}\\
\begin{split}
\nonumber=\partial_{t}c_{3}(t)\exp\left(-c_{2}(t)\right){\color{magenta}\exp\left(c_{1}(t)\widehat{J}_{+}\right){\color{magenta}\widehat{J}_{-}}\exp\left(-c_{1}(t)\widehat{J}_{+}\right)}{\color{green}{\color{teal}\widehat{U}\left(t,0\right)}}
\end{split}\\
\begin{split}
\nonumber=\partial_{t}c_{3}(t)\exp\left(-c_{2}(t)\right){\color{magenta}{\color{magenta}\left\{ \widehat{J}_{-}+c_{1}(t)\left[\widehat{J}_{+},\widehat{J}_{-}\right]+\frac{c_{1}^{2}(t)}{2!}\left[\widehat{J}_{+},\left[\widehat{J}_{+},\widehat{J}_{-}\right]\right]+\frac{c_{1}^{3}(t)}{3!}\left[\widehat{J}_{+},\left[\widehat{J}_{+},\left[\widehat{J}_{+},\widehat{J}_{-}\right]\right]\right]+...\right\} }}{\color{green}{\color{teal}\widehat{U}\left(t,0\right)}}
\end{split}\\
\begin{split}
\nonumber=\partial_{t}c_{3}(t)\exp\left(-c_{2}(t)\right){\color{magenta}\left\{ \widehat{J}_{-}+2c_{1}(t)\widehat{J}_{0}-\cancel{2}\frac{c_{1}^{2}(t)}{\cancel{2!}}\widehat{J}_{+}+\frac{c_{1}^{3}(t)}{3!}\cancelto{0}{\left[\widehat{J}_{+},\left[\widehat{J}_{+},2\widehat{J}_{0}\right]\right]}+\cancelto{0}{...}\right\} }{\color{green}{\color{teal}\widehat{U}\left(t,0\right)}}
\end{split}\\
\begin{split}
\nonumber=\partial_{t}c_{3}(t)\exp\left(-c_{2}(t)\right){\color{magenta}{\color{magenta}\left\{ \widehat{J}_{-}+2c_{1}(t)\widehat{J}_{0}-c_{1}^{2}(t)\widehat{J}_{+}\right\} }{\color{green}{\color{teal}\widehat{U}\left(t,0\right)}}},
\end{split}\\
\begin{split}
=\partial_{t}c_{3}(t)\exp\left(-c_{2}(t)\right)\widehat{J}_{-}{\color{green}{\color{teal}\widehat{U}\left(t,0\right)}}+2c_{1}(t)\partial_{t}c_{3}(t)\exp\left(-c_{2}(t)\right)\widehat{J}_{0}{\color{green}{\color{teal}\widehat{U}\left(t,0\right)}}-c_{1}^{2}(t)\partial_{t}c_{3}(t)\exp\left(-c_{2}(t)\right)\widehat{J}_{+}{\color{green}{\color{teal}\widehat{U}\left(t,0\right)}},
\end{split}
\label{S214}
\end{gather}

where we have made use of the following identity:

\begin{align}
	\exp\left(\widehat{A}\right)\widehat{B}\exp\left(-\widehat{A}\right)=\widehat{B}+\left[\widehat{A},\widehat{B}\right]+\frac{1}{2!}\left[\widehat{A},\left[\widehat{A},\widehat{B}\right]\right]+\frac{1}{3!}\left[\widehat{A},\left[\widehat{A},\left[\widehat{A},\widehat{B}\right]\right]\right]+...
\label{S215}
\end{align}

Inserting the results obtained in Eqs. (\ref{S212}), (\ref{S213}), and (\ref{S214}), into  (\ref{S211}), leads to:

\begin{gather}
	\begin{split}
		\nonumber{\color{green}{\color{teal}\partial_{t}\widehat{U}\left(t,0\right)}}= \left\{ \left(\partial_{t}c_{1}(t)\widehat{J}_{+}\right)+\left(\partial_{t}c_{2}(t)\widehat{J}_{0}-\partial_{t}c_{2}(t)c_{1}(t)\widehat{J}_{+}\right)+\right.
	\end{split}\\
	\begin{split}
		\nonumber\quad\left.\exp\left(-c_{2}(t)\right)\left(\partial_{t}c_{3}(t)\widehat{J}_{-}+2c_{1}(t)\partial_{t}c_{3}(t)\widehat{J}_{0}-c_{1}^{2}(t)\partial_{t}c_{3}(t)\widehat{J}_{+}\right)\right\} {\color{green}{\color{teal}\widehat{U}\left(t,0\right)}}
	\end{split}\\
	\begin{split}
	\nonumber= \left\{ \left(\partial_{t}c_{1}(t)-\partial_{t}c_{2}(t)c_{1}(t)-\exp\left(-c_{2}(t)\right)c_{1}^{2}(t)\partial_{t}c_{3}(t)\right)\widehat{J}_{+}\right.
	\end{split}\\
	\begin{split}
	\quad \left.+\left(\partial_{t}c_{2}(t)+2\exp\left(-c_{2}(t)\right)c_{1}(t)\partial_{t}c_{3}(t)\right)\widehat{J}_{0}+\left(\exp\left(-c_{2}(t)\right)\partial_{t}c_{3}(t)\right)\widehat{J}_{-}\right\} {\color{green}{\color{teal}\widehat{U}\left(t,0\right)}}.
	\end{split}
	\label{S216}
\end{gather}

This is the formal (non-elaborated) demonstration of the Eq. (2.5) presented by Cheng. Using  (\ref{S211}) within the Schrödinger equation, and comparing the two sides, we find three ordinary differential equations:
\begin{align}
	i\hbar\left(\partial_{t}c_{1}(t)-\partial_{t}c_{2}(t)c_{1}(t)-\exp\left(-c_{2}(t)\right)c_{1}^{2}(t)\partial_{t}c_{3}(t)\right)=a_{1}(t),
	\label{S217}
\end{align}
\begin{align}
	i\hbar\left(\partial_{t}c_{2}(t)+2\exp\left(-c_{2}(t)\right)c_{1}(t)\partial_{t}c_{3}(t)\right)=a_{2}(t),
	\label{S218}
\end{align}
\begin{align}
	i\hbar\left(\exp\left(-c_{2}(t)\right)\partial_{t}c_{3}(t)\right)=a_{3}(t).
	\label{S219}
\end{align}

From these expressions we finally obtain:

\begin{align}
	\partial_{t}c_{1}(t)=\frac{1}{i\hbar}\left[a_{1}(t)+a_{2}(t)c_{1}(t)-a_{3}(t)c_{1}^{2}(t)\right],
	\label{SM220}
\end{align}

\begin{align}
	\partial_{t}c_{2}(t)=\frac{1}{i\hbar}\left[a_{2}(t)-2a_{3}(t)c_{1}(t)\right],
	\label{SM221}
\end{align}

\begin{align}
	\partial_{t}c_{3}(t)=\frac{1}{i\hbar}\left[a_{3}(t)\exp\left(c_{2}(t)\right)\right].
	\label{SM222}
\end{align}

\part{B. Creation and annihilation operators}
\section{B1. Approaching the equations of motion}
We start with the quantum promoted Hamiltonian of a generalized parametric oscillator:
\begin{align}
	\mathcal{\widehat{H}}\left(t\right)=\frac{\widehat{p}^{2}\left(t\right)}{2m\left(t\right)}+\frac{1}{2}\,m\left(t\right)\omega^{2}\left(t\right)\widehat{q}^{2}\left(t\right).
	\label{SM223}
\end{align}
By imposing the following commutation relation:
\begin{align}
	\left[\widehat{q}\left(t\right),\widehat{p}\left(t\right)\right]=i\hslash,
	\label{SM224}
\end{align}
it can be derived that,
\begin{align}
	\left[\widehat{q}\left(t\right),\widehat{p}^{2}\left(t\right)\right]=i\hslash\,2\widehat{p}(t),
	\label{SM225}
\end{align}
\begin{align}
	\left[\widehat{p}\left(t\right),\widehat{q}^{2}\left(t\right)\right]=-i\hslash\,2\widehat{q}(t).
	\label{SM226}
\end{align}

Heisenberg equations of motion dictates the dynamics of a given quantum operator:

\begin{align}
	\frac{d}{dt}\widehat{O}\left(t\right)=\frac{1}{i\hslash}\,\left[\widehat{O}\left(t\right),\widehat{H}\left(t\right)\right].
	\label{SM227}
\end{align}
Specifically, for the position and momentum operators:

\begin{align}
	\frac{d\widehat{q}\left(t\right)}{dt}=\frac{\widehat{p}\left(t\right)}{m\left(t\right)},
	\label{SM228}
\end{align}
\begin{align}
	\frac{d\widehat{p}\left(t\right)}{dt}=-m\left(t\right)\omega^{2}\left(t\right)\widehat{q}\left(t\right).
	\label{SM229}
\end{align}

We define the most general form for the creation and annihilation quantum operators as linear combination of the above position and momentum operators:

\begin{align}
	\widehat{a}\left(t\right)=A\left(t\right)\,\widehat{q}\left(t\right)+i\,B\left(t\right)\,\widehat{p}\left(t\right),
	\label{SM230}
\end{align}
\begin{align}
	\widehat{a}^{\dagger}\left(t\right)=A\left(t\right)\,\widehat{q}\left(t\right)-i\,B\left(t\right)\,\widehat{p}\left(t\right).
	\label{SM231}
\end{align}

The canonical bosonic commutation relation will impose on the functions $A\left(t\right)$ and $B\left(t\right)$ the following condition:
\begin{align}
	\left[\widehat{a}\left(t\right),\widehat{a}^{\dagger}\left(t\right)\right]=1=2\hslash\,A\left(t\right)B\left(t\right)\rightarrow A\left(t\right)B\left(t\right)=\frac{1}{2\hslash}.
	\label{SM232}
\end{align}

By employing this set of useful commutation relations:
\begin{align}
	\left[\widehat{a}\left(t\right),\widehat{q}\left(t\right)\right]=\hslash\,B\left(t\right),
	\label{SM233}
\end{align}
\begin{align}
	\left[\widehat{a}\left(t\right),\widehat{q}^{2}\left(t\right)\right]=\hslash\,2B\left(t\right)\,\widehat{q}(t),
	\label{SM234}
\end{align}
\begin{align}
	\left[\widehat{a}\left(t\right),\widehat{p}\left(t\right)\right]=i\hslash\,A\left(t\right),
	\label{SM235}
\end{align}
\begin{align}
	\left[\widehat{a}\left(t\right),\widehat{p}^{2}\left(t\right)\right]=i\hslash\,2A\left(t\right)\,\widehat{p}(t),
	\label{SM236}
\end{align}

the equation of motion for the annihilation operator can be written as,
\begin{align}
	\frac{d\widehat{a}\left(t\right)}{dt}=\frac{1}{i\hslash}\,\left[\widehat{a}\left(t\right),\widehat{H}\left(t\right)\right]=\frac{A\left(t\right)}{m\left(t\right)}\,\widehat{p}\left(t\right)-i\,m\left(t\right)\omega^{2}\left(t\right)B\left(t\right)\,\widehat{q}\left(t\right).
	\label{SM237}
\end{align}

We can also independently derive the equation of motion as
\begin{align}
\nonumber\frac{d\widehat{a}\left(t\right)}{dt}=\frac{d}{dt}\,\left\{ A\left(t\right)\,\widehat{q}\left(t\right)+i\,B\left(t\right)\,\widehat{p}\left(t\right)\right\} &\\\nonumber=\frac{dA\left(t\right)}{dt}\,\widehat{q}\left(t\right)+A\left(t\right)\,\frac{d\widehat{q}\left(t\right)}{dt}+i\frac{dB\left(t\right)}{dt}\,\widehat{p}\left(t\right)+i\,B\left(t\right)\,\frac{d\widehat{p}\left(t\right)}{dt}&\\=\frac{dA\left(t\right)}{dt}\,\widehat{q}\left(t\right)+i\frac{dB\left(t\right)}{dt}\,\widehat{p}\left(t\right)+\,\frac{A\left(t\right)}{m\left(t\right)}\widehat{p}\left(t\right)-i\,m\left(t\right)\omega^{2}\left(t\right)B\left(t\right)\widehat{q}\left(t\right).
	\label{SM238}
\end{align}
So, one option to satisfy both equations of motion is,
\begin{align}
	\boxed{\frac{dA\left(t\right)}{dt}=\frac{dB\left(t\right)}{dt}=0.}
	\label{SM239}
\end{align}
Thus,
\begin{align}
	\widehat{a}\left(t\right)=A\,\widehat{q}\left(t\right)+i\,B\,\widehat{p}\left(t\right),
	\label{SM240}
\end{align}
\begin{align}
	\widehat{a}^{\dagger}\left(t\right)=A\,\widehat{q}\left(t\right)-i\,B\,\widehat{p}\left(t\right).
	\label{SM241}
\end{align}
And the inverse relations are,
\begin{align}
	\widehat{q}\left(t\right)=\frac{1}{2A}\,\left(\widehat{a}^{\dagger}\left(t\right)+\widehat{a}\left(t\right)\right),
	\label{SM242}
\end{align}
\begin{align}
	\widehat{p}\left(t\right)=\frac{i}{2B}\,\left(\widehat{a}^{\dagger}\left(t\right)-\widehat{a}\left(t\right)\right).
	\label{SM243}
\end{align}
We can now substitute this into the equation of motion to rewrite it just in terms of creation and annihilation operators:
\begin{gather}
\begin{split}
\nonumber\frac{d\widehat{a}\left(t\right)}{dt}=\frac{A}{m\left(t\right)}\,\widehat{p}\left(t\right)-im\left(t\right)\omega^{2}\left(t\right)B\,\widehat{q}\left(t\right)
\end{split}\\
\begin{split}
=-i\left(\frac{A}{2m\left(t\right)B}+\frac{m\left(t\right)\omega^{2}\left(t\right)B}{2A}\right)\widehat{a}\left(t\right)+i\left(\frac{A}{2m\left(t\right)B}-\frac{m\left(t\right)\omega^{2}\left(t\right)B}{2A}\right)\widehat{a}^{\dagger}\left(t\right),
\end{split}
	\label{SM244}
\end{gather}
which is a equation of motion in the form of a differential Bogoliubov transformation, which can also be compactly written as,
\begin{align}
	\frac{d\widehat{a}\left(t\right)}{dt}=-if(t)\widehat{a}\left(t\right)+ig(t)\widehat{a}^{\dagger}\left(t\right),
	\label{SM245}
\end{align}
and similarly for the creation operator,
\begin{align}
	\frac{d\widehat{a}^{\dagger}\left(t\right)}{dt}=-ig(t)\widehat{a}\left(t\right)+if(t)\widehat{a}^{\dagger}\left(t\right),
	\label{SM246}
\end{align}
so that
\begin{align}
	\frac{d}{dt}\,\left[\begin{array}{c}
		\widehat{a}\left(t\right)\\
		\widehat{a}^{\dagger}\left(t\right)
	\end{array}\right]=i\,\left[\begin{array}{cc}
		-f(t) & g(t)\\
		-g(t) & f(t)
	\end{array}\right]\left[\begin{array}{c}
		\widehat{a}\left(t\right)\\
		\widehat{a}^{\dagger}\left(t\right)
	\end{array}\right].
	\label{SM247}
\end{align}

\section{B2. Electromagnetic case}
Now, we move on to the electromagnetic Hamiltonian of a temporal metamaterial:
\begin{align}
	\mathcal{\widehat{H}}\left(t\right)=\frac{1}{2}\,\int_{0}^{\infty}dk\,\left(\frac{1}{\varepsilon\left(t\right)}\,\left(\widehat{p}_{k}\left(t\right)\,\widehat{p}_{k}^{\dagger}\left(t\right)+\widehat{p}_{k}^{\dagger}\left(t\right)\,\widehat{p}_{k}\left(t\right)\right)+\varepsilon\left(t\right)\omega_{k}^{2}\left(t\right)\,\left(\widehat{q}_{k}\left(t\right)\,\widehat{q}_{k}^{\dagger}\left(t\right)+\widehat{q}_{k}^{\dagger}\left(t\right)\widehat{q}_{k}\left(t\right)\right)\right).
	\label{SM248}
\end{align}
The corresponding commutation relation for the dynamical variable and its conjugated momentum reads as,
\begin{align}
	\left[\widehat{A}_{x}\left(z,t\right),\widehat{\Pi}_{x}^{\dagger}\left(z',t\right)\right]=i\hslash\,\delta\left(z-z'\right)\rightarrow\left[\widehat{q}_{k}\left(t\right),\widehat{p}_{k'}^{\dagger}\left(t\right)\right]=i\hslash\,\delta\left(k-k'\right)=-\left[\widehat{p}_{k}\left(t\right),\widehat{q}_{k'}^{\dagger}\left(t\right)\right],
	\label{SM249}
\end{align}
\begin{align}
\left[\widehat{q}_{k}\left(t\right),\widehat{p}_{k'}\left(t\right)\,\widehat{p}_{k'}^{\dagger}\left(t\right)+\widehat{p}_{k'}^{\dagger}\left(t\right)\,\widehat{p}_{k'}\left(t\right)\right]=i\hslash\,2\widehat{p}_{k}\left(t\right)\delta\left(k-k'\right),
	\label{SM250}
\end{align}
\begin{align}
	\left[\widehat{p}_{k}\left(t\right),\widehat{q}_{k'}\left(t\right)\,\widehat{p}_{k'}^{\dagger}\left(t\right)+\widehat{q}_{k'}^{\dagger}\left(t\right)\,\widehat{q}_{k'}\left(t\right)\right]=-i\hslash\,2\widehat{q}_{k}\left(t\right)\delta\left(k-k'\right).
	\label{SM251}
\end{align}
Notice that all other commutation pairs are zero. Once again, Heisenberg equations of motion lead to,
\begin{align}
	\frac{d\widehat{q}_{k}\left(t\right)}{dt}=\frac{1}{i\hslash}\,\left[\widehat{q}_{k}\left(t\right),\widehat{H}\left(t\right)\right]=\frac{\widehat{p}_{k}\left(t\right)}{\varepsilon\left(t\right)},
	\label{SM252}
\end{align}
\begin{align}
	\frac{d\widehat{p}_{k}\left(t\right)}{dt}=\frac{1}{i\hslash}\,\left[\widehat{p}_{k}\left(t\right),\widehat{H}\left(t\right)\right]=-\varepsilon\left(t\right)\omega_{k}^{2}\left(t\right)\widehat{q}_{k}\left(t\right).
	\label{SM253}
\end{align}
Regarding the creation and annihilation quantum operators, and noticing that they can be defined as linear combinations of certain initial coefficients that evolves in time in terms of position and momentum operators, it follows that:
\begin{align}
	\widehat{a}_{k}\left(t\right)=\sqrt{\frac{1}{2\hslash\varepsilon\left(0\right)\omega_{k}\left(0\right)}}\,\left(\varepsilon\left(0\right)\omega_{k}\left(0\right)\widehat{q}_{k}\left(t\right)+i\,\widehat{p}_{k}\left(t\right)\right),
	\label{SM254}
\end{align}
\begin{align}
	\widehat{a}_{-k}^{\dagger}\left(t\right)=\sqrt{\frac{1}{2\hslash\varepsilon\left(0\right)\omega_{k}\left(0\right)}}\,\left(\varepsilon\left(0\right)\omega_{k}\left(0\right)\widehat{q}_{k}\left(t\right)-i\,\widehat{p}_{k}\left(t\right)\right).
	\label{SM255}
\end{align}
The inverse relations are,
\begin{align}
	\widehat{q}_{k}\left(t\right)=\sqrt{\frac{2\hslash}{\varepsilon\left(0\right)\omega_{k}\left(0\right)}}\,\frac{1}{2}\,\left(\widehat{a}_{-k}^{\dagger}\left(t\right)+\widehat{a}_{k}\left(t\right)\right),
	\label{SM256}
\end{align}
\begin{align}
	\widehat{p}_{k}\left(t\right)=\sqrt{2\hslash\varepsilon\left(0\right)\omega_{k}\left(0\right)}\,\frac{i}{2}\left(\widehat{a}_{-k}^{\dagger}\left(t\right)-\widehat{a}_{k}\left(t\right)\right).
	\label{SM257}
\end{align}
The equation of motion of the annihilation operator is then given by 
\begin{align}
\nonumber\frac{d\widehat{a}_{k}\left(t\right)}{dt}=\frac{1}{i\hslash}\left[\widehat{a}_{k}\left(t\right),\widehat{H}\left(t\right)\right]=\sqrt{\frac{1}{2\hslash\varepsilon\left(0\right)\omega_{k}\left(0\right)}}\,\left(\frac{\varepsilon\left(0\right)\omega_{k}\left(0\right)}{\varepsilon\left(t\right)}\,\widehat{p}_{k}\left(t\right)-i\varepsilon\left(t\right)\omega_{k}^{2}\left(t\right)\widehat{q}_{k}\left(t\right)\right)&\\\nonumber=\frac{i}{2}\left(\frac{\varepsilon\left(0\right)\omega_{k}\left(0\right)}{\varepsilon\left(t\right)}\left(\widehat{a}_{-k}^{\dagger}\left(t\right)-\widehat{a}_{k}\left(t\right)\right)-\frac{\varepsilon\left(t\right)\omega_{k}^{2}\left(t\right)}{\varepsilon\left(0\right)\omega_{k}\left(0\right)}\left(\widehat{a}_{-k}^{\dagger}\left(t\right)+\widehat{a}_{k}\left(t\right)\right)\right)&\\-\frac{i}{2}\left(\frac{\varepsilon\left(0\right)\omega_{k}\left(0\right)}{\varepsilon\left(t\right)}+\frac{\varepsilon\left(t\right)\omega_{k}^{2}\left(t\right)}{\varepsilon\left(0\right)\omega_{k}\left(0\right)}\right)\widehat{a}_{k}\left(t\right)+\frac{i}{2}\left(\frac{\varepsilon\left(0\right)\omega_{k}\left(0\right)}{\varepsilon\left(t\right)}-\frac{\varepsilon\left(t\right)\omega_{k}^{2}\left(t\right)}{\varepsilon\left(0\right)\omega_{k}\left(0\right)}\right)\widehat{a}_{-k}^{\dagger}\left(t\right).
	\label{SM258}
\end{align}
Introducing the Hamiltonian and, through Heisenberg equations, the bosonic commutation relations, the preservation of those equations of motion leads to a system of coupled differential equations with time-varying coefficients, which can be compactly written in matrix form: 
\begin{align}
	\frac{d}{dt}\,\left[\begin{array}{c}
		\widehat{a}_{k}\left(t\right)\\
		\widehat{a}_{-k}^{\dagger}\left(t\right)
	\end{array}\right]=i\left[\begin{array}{cc}
		-f_{k}(t) & g_{k}(t)\\
		-g_{k}(t) & f_{k}(t)
	\end{array}\right]\left[\begin{array}{c}
		\widehat{a}_{k}\left(t\right)\\
		\widehat{a}_{-k}^{\dagger}\left(t\right)
	\end{array}\right],
	\label{SM259}
\end{align}
where the time-dependent coefficients are:
\begin{align}
	f_{k}\left(t\right)=\frac{1}{2}\left(\frac{\varepsilon\left(0\right)\omega_{k}\left(0\right)}{\varepsilon\left(t\right)}+\frac{\varepsilon\left(t\right)\omega_{k}^{2}\left(t\right)}{\varepsilon\left(0\right)\omega_{k}\left(0\right)}\right),
	\label{SM260}
\end{align}
\begin{align}
	g_{k}\left(t\right)=\frac{1}{2}\left(\frac{\varepsilon\left(0\right)\omega_{k}\left(0\right)}{\varepsilon\left(t\right)}-\frac{\varepsilon\left(t\right)\omega_{k}^{2}\left(t\right)}{\varepsilon\left(0\right)\omega_{k}\left(0\right)}\right).
	\label{SM261}
\end{align}
Since such a system of coupled differential equations correspond to a non-autonomous system $\left(dX/dt=f(X(t),t)\right)$, in which the law governing the evolution of the system does not depend solely on the system's current state but also on time itself, the eigen-decomposition might not be the best option to solve it, since both eigenvectors and eigenvalues vary in time. 
\section{B3. Frequency shifts}
As a brief double-check remark, we note that the coefficients given in (\ref{SM260}) and (\ref{SM261}) are related with prefactors obtained in the Hamiltonian (\ref{SM48}), 
\begin{align*}
\mathcal{\widehat{H}}\left(t\right)=\frac{1}{2}\,\int dk\,\left(\frac{1}{\varepsilon\left(t\right)}\,\left(\widehat{p}_{k}\left(t\right)\widehat{p}_{k}^{\dagger}\left(t\right)+\widehat{p}_{k}^{\dagger}\left(t\right)\widehat{p}_{k}\left(t\right)\right)+\frac{k^{2}c^{2}}{\mu\left(t\right)}\,\left(\widehat{q}_{k}\left(t\right)\widehat{q}_{k}^{\dagger}\left(t\right)+\widehat{q}_{k}^{\dagger}\left(t\right)\widehat{q}_{k}\left(t\right)\right)\right)
\end{align*}
using (\ref{SM254}) and (\ref{SM255}), and $\left[\widehat{a}_{k}\left(t\right),\widehat{a}_{-k}\left(t\right)\right]=\left[\widehat{a}_{-k}\left(t\right),\widehat{a}_{-k}^{\dagger}\left(t\right)\right]=0$, the Hamiltonian for a temporal metamaterial is finally written in terms of the creation and annihilation quantum operators:
\begin{align}
	\mathcal{\widehat{H}}\left(t\right)=\frac{\hbar}{2}\int dk\,\left[\omega_{k}(t)+\Omega_{k}(t)\right]\left(\widehat{a}_{k}\left(t\right)\widehat{a}_{k}^{\dagger}\left(t\right)+\widehat{a}_{k}^{\dagger}\left(t\right)\widehat{a}_{k}\left(t\right)\right)+\frac{\hbar}{2}\int dk\,\xi_{k}(t)\left(\widehat{a}_{k}\left(t\right)\widehat{a}_{-k}\left(t\right)+\widehat{a}_{-k}^{\dagger}\left(t\right)\widehat{a}_{k}^{\dagger}\left(t\right)\right),
	\label{SM262}
\end{align}
where,
\begin{align}
	\omega_{k}(t)=\frac{kc}{\sqrt{\varepsilon(t)\mu(t)}},
	\label{SM263}
\end{align}
\begin{align}
	\Omega_{k}(t)=\left[\frac{1}{2}\left(\sqrt{\frac{\mu(0)\varepsilon(t)}{\varepsilon(0)\mu(t)}}+\sqrt{\frac{\varepsilon(0)\mu(t)}{\mu(0)\varepsilon(t)}}\right)-1\right]\omega_{k}(t),
	\label{SM264}
\end{align}
\begin{align}
	\xi_{k}(t)=\frac{1}{2}\left(\sqrt{\frac{\mu(0)\varepsilon(t)}{\varepsilon(0)\mu(t)}}-\sqrt{\frac{\varepsilon(0)\mu(t)}{\mu(0)\varepsilon(t)}}\right)\omega_{k}(t).
	\label{SM265}
\end{align}
Performing the substitution $m(t)\rightarrow\varepsilon(t)$ and $\omega_{k}(t)=kc/\sqrt{\varepsilon(t)\mu(t)}$, equations (\ref{SM260}) and (\ref{SM261}) can be written as follows,
\begin{align}
	f_{k}\left(t\right)=\frac{1}{2}\left(\frac{\varepsilon\left(0\right)\mu\left(t\right)+\varepsilon\left(t\right)\mu\left(0\right)}{\sqrt{\varepsilon\left(0\right)\mu\left(0\right)}}\right)\frac{kc}{\varepsilon\left(t\right)\mu\left(t\right)},
	\label{SM266}
\end{align}
\begin{align}
	g_{k}\left(t\right)=\frac{1}{2}\left(\frac{\varepsilon\left(0\right)\mu\left(t\right)-\varepsilon\left(t\right)\mu\left(0\right)}{\sqrt{\varepsilon\left(0\right)\mu\left(0\right)}}\right)\frac{kc}{\varepsilon\left(t\right)\mu\left(t\right)},
	\label{SM267}
\end{align}
so that they yield the following relations:
\begin{align}
	\omega_{k}(t)+\Omega_{k}(t)=\frac{1}{2}\left(\frac{\varepsilon(0)\mu(t)+\varepsilon(t)\mu(0)}{\sqrt{\varepsilon(0)\mu(0)}}\right)\frac{kc}{\varepsilon(t)\mu(t)}=f_{k}\left(t\right),
	\label{SM268}
\end{align}
\begin{align}
	\xi_{k}(t)=\frac{1}{2}\left(\frac{\varepsilon(t)\mu(0)-\varepsilon(0)\mu(t)}{\sqrt{\varepsilon(0)\mu(0)}}\right)\frac{kc}{\varepsilon(t)\mu(t)}=-g_{k}(t).
	\label{SM269}
\end{align}
Noteworthily, we note that the sum of time-dependent frequency $\omega_{k}(t)$ and time-matching $\Omega_{k}(t)$ expressed in (\ref{SM268}) takes always positive values for $\varepsilon(t)>1$ and $\mu(t)>1$,
\begin{align}
	f_{k}\left(t\right)=\frac{1}{2}\left(\frac{\varepsilon(0)}{\varepsilon(t)}+\frac{\mu(0)}{\mu(t)}\right)\frac{kc}{\sqrt{\varepsilon(0)\mu(0)}},
	\label{SM270}
\end{align}
so $0<f_{k}\left(t\right)\leq\omega_{k}\left(0\right)$. On the other hand, the squeezing time-modulation matching in (\ref{SM269}),
\begin{align}
	g_{k}(t)=\frac{1}{2}\left(\frac{\varepsilon(0)}{\varepsilon(t)}-\frac{\mu(0)}{\mu(t)}\right)\frac{kc}{\sqrt{\varepsilon(0)\mu(0)}},
	\label{SM271}
\end{align}
will be positive for all those time instants where impedance is greater than its initial value ($Z(t)>Z(0)$), negative when it is lesser than the initial value ($Z(t)<Z(0)$), and zero when it matches the initial value ($Z(t)=Z(0)$).
\newpage
\part{C. Delta properties}
From the fundamental property:
\begin{align}
	\int_{-\infty}^{+\infty}dt\thinspace f(t)\thinspace\delta(t-t_{0})\equiv f(t_{0}),
	\label{SM272}
\end{align}
the equation that defines derivatives of the delta function is,
\begin{align}
	\int_{-\infty}^{+\infty}dt\thinspace f(t)\thinspace\delta^{(n)}(t-t_{0})\equiv-\int_{-\infty}^{+\infty}dt\thinspace\frac{\partial f(t)}{\partial t}\thinspace\delta^{(n-1)}(t-t_{0}),
	\label{SM273}
\end{align}
where (n) indicates the order of the derivative. Letting $f(t)=t\thinspace g(t)$, it follows that,
\begin{gather}
\begin{split}
\nonumber\int_{-\infty}^{+\infty}dt\thinspace\left[t\thinspace g(t)\right]\thinspace\delta^{(n)}(t-t_{0})\equiv-\int_{-\infty}^{+\infty}dt\thinspace\frac{\partial\left[t\thinspace g(t)\right]}{\partial t}\thinspace\delta^{(n-1)}(t-t_{0})
\end{split}\\
\begin{split}
	\nonumber=-\int_{-\infty}^{+\infty}dt\thinspace\left[g(t)+t\thinspace\frac{\partial g(t)}{\partial t}\right]\thinspace\delta^{(n-1)}(t-t_{0})
\end{split}\\
\begin{split}
	=-\left[\int_{-\infty}^{+\infty}dt\thinspace g(t)\thinspace\delta^{(n-1)}(t-t_{0})+\int_{-\infty}^{+\infty}dt\thinspace t\thinspace\frac{\partial g(t)}{\partial t}\thinspace\delta^{(n-1)}(t-t_{0})\right].
\end{split}
\label{S274}
\end{gather}

Hence, first order derivatives, i.e., for n=1, are given by,
\begin{gather}
\begin{split}
\nonumber\int_{-\infty}^{+\infty}dt\thinspace\left[t\thinspace g(t)\right]\thinspace\frac{\partial}{\partial t}\delta(t-t_{0})\equiv-\left[\int_{-\infty}^{+\infty}dt\thinspace g(t)\thinspace\delta(t-t_{0})+\cancelto{0}{\int_{-\infty}^{+\infty}dt\thinspace t\thinspace\frac{\partial g(t)}{\partial t}\thinspace\delta(t-t_{0})}\right]
\end{split}\\
\begin{split}
\int_{-\infty}^{+\infty}dt\thinspace\left[t\thinspace g(t)\right]\thinspace\frac{\partial}{\partial t}\delta(t-t_{0})\equiv-\int_{-\infty}^{+\infty}dt\thinspace g(t)\thinspace\delta(t-t_{0}).
\end{split}
\end{gather}
Thus,
\begin{align}
\thinspace\frac{\partial}{\partial t}\delta(t-t_{0})=-\delta(t-t_{0}).
\end{align}

\end{document}